\documentclass{aa}

\usepackage{graphicx}
\usepackage{txfonts}
\usepackage{natbib}
\bibpunct{(}{)}{;}{a}{}{,} 

\usepackage[bookmarks=false]{hyperref}
\hypersetup{colorlinks=true,urlcolor=blue,citecolor=blue,pdfborder= 0 0 0}

\begin{document}

   \title{Alignment of galaxies relative to their local environment in SDSS-DR8}

   \author{A.~Hirv\inst{\ref{inst1}}
   \and J.~Pelt\inst{\ref{inst1}}
   \and E.~Saar\inst{\ref{inst1},\ref{inst2}}
   \and E.~Tago\inst{\ref{inst1}}
   \and A.~Tamm\inst{\ref{inst1}}
   \and E.~Tempel\inst{\ref{inst1},\ref{inst3}}
   \and M.~Einasto\inst{\ref{inst1}}
   }

   \institute{Tartu Observatory, Observatooriumi 1, 61602 T\~{o}ravere, Estonia\label{inst1}
   \and Estonian Academy of Sciences, EE-10130 Tallinn, Estonia\label{inst2}
   \and Leibniz-Institut f\"ur Astrophysik Potsdam (AIP), An der Sternwarte 16, D-14482 Potsdam, Germany\label{inst3}
   }

   \titlerunning{Alignment of galaxies in SDSS-DR8}

   \date{Received ---; accepted ---}

 
  \abstract
   {}
   {We study the alignment of galaxies relative to their local environment in SDSS-DR8 and, using these data, we discuss evolution scenarios for different types of galaxies.}
   {We defined a vector field of the direction of anisotropy of the local environment of galaxies. We summed the unit direction vectors of all close neighbours of a given galaxy
   in a particular way to estimate this field. We found the alignment angles between the spin axes of disc galaxies, or the minor axes of elliptical galaxies, and the direction of
   anisotropy.
   The distributions of cosines of these angles are compared to the random distributions to analyse the alignment of galaxies.}
   {Sab galaxies show perpendicular alignment relative to the direction of anisotropy in a sparse environment, for single galaxies and galaxies of low luminosity.
   Most of the parallel alignment of Scd galaxies comes from dense regions, from $2 \dots 3$ member groups and from galaxies with low
   luminosity. The perpendicular alignment of S0 galaxies does not depend strongly on environmental density nor luminosity; it is detected
   for single and $2 \dots 3$ member group galaxies, and for main galaxies of $4 \dots 10$ member groups.
   The perpendicular alignment of elliptical galaxies is clearly detected for single galaxies and for members of $\le 10$ member groups;
   the alignment increases with environmental density and luminosity.}
   {We confirm the existence of fossil tidally induced alignment of Sab galaxies at low $z$. The alignment of Scd galaxies can be explained via the infall of matter to filaments.
   S0 galaxies may have encountered relatively massive mergers along the direction of anisotropy. Major mergers along this direction can explain the alignment of elliptical galaxies.
   Less massive, but repeated mergers are possibly responsible for the formation of elliptical galaxies in sparser areas and for less luminous elliptical galaxies.}


   \maketitle
%

\section{Introduction}

Observations show that the large-scale distribution of galaxies is strongly anisotropic.
Galaxy clusters, filaments of field galaxies, and superclusters form the hierarchical cosmic web, in which galaxy systems are separated by almost empty voids
\citep[see e.g.][]{Gregory1978,Joeveer1978,Chincarini1979}.

\citet{Bond1996} explained that the present cosmic web already exists in a pattern of primordial density perturbations and later evolution just sharpens this image. 
\citet{Einasto2011a,Einasto2011b} and \citet{Suhhonenko2011} showed that the skeleton of the cosmic web is visible at very high redshifts ($z = 100$ for a particular model).
Their simulations revealed that the highest density peaks of the cosmic web form where medium- and large-scale density perturbations have maxima, as proposed by \citet{Einasto2005}.
These density peaks correspond to rich galaxy
clusters located in superclusters in the local Universe \citep[][]{Einasto2012}. Small-scale density perturbations are seeds for future galaxies, but they can withstand the cosmic
expansion and grow only in locations where the maxima of larger scale perturbations provide a high enough overall density.
Therefore, galaxies form in a pre-existing large-scale structure (LSS), as already recognised by \citet{Einasto1980}.

There is general agreement that tidal forces due to inhomogeneities in the LSS acting at the epoch of galaxy formation caused the initial alignment of galaxies. According to the linear
alignment model,
dark matter (DM) haloes obtained their elongated shape and orientation as a result of gravitational collapse in the primordial tidal field. This aligned their major axes preferentially
with the direction of the field. And the baryonic component of an elliptical galaxy in equilibrium follows the shape of its DM halo \citep[see][]{Catelan2001,Hirata2004}.
\citet{Camelio2015} showed that tidal stretching at the present epoch cannot generate the observed ellipticities (and alignment) of elliptical galaxies.

According to linear tidal torque theory, disc galaxies and their DM haloes got initial spins from the influence of tidal torques on expanding protogalactic material
\citep[see][]{Peebles1969,Doroshkevich1970,White1984}.
\citet{Catelan2001} demonstrated that the resulting large-scale alignment of projected shapes of disc galaxies is similar to, but weaker than, elliptical galaxies. 
This explains the preferred perpendicular alignment of the spins of disc galaxies with the direction of a filament and spins laying mostly parallel to the surface of a void.
On the other hand, \citet{Porciani2002a,Porciani2002b} found that non-linear processes at later times may significantly change spin directions predicted by linear tidal
torque theory.
\citet{Codis2015b} developed a theory of constrained tidal torques near filaments. Their treatment predicts parallel alignment of spins of DM haloes and a filament in the vicinity of a
saddle point, while perpendicular alignment is favoured apart from the saddle.

Cosmological N-body simulations support the ideas of correlated alignment of DM haloes in filaments and sheets.
\citet{Altay2006} found that minor axes of galaxy-size haloes are dominantly perpendicular to the line connecting a pair of cluster-size haloes if distances up to
$4$~$h^{-1}\mathrm{Mpc}$ from that line were considered. \citet{Brunino2007} showed that spins of haloes of disc galaxies laying on the surface of a void tend to be parallel to that
surface.
\citet{Trowland2013} found a weak perpendicular alignment of spins of DM haloes with directions of filaments at $z \ge 1$, and with massive haloes showing a stronger signal.
However, at $z \sim 0$ spins of less massive haloes showed dominantly parallel alignment, while spins of more massive haloes kept their perpendicular orientation.
\citet{Libeskind2012} reported parallel alignment of DM halo spins and filaments. \citet{Aragon-Calvo2007}, \citet{Codis2012}, and \citet{Libeskind2013}
showed that spins of massive DM haloes are preferentially perpendicular to directions of filaments, while spins of less massive haloes have a dominantly parallel alignment.
The parallel alignment of spins of less massive haloes with filaments was also reported by \citet{Zhang2009}. These authors showed that the preferential spin alignment
relative to the direction to the most massive DM haloes (clusters) changes from parallel to perpendicular in a close neighbourhood of clusters. Results of \citet{Libeskind2014} indicate
that accretion of subhaloes in filaments takes place along the spines of filaments, while more massive haloes show more coherent paths of mergers.
Results of \citet{Codis2012} and \citet{Trowland2013} show bulk flows in filaments towards clusters and infall of matter from walls to filaments. Flows parallel to walls and filaments
were also detected by \citet{Forero-Romero2014}. Therefore, besides the tides, infall of matter from sheets to filaments can explain the parallel alignment of spins with filaments.
The perpendicular alignment of spins of massive haloes can also arise from secondary infall and collisions along filaments, which is a common scenario in simulations.

Cosmological hydrodynamical simulations agree with the N-body results.
\citet{Dubois2014} and \citet{Welker2014} found, for $z > 1.2$, that mergers along the filamentary structure of LSS set the spin axes of 
massive red, early-type galaxies mostly perpendicular to their nearest filament.
Their results indicate that the absence of mergers allows low-mass, blue disc galaxies to keep 
spins parallel to the filament. They reported also that steady accretion of gas can build up the spin of a low-mass disc galaxy and (re)align it to be parallel to the filament.
Parallel alignment of spins of
low-mass galaxies and perpendicular alignment of spins of high-mass galaxies with filaments at $z = 1.2$ was also found by \citet{Codis2015a}.
Simulations by \citet{Hahn2010} reveal that spins of low-mass disc galaxies are initially aligned perpendicular to the direction of a filament. This is consistent with
the tidal torque theory in regions away from a saddle point. These authors reported that the initial spin alignment of low-mass disc galaxies can be clearly detected in low density
regions at $z \gtrsim 0.5$; but this spin alignment vanishes in high density regions and at later epochs, possibly owing to mergers and other non-linear effects. Hahn et al. found that
spins of massive disc galaxies are aligned preferentially
with the direction of a filament. They also proposed that the initial spin orientations of disc galaxies may change to parallel alignment, as the galaxies gain mass through
mergers and gas infall. One should still assume that there must be relatively few significant mergers along the filament to achieve this
kind of alignment. To make things even more complicated, simulations by \citet{Aumer2014} point out that these disc galaxies, which do not have very quiescent formation histories,
can still be photometrically confirmed as late-type disc galaxies. These galaxies may have a high percentage of stars in kinematically non-disc
components, which may remain photometrically undetectable. This means that mergers may have changed the initial spin alignments of late-type disc galaxies by $z \sim 0$.

The importance of mergers in galaxy evolution was already realised by \citet{Toomre1972}, who proposed that elliptical galaxies may form via the merging of spiral galaxies.
Simulations \citep[e.g.][]{Bournaud2005,Hopkins2008a,Hopkins2008b} show that massive elliptical galaxies can be products of major mergers of disc galaxies.
\citet{Bournaud2005} also proposed an alternative scenario with repeated mergers that have higher mass ratios.

Measuring the alignment of galaxies relative to their local environment in the LSS helps to verify galaxy formation
schemes. Moreover, knowledge of intrinsic alignment properties of galaxies can be used to locate filaments of galaxies \citep[][]{Pimbblet2005b,Rong2016}.
Modelling of this alignment is needed in weak-lensing studies to quantify the intrinsic alignment bias \citep[][]{Joachimi2011,Blazek2015,Codis2015a};
see also \citet{Joachimi2015}, \citet{Kiessling2015}, and \citet{Kirk2015} for an excellent overview of galaxy alignments.

Three-dimensional analysis of galaxy spin alignments was already used by \citet{Jaaniste1977}, \citet{Kapranidis1983}, \citet{Flin1986}, \citet{Kashikawa1992}, and \citet{Navarro2004},
who studied orientations of galaxies in the local supercluster.
Spatial studies of spin orientations in much larger volumes were made possible by present galaxy redshift surveys.
\citet{Trujillo2006} analysed the Two Degree Field Galaxy Redshift Survey and SDSS-DR3 and reported that spins of spiral galaxies located on the shells of the largest cosmic
voids are preferentially parallel to the surface of a void.
Using the Tully galaxy catalogue and the tidal field calculated from the Two Mass Redshift Survey, \citet{LeeErdogdu2007} found that spin axes of disc galaxies align with the intermediate
principal axes of the local tidal tensor; i.e. these spins are perpendicular to filaments.
\citet{Jones2010} analysed a sample of edge-on SDSS-DR5 spiral galaxies and found evidence that spins of less massive spirals in areas with
lower environmental density are preferentially perpendicular to the direction of filaments. Using the SDSS-DR8 data \citet{Tempel2013a,Tempel2013b}
confirmed that spins or minor axes of early-type galaxies have a predominantly perpendicular alignment with filaments and spins of bright late-type galaxies show
parallel alignment. Results by \citet{Zhang2015} indicate perpendicular alignment of spins of disc galaxies and filaments.

One can obtain useful information on galaxy alignments even when the inclination angles of galaxies are unknown. We refer here to some recent results.
\citet{Faltenbacher2009} analysed the SDSS-DR6 data and found that apparent major axes of red, luminous galaxies align with the local direction of the LSS up to a projected
distance $\sim 60$~$h^{-1}\mathrm{Mpc}$,
but no alignment was detected for blue galaxies. \citet{Okumura2009} studied luminous red galaxies from SDSS-DR6 and
reported alignment between galaxy pairs up to $\sim 30$~$h^{-1}\mathrm{Mpc}$ scale with a stronger signal from brighter galaxies.
\citet{Wang2009} found, from SDSS-DR4 data, that apparent major axes of central galaxies of groups align with the (projected) direction to the
nearest neighbouring group. \citet{Li2013} studied a sample of massive galaxies in SDSS-DR9 and found that their apparent major axes show alignment with the
large-scale distribution of galaxies up to the $\sim 70$~$h^{-1}\mathrm{Mpc}$ scale, and there is a stronger signal from more massive galaxies.
\citet{Zhang2013} analysed SDSS-DR7 and reported a parallel alignment of apparent major axes of galaxies with the projected direction of filaments, where bright, red
main galaxies of groups show the strongest signal and blue satellites show no signal.
\citet{Libeskind2015} used the Cosmicflows-2 data and confirmed that satellites of the Local Group and Centaurus~A are mostly in planes that are parallel with the
surface of the Local Void.

To measure the alignment of galaxy spins with respect to the cosmic web, one may expect that we need to carefully locate and model that structure first and then estimate its direction.
Various methods for finding filaments of galaxies from observed or simulated data are listed (with references) by
\citet{Pimbblet2005a}, \citet{Zhang2009}, \citet{Cautun2013}, and \citet{Tempel2014b}. Another way is to study the alignment with the principal axes of the tidal or velocity shear tensor
\citep[see e.g.][]{LeeErdogdu2007,Libeskind2014}.

We choose a different approach. We define and estimate only the direction of anisotropy of the local environment (DALE) without any careful modelling of the filamentary network.
We develop and apply an algorithm for estimating DALE, which makes use of the fact that the spatial distribution of galaxies is anisotropic up to a considerable scale, and cosmic
structures of galaxies are usually locally elongated in some direction. For a straight isolated filament, DALE is equivalent to
the direction of the spine. In principle, our approach is similar to the alignment correlation function method developed by \citet{Faltenbacher2009} and \citet{Li2013}.

We do not need the classification of the cosmic web into filaments, clusters, and voids for a general galaxy alignment analysis.
However, to get more detailed results, we use the data of galaxy groups and clusters of the SDSS data release~8.

The previous observational evidence of galaxy alignments, along with different evolution scenarios of different types of galaxies as described above, clearly needs to be confirmed
and clarified.
Our paper is a step in this direction; we estimate and interpret the alignment properties of different morphological types of galaxies and try to constrain their possible evolution
scenarios.

The paper is organised as follows: a brief description of the data used is given in Sect.~\ref{data_desc}. We describe our method, related problems, and possible
modifications in Sect.~\ref{method_desc}. In Sect.~\ref{results_sect} we present the results of alignment analysis for different samples of galaxies selected by morphological type,
group richness, luminosity, and environmental density. We start the discussion in Sect.~\ref{disc_sect} with some technical considerations about our method, and continue with a
comparison of our results and other observational and simulation studies. In Sect.~\ref{concl_sect} we list the results, give their interpretation, and discuss future prospects.
In Appendix~\ref{Bisous_DALE_sect} DALE is compared to directions of filaments extracted with the Bisous model \citep[see][]{Stoica2005,Tempel2016,Tempel2014b}.


\section{Data}\label{data_desc}

We use the catalogue of groups and clusters of galaxies compiled by \citet{Tempel2012} from the SDSS data release~8 \citep[][]{Aihara2011}.
The catalogue assumes the standard cosmological parameters: the Hubble constant $H_{0} = 100\,h\,\mathrm{km\,s^{-1}Mpc^{-1}}$, matter density $\Omega_{\mathrm{m}} = 0.27$, and
dark energy density $\Omega_{\Lambda} = 0.73$. This magnitude limited sample contains $576493$
galaxies with the Petrosian $r$-band  magnitude $m \le 17.77$; the observed redshift $z = 0.0078 \dots 0.2$. There are $77858$ groups with richness $N_{\mathrm{rich}} = 2 \dots 878$.
A modified friends-of-friends method with a variable linking length has been used to compose the catalogue of groups \citep[see for details][]{Tago2008,Tago2010}.
We exploit the value of the normalised
luminosity density field ($D$) smoothed for $4$~$h^{-1}\mathrm{Mpc}$ scale as a measure of the environmental density for a galaxy. In the catalogue, $D$ has been normalised by
$1.659 \cdot 10^{8}\,h L_{\mathrm{\odot}} \mathrm{Mpc}^{-3}$, which is the average smoothed luminosity density for the whole field;
see \citet{Liivamagi2012} and \citet{Tempel2014c} for details of calculation of luminosity density.
To describe the brightness of a galaxy, we use its $r$-band luminosity ($L$) in the units of $10^{10}\,h^{-2}L_{\mathrm{\odot}}$.
Co-moving distances corrected for the finger-of-God effect, right ascension, and declination are adopted for positions of galaxies.
To exclude galaxies in the border area of the survey, we use the co-moving distances from the survey mask.
We define a main galaxy as the most luminous galaxy in a group. Other galaxies in the group are referred to as satellites.

Orientations of spin axes for $569266$ galaxies are calculated using the position and inclination angles modelled by \citet{Tempel2013a,Tempel2015a}.
Two different morphological classifications of galaxies are used. First, the classification into early and late types based on \citet{Tempel2012}. Second, the classification into
spiral~-- Sab and Scd, lenticular~-- S0, and elliptical~-- E types based on \citet{Huertas-Company2011}.
Sizes of subsamples of galaxies analysed in the present work are given in Tables~\ref{sample_sizes} and~\ref{sample_sizes_ny70}.

\section{Methods}\label{method_desc}

\subsection{Directions of galaxy spins}

The orientation of a galaxy can be described by the unit direction vector of spin or minor axis.
We compute the spin vectors using a standard procedure presented in detail by \citet{LeeErdogdu2007}. We give here a short description of this procedure.

First, we compute the components $\hat{L}_{\mathrm{x}}, \hat{L}_{\mathrm{y}}, \hat{L}_{\mathrm{z}}$ of the unit spin vector of a galaxy in its local Cartesian coordinate system,
with the origin at the position of the galaxy, the
$\mathrm{y}$-axis towards the north celestial pole, and the $\mathrm{z}$-axis in the direction of line-of-sight, using the \citet{Tempel2013a,Tempel2015a} inclination and position angles.

Second, the unknown direction of rotation of a galaxy and the true sign of the spin vector do not affect our analysis. But as we do not know which edge of a galaxy is
closer to us, we can fix the direction of the spin axis up to the sign ambiguity of $\hat{L}_{\mathrm{z}}$. This gives us two equally possible unit spin vectors for every galaxy
and weakens the measured intrinsic alignment signal in most cases.

Third, we convert the local coordinates of the two unit spin vectors into a geocentric equatorial rectangular coordinate system. One can find the appropriate conversion equations
from \citet{LeeErdogdu2007} and \citet{OGP2011}.

If the galaxy is face- or edge-on, we still calculate two unit spin vectors, although these give an identical unsigned direction in this case. The procedure of the spin vector
calculation is also applied to elliptical galaxies, where the orientation data is coming mainly from the position angles of minor axes. The uncertainty of the estimated inclination angles
is strongest for E galaxies but it weakens the measured intrinsic alignment signal for all morphological types.

\subsection{Direction of anisotropy of local environment (DALE)}

\subsubsection{Definition}\label{dale_def}

\begin{figure}
\centering
\includegraphics[width=\hsize,clip]{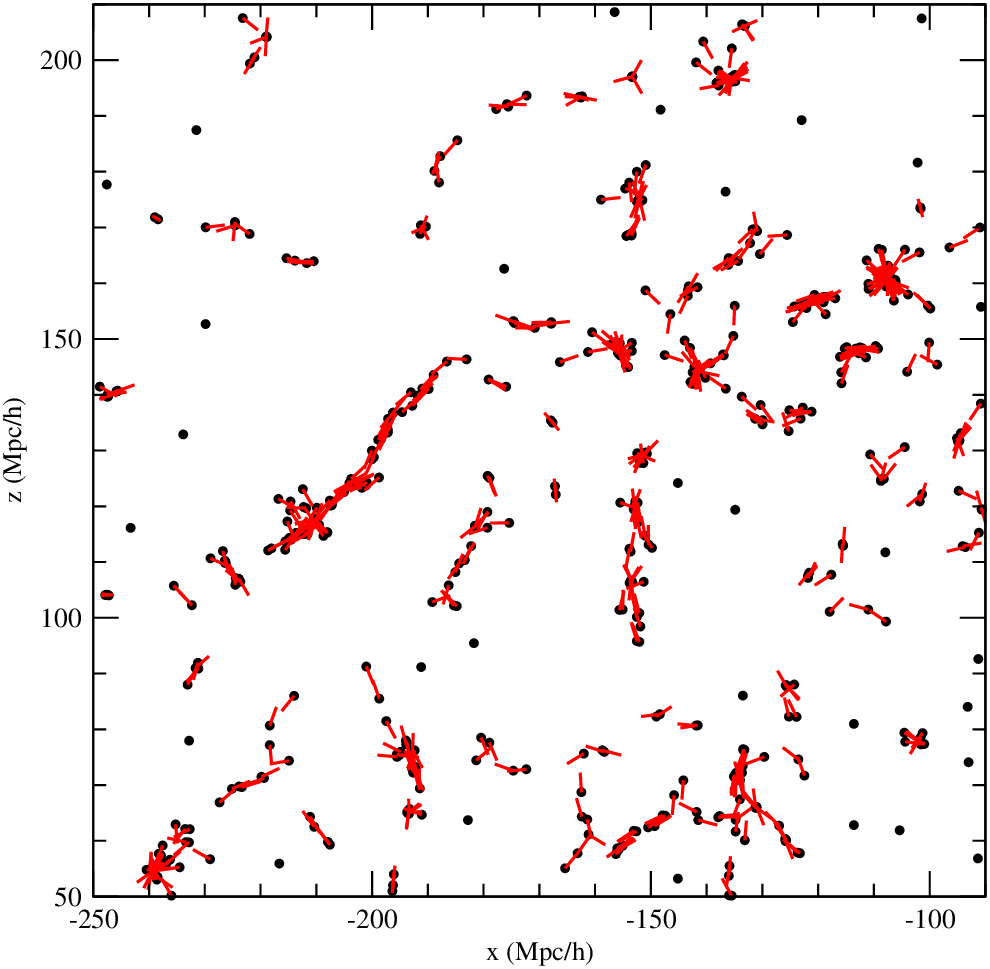}
\caption{Two-dimensional projection of a $\sim 2.1$~$h^{-1}\mathrm{Mpc}$ deep test field for DALE computation. The red line segments show the local direction of anisotropy calculated
in the positions of galaxies, shown with black dots. There are $662$ galaxies; the neighbourhood radius $\nu = 7.0$~$h^{-1}\mathrm{Mpc}$. The signs of direction vectors are
random, which results from the definition of our method. Several directions in a single point are due to overlapping galaxies in projection.}
\label{test_picture}
\end{figure}

We assume that the spatial distribution of galaxies may be locally anisotropic and galaxies may form elongated cosmic structures. The orientation of an elongated structure at the
location of a given galaxy can be described by DALE, which is defined as follows:
\begin{enumerate}
\item We fix a single parameter, the maximum radius $\nu$, which we use to search for neighbouring galaxies for the given galaxy.
\item We find the unit direction vectors from the position of the given galaxy to all its neighbours within the $\nu$ radius. 
\item\label{sum_def} We next sum these normalised vectors in a random order taking the angle $\beta$ between the running sum and the vector 
to be added into account. This angle is in the range $0\degr \le \beta \le 180\degr$. If $\beta > 90\degr$, we add the opposite unit vector instead of the original vector.
\item\label{norm_def} The final sum is normalised to unity and can be used as a vector of DALE.
\end{enumerate}
We estimate DALE in the position of a galaxy, if there is at least one neighbour in the $\nu$ radius.
To get an idea of how our method works, a test field with estimated DALE vectors is shown in Fig.~\ref{test_picture}. We selected a $\sim 2.1$~$h^{-1}\mathrm{Mpc}$ thick
slice from the galaxy catalogue and applied the above defined algorithm using $\nu = 7.0$~$h^{-1}\mathrm{Mpc}$. Every galaxy with at least one neighbour has an estimate of DALE.
There may be several direction estimates in a single point due to overlapping galaxies in projection. The signs of DALE vectors are random, but this does not matter in our analysis of
galaxy alignment (see Sect.~\ref{emagl}).

\subsubsection{A closer look at the computation of DALE}

In addition to the sign, an unsigned direction of a DALE vector may also depend on the order in which we choose the unit vectors for summing. This may happen if the number of neighbours
for a given galaxy is $3$ or greater. In order to suppress shot noise, we calculate an average estimate of DALE.
We repeat the computation of DALE, selecting the neighbours of a galaxy in random order many times. We calculate the normalised sum of the resulting DALE vectors as described in
Sect.~\ref{dale_def},~points~\ref{sum_def} and~\ref{norm_def}.
The upper limit for repetitions is set to $1000$,
but we may stop the loop if the number of repetitions reaches the factorial of the number of neighbours for a given galaxy.
Using the factorial here may also cause skipping of some permutations, as the neighbours are selected randomly. This drawback is eliminated by the next step of our algorithm.

\begin{figure}
\centering
\includegraphics[width=0.49\hsize,clip]{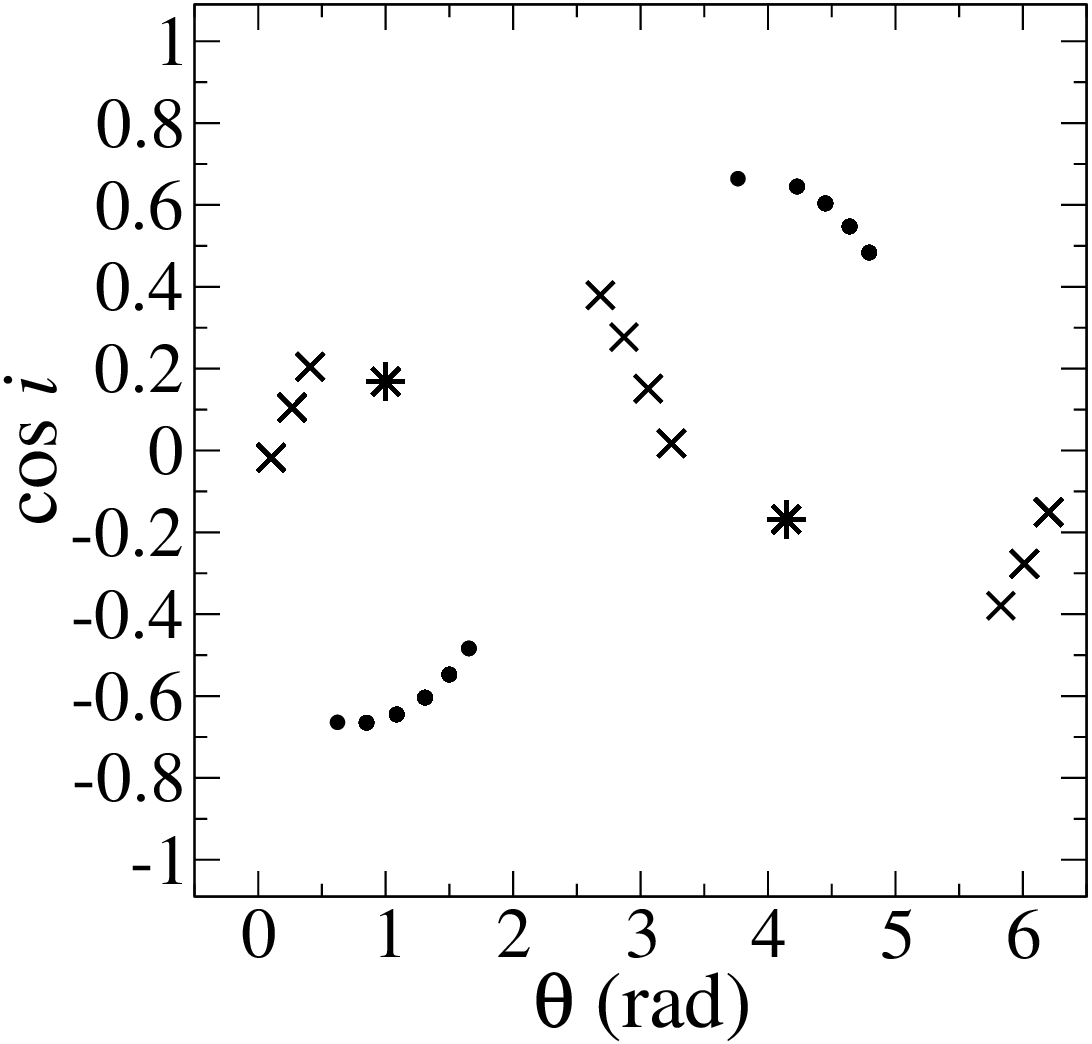}
\includegraphics[width=0.49\hsize,clip]{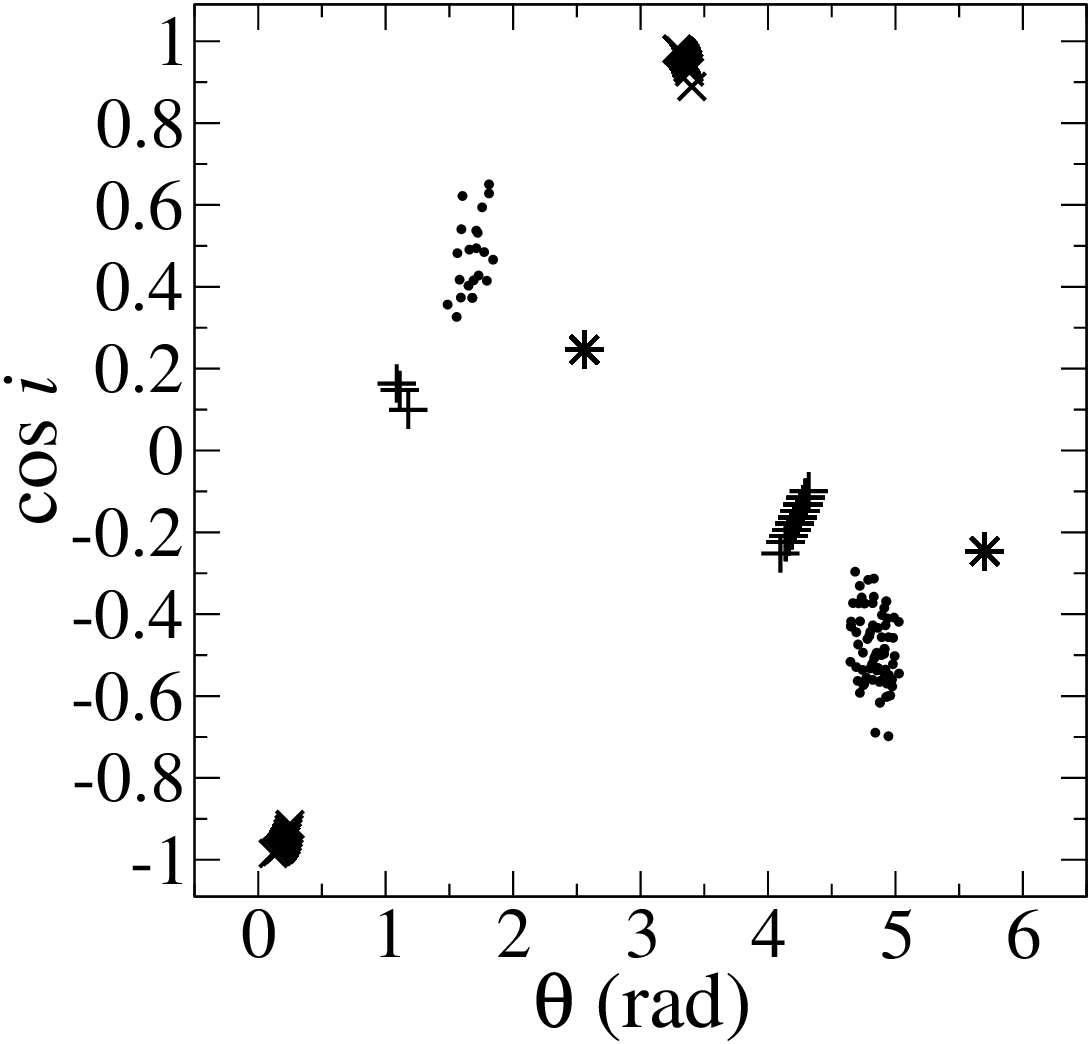}
\caption{Some examples of the distribution of average DALE vectors. The vectors are shown with different symbols for $3$ different galaxies with $3$ neighbours in the left panel
and for $4$ different galaxies with $4$ neighbours in the right panel. Here $i$ is the polar angle and $\theta$ is the azimuthal angle of a normalised vector in spherical coordinates;
$\nu = 7.0$~$h^{-1}\mathrm{Mpc}$, $n = 100$.}
\label{3_4_hajumine}
\end{figure}

We can run the whole procedure of randomly choosing neighbours and averaging over the DALE vectors several ($n$) times.
One can see that the unsigned average DALE is definitely fixed for a given point if all accounted neighbours are on a straight line.
Things are different if the distribution of selected neighbours is isotropic; there is no direction of anisotropy and subsequent estimates of average DALE may be totally different.
An average DALE may get very different directions even for a galaxy with three neighbours.
Some examples of the distribution of
average DALE vectors are shown in Fig.~\ref{3_4_hajumine}. As the signs of the vectors are arbitrary, there can be two equivalent concentrations of directions for every galaxy.

Measuring and registering the accuracy of estimated directions is not our aim, but the fact that the average DALE for a galaxy may get different values must be taken into account.
We use $n$ average DALE vectors for a given point simultaneously in the analysis of mutual alignment of galaxies and DALE.
We describe in Sect.~\ref{emagl} how this approach automatically eliminates the effects due to unstable direction estimates.
In the case of our data, the variability of the results due to different runs of averaged DALE estimation is practically cancelled out if $n = 100$.

\subsubsection{The algorithm for DALE}

For clarity, we list here the full sequence of calculations needed to compute DALE in a position of a galaxy as follows:
\begin{enumerate}
\item Select all neighbours for the given galaxy in the $\nu$ radius.
\item Find the unit direction vectors from the position of the galaxy to all selected neighbours.
\item\label{sum_lab} Sum these normalised direction vectors in a random order, as in Sect.~\ref{dale_def},~point~\ref{sum_def}.
\item\label{norm_lab} Normalise the sum to unity and save it as a DALE vector.
\item\label{fac_lab} Repeat Steps~\ref{sum_lab}~and~\ref{norm_lab} until the factorial of the number of selected neighbours or $1000$ repetitions is reached.
\item\label{ave_lab} Calculate the average of DALE vectors in the given position as in Sect.~\ref{dale_def},~points~\ref{sum_def},~\ref{norm_def}.
\item Repeat $n$ times since Step~\ref{sum_lab} and store all average DALE vectors.
\end{enumerate}
Steps~\ref{fac_lab} and~\ref{ave_lab} are not principally essential, but the computation of average DALE vectors allows us to apply an optimal number of summations independently
for every galaxy. This helps also to keep $n$ small to reduce the amount of data processed by the kernel density estimation procedure in Sect.~\ref{emagl}.

\subsubsection{Some problems}\label{some_problems}

The estimated DALE may be less correct at the border area of the galaxy survey. We tested some of our samples with different values of $\nu = 5.0 \dots 12.0$~$h^{-1}\mathrm{Mpc}$;
inclusion of the galaxies at the border did not notably change the results of the analysis of mutual alignment of galaxies and DALE. However, we cannot exclude spurious effects
for all used samples and scales. Therefore, after estimating DALE for all available galaxies at a particular $\nu$ value, we exclude from further analysis galaxies with distances from the
edge of the survey that are smaller than $\nu$.

One can find more situations where DALE is poorly defined. Consider for example a single galaxy positioned just at 
the distance $\nu$ away from a cluster or filament of other galaxies. It can be argued whether our measured DALE at the position of 
this galaxy is simply noise or reflects the direction of a real but badly resolved filament. We can expect that such cases do not affect the results if
we analyse a big survey of galaxies.

Maximum and minimum useful values of $\nu$ exist. Our method measures pure noise if $\nu$ is of the order of the scale of 
homogeneity of the cosmic web. On the other hand, to avoid noisy estimates, we should use a value of $\nu$ that is larger than the typical diameter of 
the elongated structure in which we are going to estimate DALE. The neighbourhood radius $\nu$ can be interpreted as a smoothing parameter: lower values 
give meaningful DALE for smaller scales, but noise for larger objects; higher values of $\nu$ smooth out the direction data for smaller 
scales, but give useful directions for larger objects. Consequently, the most natural way to proceed is to try different values 
of $\nu$ and to study the LSS at several scales.

\subsubsection{Possible modifications and improvements}

Although we use the method described above, there are some extensions and improvements possible as follows:

First, our method can be used for a fast estimation of the direction of the local number density gradient. For that, we have to sum the original unit direction
vectors to the neighbours despite the angle between the running sum and the given unit vector. The algorithm is fast, as there is no dependence on the summing order in
this case. The total (un-normalised) length of the sum can be used to estimate the value of the number density gradient at a given location.

Second, we may filter or select galaxies before estimating DALE.
We can require more than one neighbour inside the $\nu$ radius to allow the DALE estimation.

Third, after reflecting the unit direction vectors of neighbours as needed, we can sum the variances of their coordinates and use this sum
to estimate the accuracy of the given run of DALE estimation. For an averaged DALE, we can use the mean of summed variances as an estimate of accuracy.
The accuracy of the estimated DALE in a single step can be also described
by the ratio of the total length of the un-normalised sum vector to the number of neighbours.
According to our experience, using accuracy based weights in the analysis of the mutual alignment of galaxies and the DALE may not improve the sensitivity of detecting the alignment signal.

Fourth, our method can be extended naturally to a multi-scale algorithm. When trying different values of $\nu$, we did not observe any flip of the mutual alignment signal of
galaxies and DALE. The $\nu$ value used for a given run of the averaged DALE estimation can be picked randomly from a predefined range. We can
run the averaged DALE estimation many times with different values of $\nu$ and store the results. In the resulting table of directional data there may be values of $\nu$ for every
galaxy where the mutual alignment signal is stable (other $\nu$ values will give noise). It remains to be tested whether this approach has a higher sensitivity
than the standard implementation of our method.

\subsection{Estimating the mutual alignment of galaxies and DALE}\label{emagl}

After estimating the DALE using all available galaxies and excluding the border area we have $n$ sets of directional data.
Now we may select a subsample of galaxies to study their alignment relative to the DALE. We compare the $n$ average DALE vectors
at the position of a galaxy to the orientation of the spin or minor axis of the galaxy.

We let $\gamma$ be the angle between DALE and the spin axis or the shortest axis of a galaxy. We may fix $0\degr \le \gamma \le 90\degr$, as the signs of the corresponding direction vectors
are arbitrary in our treatment.
We say that a galaxy has a rather parallel alignment with DALE, if $0\degr \le \gamma \le 45\degr$ and rather perpendicular alignment if
$45\degr < \gamma \le 90\degr$. Technically,
$\cos \gamma = |\vec{\hat{L}_1} \cdot \vec{\hat{L}_2}|$, where $\vec{\hat{L}_1}$ and $\vec{\hat{L}_2}$ are the unit direction vectors of spin and DALE.
As there are two possible spin axes for every galaxy, we also have two values of $\gamma$ for the given estimate of DALE. Hence, for the given
galaxy we have $2n$ estimates of $\cos \gamma$.

We compose the kernel density estimate of the probability density function (PDF) of $\cos \gamma$ using the two possible spin axes and $n$ estimates
of DALE for every galaxy in our selected sample. We use a Gaussian kernel with a fixed bandwidth $h = 0.045$.

We can expect that a galaxy with $n$ totally random and isotropic estimates of DALE at its location spreads its $2n$ $\cos \gamma$ values uniformly in the range $0 \dots 1$.
A galaxy with more stable direction estimates concentrates its $\cos \gamma$ values in some part of the range according to the distribution of the $n$ DALE vectors.
This removes the variability of the total PDF because of the unstable DALE estimates.

If the intrinsic alignment of galaxies and DALE were entirely random and the distributions of all position and inclination angles of 
measured galaxies (or all direction vectors of DALE) were isotropic, the observed PDF of $\cos \gamma$ would show uniform 
distribution. Because of selection effects, this is not the case and the PDF of $\cos \gamma$ is not uniform, even for 
intrinsic random alignment of galaxies and the DALE. To study the intrinsic alignment, we have to compare our observed PDF 
of $\cos \gamma$ to the PDF arising from pure selection effects. Here we make use of the randomisation scheme proposed by 
\citet{Tempel2013a}. We randomise the inclination and position angle pairs between the positions of the selected galaxies and
compute new artificial spins. Next we compute the PDF of $\cos \gamma$ corresponding to these artificial spins. The randomisation of
the inclination and position angle pairs and subsequent recalculation of artificial spins and PDF of $\cos \gamma$ is repeated $1000$
times. We use the mean of the PDFs of $\cos \gamma$ of these randomised samples as the estimate of the PDF due to selection effects. One thousand runs of randomisation was sufficient to
get a stable mean with a negligible formal confidence interval.
We say that the parallel (or perpendicular) alignment signal is detectable if the PDF of selection effects stays clearly outside of the confidence interval of the
observed PDF.
See Sect.~\ref{about_ci} for details how the confidence intervals are estimated.


\subsubsection{Confidence intervals}\label{about_ci}

As we have $2n$ values of $\cos \gamma$ for every galaxy, we have balanced clustered data. We need to use bootstrap
methods suitable for hierarchical data to compute the confidence interval for the observed PDF. As discussed by \citet[pages~100--101]{DavisonHinkley1997} and \citet{Ren2010},
resampling with replacement at the highest level and resampling without replacement at lower levels is the preferred resampling strategy for hierarchical data.
As our PDFs do not depend on the ordering of $\cos \gamma$ values related to a given galaxy, we may stop at resampling with replacement at the highest level.
We implement cluster bootstrap, where the cluster of all $2n$ values of $\cos \gamma$ related to a given galaxy is kept intact.
Because of the large number of clusters in our data, we may use the na\"{\i}ve implementation of cluster bootstrap where sizes of the original and bootstrap samples
are equal \citep[see][]{Kolenikov2010}. The cluster bootstrap must also be used in case we only had one DALE estimate for every galaxy, as there are two possible spin axes and
two values of $\cos \gamma$. 

We use undersmoothing kernel \citep[as suggested by][]{DavisonHinkley1997,Fiorio2004} in the kernel density estimation procedure, while computing the bootstrap 
confidence interval for the observed PDF. The bandwidth of the undersmoothing kernel is calculated as
$h_{\mathrm{us}} = h\,N^{1/5}/N^{1/4}$, where $N$ is the number of observed $\cos \gamma$ values and $h$ is the bandwidth used for real data \citep[][]{Jann2007,Fiorio2004}.
We note that this gives $h_{\mathrm{us}} \approx 0.5h$ for our samples.
We compared two $95$\% confidence intervals for the PDF of $\cos \gamma$ of the observed data: the percentile bootstrap interval and the standard bootstrap interval
\citep[see][]{Efron1986}. Both intervals had equal width for our data.

\citet{Tempel2013a} computed the confidence intervals for the PDF estimate of selection effects as pointwise percentiles of all PDFs from the randomisation procedure.
In our case, their approach gave confidence intervals that were approximately $1.5 \dots 1.7$ times narrower than the cluster bootstrap interval on the observed data.
We use the standard cluster bootstrap confidence interval of the observed data in all our figures, as it gave the most conservative results.


\subsection{Finding dependencies on observational parameters}\label{dep_par}

There are two alternative ways of finding critical values of the observational parameters that affect the alignment preferences of
galaxies. First, we can split the sample of selected galaxies by their $\gamma$ value into two parts: galaxies that have a rather parallel 
alignment with DALE ($0\degr \le \gamma \le 45\degr$) and galaxies with a rather perpendicular alignment ($45\degr < \gamma \le 90\degr$). Using these subsamples we can compose and 
compare two distributions of an observational parameter. If the orientational preferences of galaxies depends on
the parameter under question, the two distributions of the parameter must also differ under visual and Kolmogorov-Smirnov (K-S) tests.
An approximate critical value of the observational parameter $\alpha$ that distinguishes different alignment properties can be found from 
the plot of the two PDFs of the parameter (see Fig.~\ref{den4_pdf_hc} for an example).

Second, we can split our samples of galaxies at a trial value of an observational parameter and compare the corresponding figures of the PDFs of $\cos \gamma$.
If the alignment signal changes between the two subsamples, we found dependency on the given parameter and may further adjust the trial value.

The first method cannot be used blindly, as selection effects may also depend on an observational parameter. Because of changing selection effects, one may not detect dependence of the
alignment signal on the given parameter or one may detect spurious dependence. Therefore, we create subsamples of the selected galaxies using the estimated critical value of the parameter
and check whether it indeed has a significant effect on the orientations of galaxies.


\section{Results}\label{results_sect}

\subsection{Mutual alignment of galaxies and DALE at different scales}\label{diff_scales}

We estimated DALE for the whole catalogue of galaxies at different scales:
$\nu = 0.4$, $2.0$, $5.0$, $6.0$, $7.0$, $8.0$, $9.0$, $10.0$, $12.0$, $24.0$, $48.0$, and $96.0$~$h^{-1}\mathrm{Mpc}$. In Table~\ref{sample_sizes} one can find the numbers of galaxies
where the direction estimation was possible for some of the $\nu$ values, together with the numbers of galaxies with estimated DALE after the border area of the survey
was excluded. We note that the DALE estimation was possible for all galaxies of the
catalogue at $\nu = 24.0$ and $48.0$~$h^{-1}\mathrm{Mpc}$, but many of them resided in the border region of the survey and were excluded from alignment analysis.

\begin{figure}
\centering
\includegraphics[width=0.49\hsize,clip]{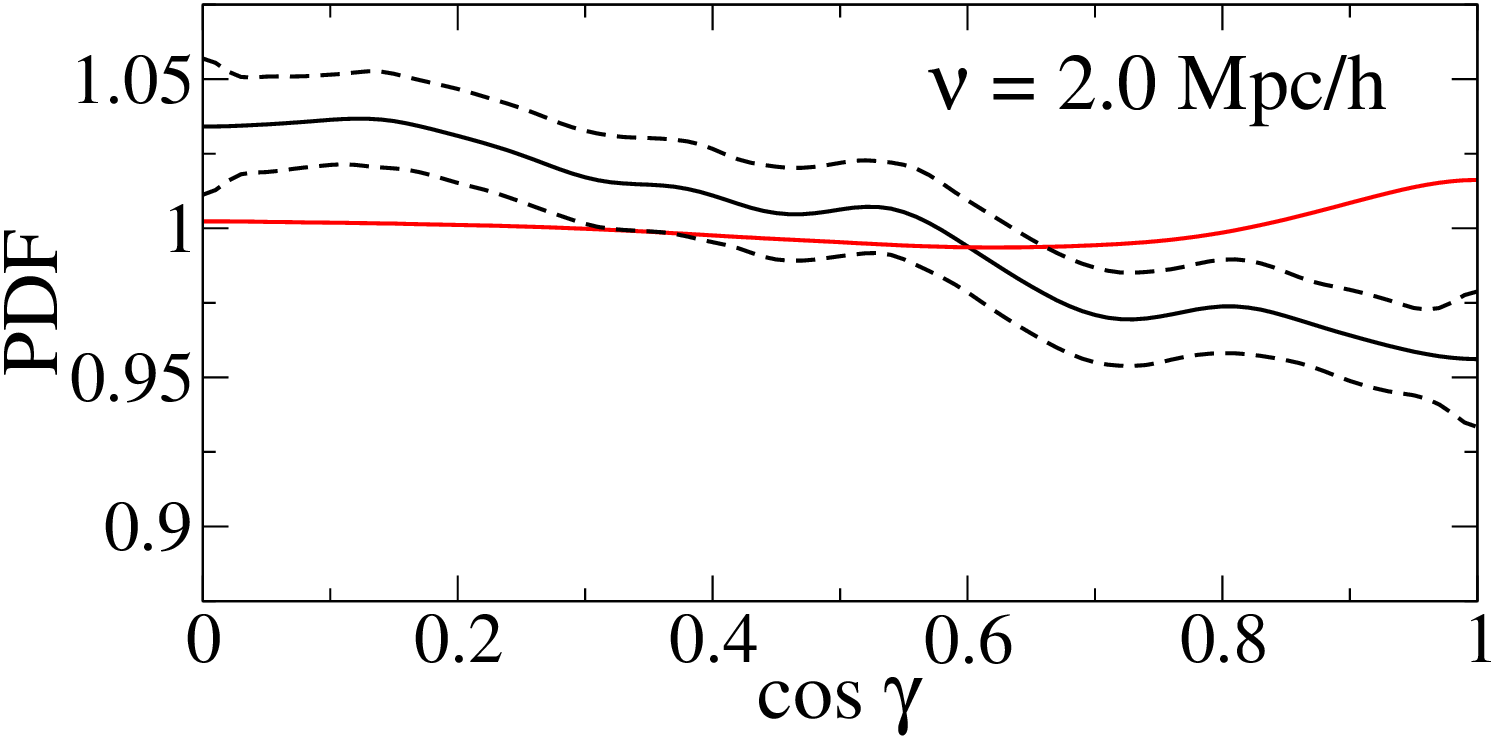}
\includegraphics[width=0.49\hsize,clip]{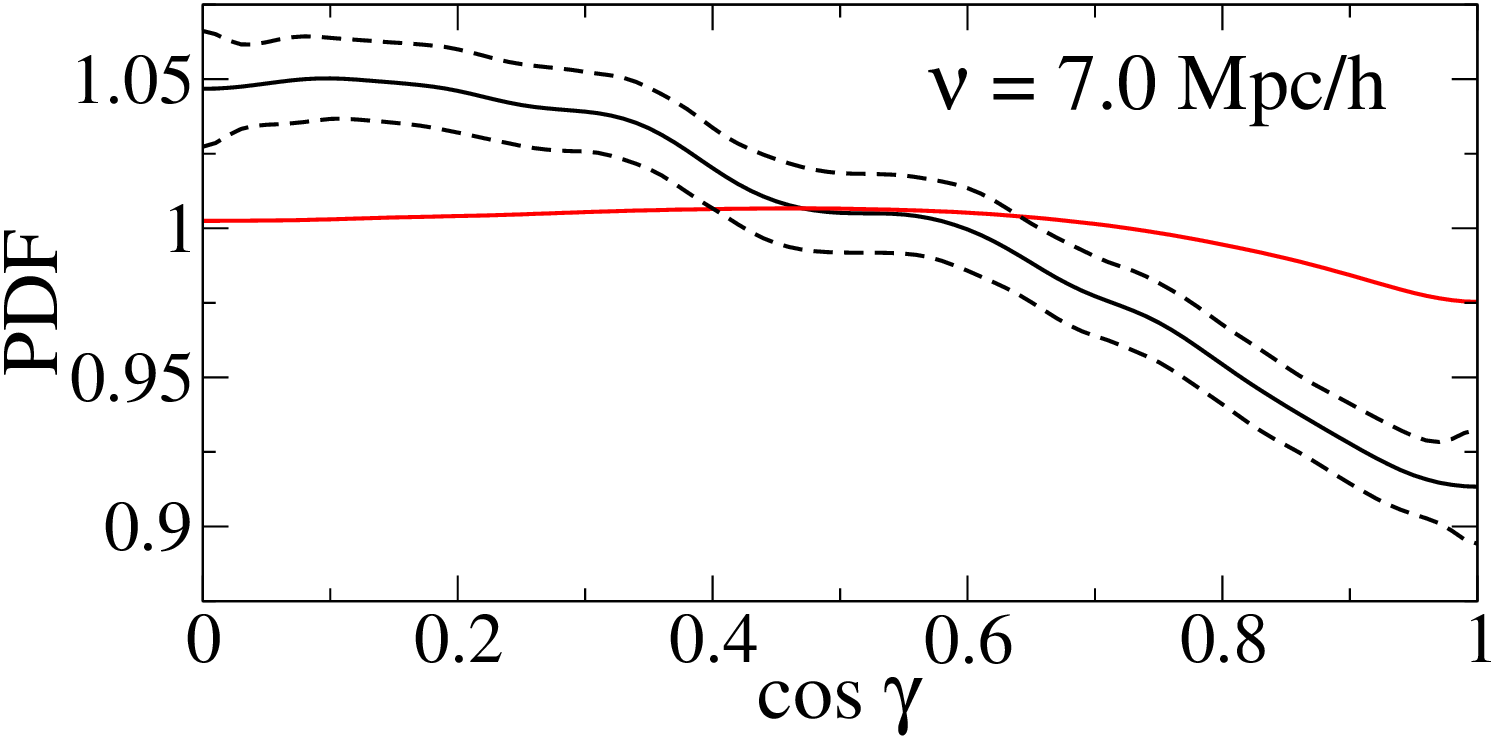}

\includegraphics[width=0.49\hsize,clip]{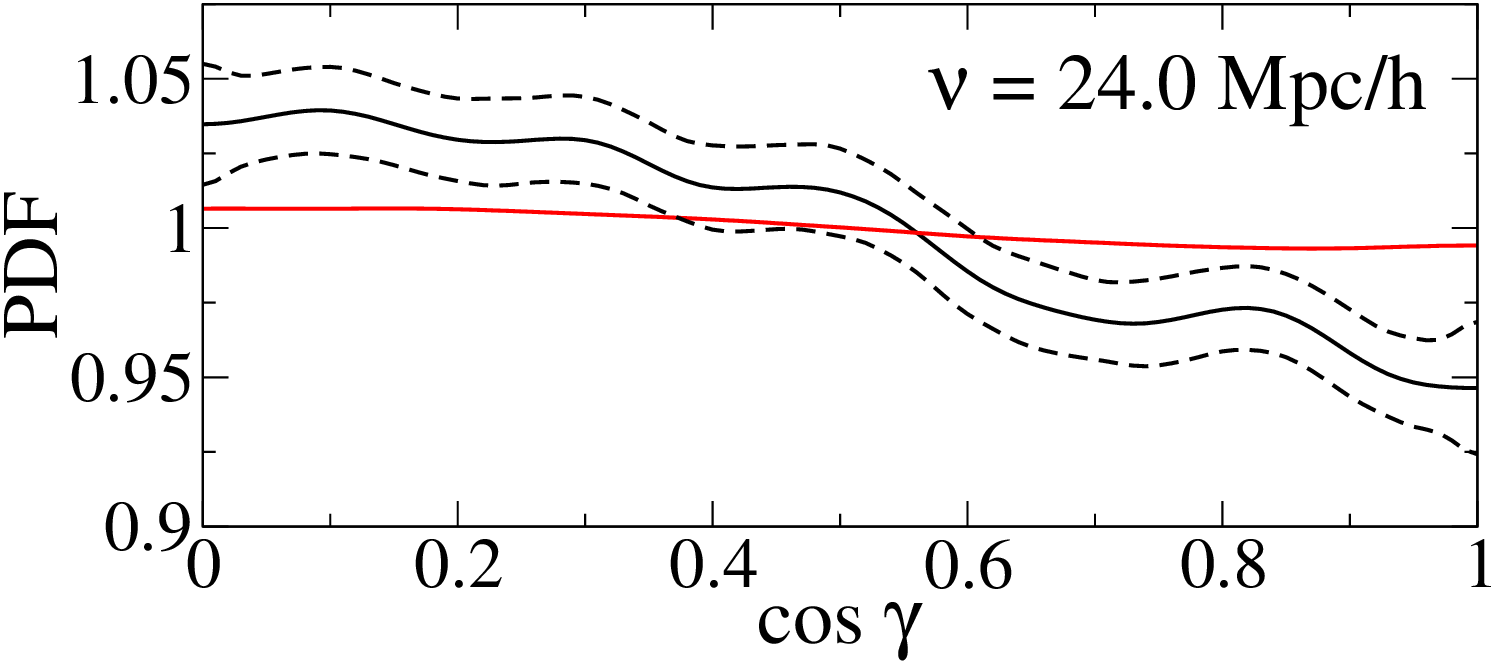}
\includegraphics[width=0.49\hsize,clip]{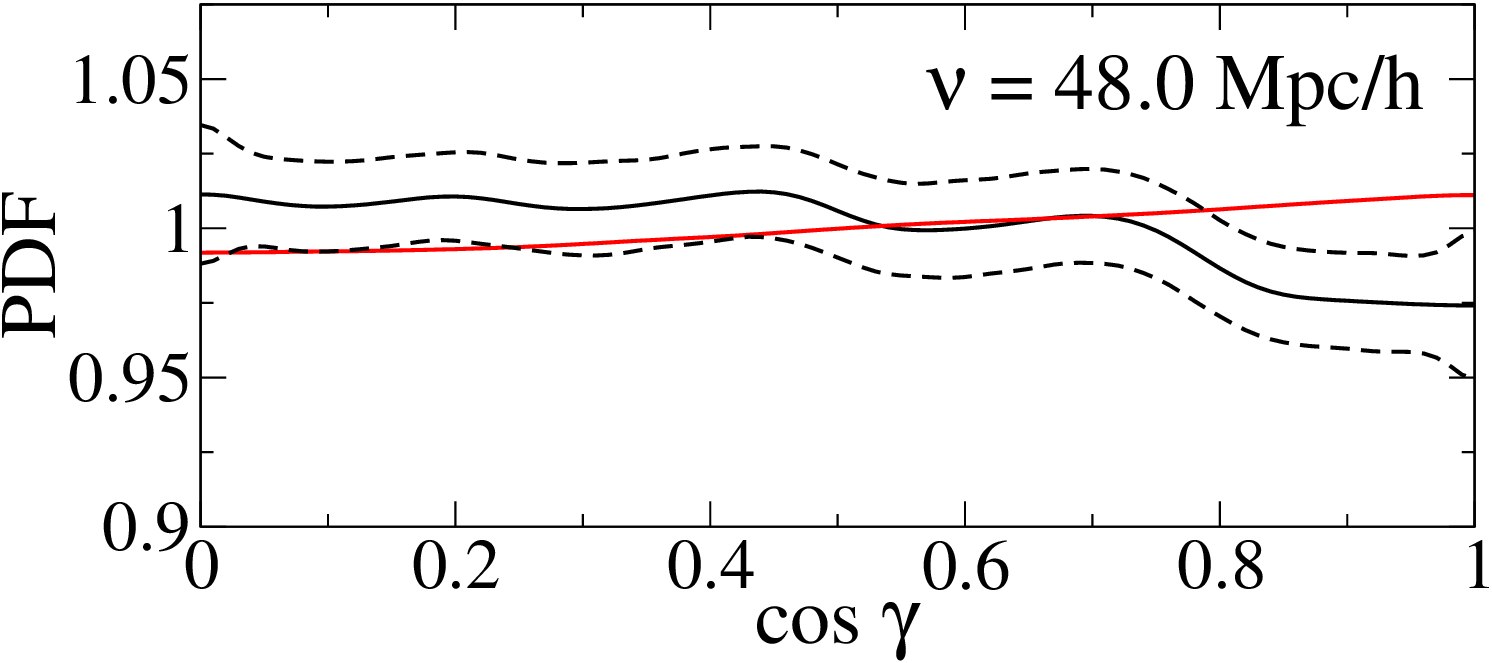}
\caption{PDFs of the cosine of the angle $\gamma$ between DALE and the spin or minor axes of early-type galaxies (\citealt{Tempel2012} classification). The neighbourhood radius
$\nu = 2.0, 7.0, 24.0, 48.0$~$h^{-1}\mathrm{Mpc}$.
The observed signal with its $95$\% confidence interval is depicted in black; the selection effect is depicted in red. Excess of low (and deficiency of high) values of observed
$\cos \gamma$ compared to the PDF of the selection effect indicates perpendicular alignment of spin or minor axes of early-type galaxies relative to DALE.}
\label{cos_gamma_ei_mny}
\end{figure}

\begin{figure}
\centering
\includegraphics[width=\hsize,clip]{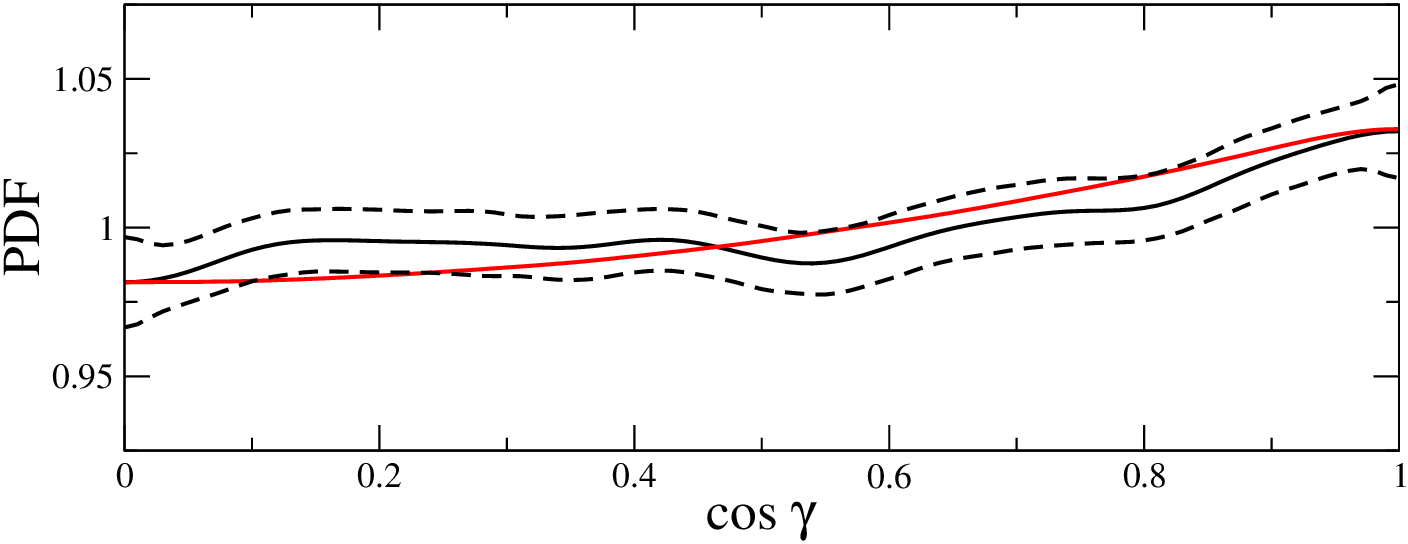}
\caption{Same as Fig.~\ref{cos_gamma_ei_mny} but for spin axes of late-type galaxies (\citealt{Tempel2012} classification); $\nu = 7.0$~$h^{-1}\mathrm{Mpc}$.}
\label{cos_gamma_si}
\end{figure}

To analyse the dependence of galaxy alignment on the value of $\nu$, we divided galaxies into early- and late-type galaxies using the \citet{Tempel2012}
classification. The early-type galaxies already show traces of preferred perpendicular alignment at $\nu = 0.4$~$h^{-1}\mathrm{Mpc}$. The signal strength
increases with $\nu$ and reaches maximum at $\nu = 7.0 \dots 12.0$~$h^{-1}\mathrm{Mpc}$, but the perpendicular alignment signal declines
gradually at higher $\nu$ values and it vanishes at $\nu = 96.0$~$h^{-1}\mathrm{Mpc}$.
Alignment signals of the early sample at $\nu = 2.0$, $7.0$, $24.0$ and $48.0$~$h^{-1}\mathrm{Mpc}$ are shown in
Fig.~\ref{cos_gamma_ei_mny}. The perpendicular alignment signal of the early-type galaxies at $\nu = 7.0 \dots 12.0$~$h^{-1}\mathrm{Mpc}$ is fully visible even if we do not
apply any selection by galaxy type in our galaxy catalogue.
The sample of late-type galaxies shows a nearly insignificant perpendicular alignment signal at $\nu = 5.0 \dots 8.0$~$h^{-1}\mathrm{Mpc}$. In Fig.~\ref{cos_gamma_si} we show the
negligible alignment of the late-type galaxy sample at $\nu = 7.0$~$h^{-1}\mathrm{Mpc}$.
\begin{table}
\caption{Sizes of samples for different values of the neighbourhood radius $\nu$.}
\label{sample_sizes}
\centering
\begin{tabular}{@{} l | l @{~~} l @{~~} l @{~~} l @{~~} l @{~~} l @{}}
\hline\hline
   $\nu$                  & $2.0$    & $7.0$    &  $12.0$  & $24.0$   & $48.0$   & $96.0$ \\
\hline
   $N_{\mathrm{d}}$       & $420575$ & $566593$ & $575746$ & $576493$ & $576493$ & $576493$ \\
   $N_{\mathrm{ob}}$      & $404918$ & $510404$ & $483716$ & $408068$ & $288188$ & $117865$ \\
   $N_{\mathrm{early}}$    & $99567$  & $131358$ & $126116$ & $109282$ & $80209$ & $34906$ \\
   $N_{\mathrm{late}}$    & $185681$ & $223631$ & $209059$ & $172299$ & $117345$ & $44149$ \\
\hline
\end{tabular}
\tablefoot{$\nu$~-- in $h^{-1}\mathrm{Mpc}$; $N_{\mathrm{d}}$~-- number of galaxies with estimated DALE; $N_{\mathrm{ob}}$~-- number of galaxies with estimated DALE outside the border area;
$N_{\mathrm{early}}$-- number of early-type galaxies used to compute PDF; and $N_{\mathrm{late}}$-- number of late-type galaxies used to compute PDF.}
\end{table}

\begin{figure}
\centering
\includegraphics[width=0.49\hsize,clip]{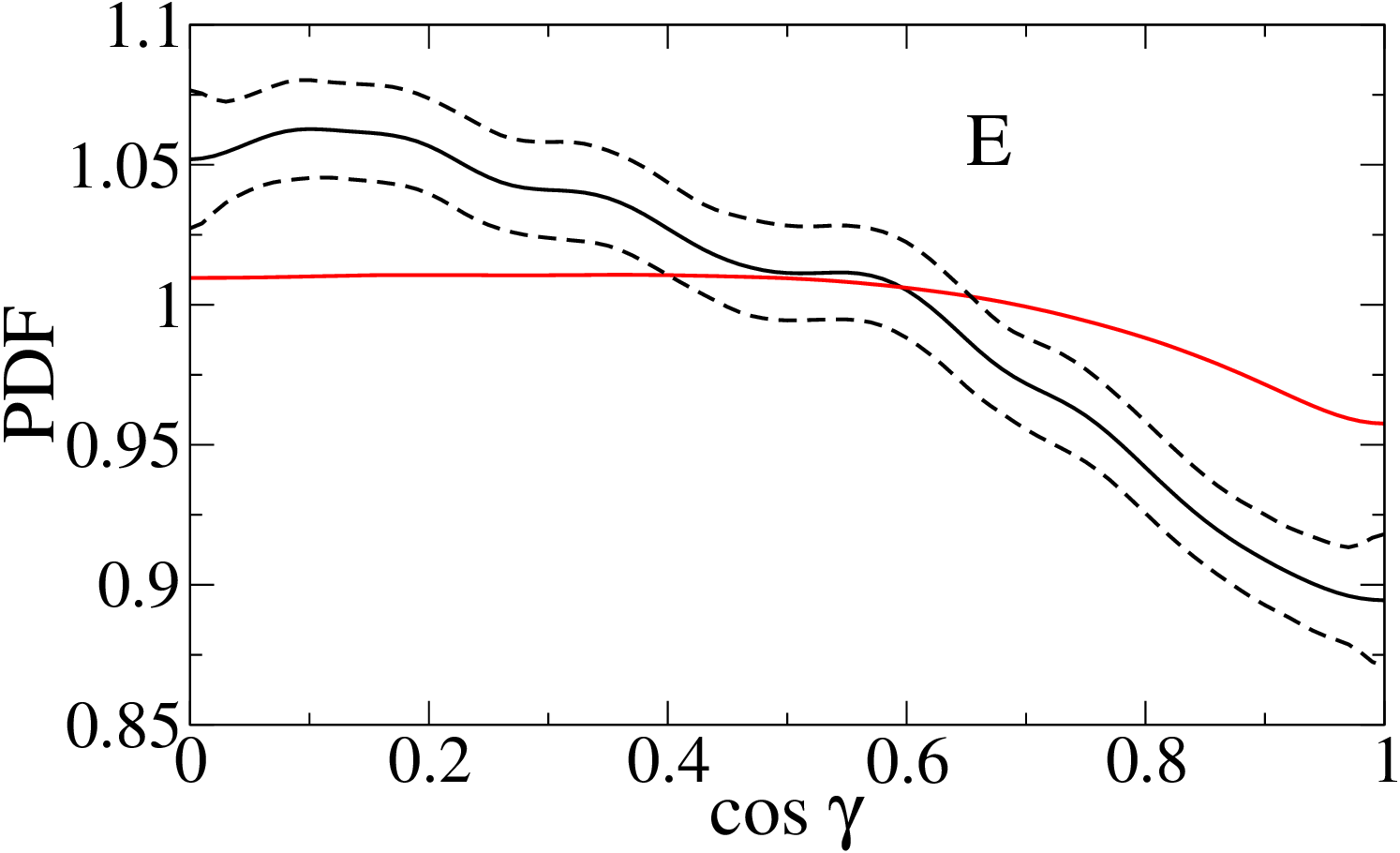}
\includegraphics[width=0.49\hsize,clip]{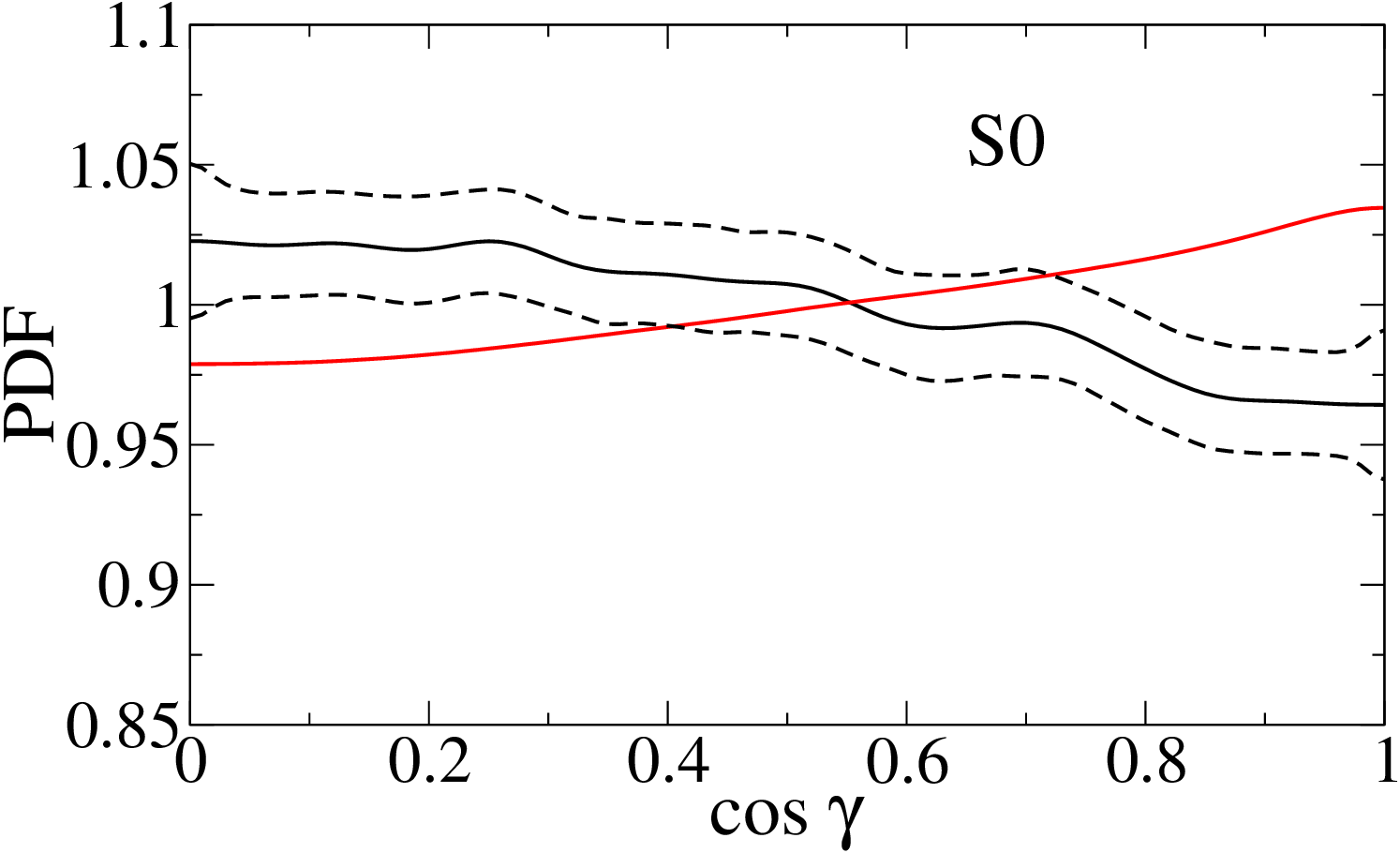}

\includegraphics[width=0.49\hsize,clip]{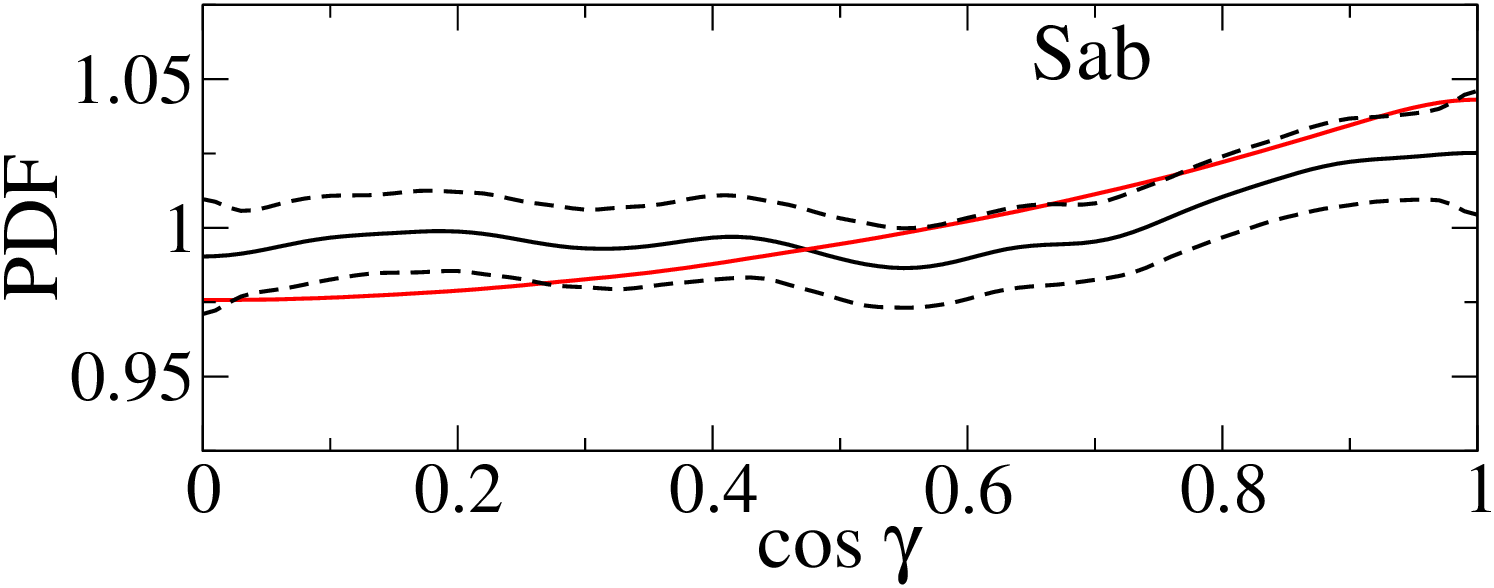}
\includegraphics[width=0.49\hsize,clip]{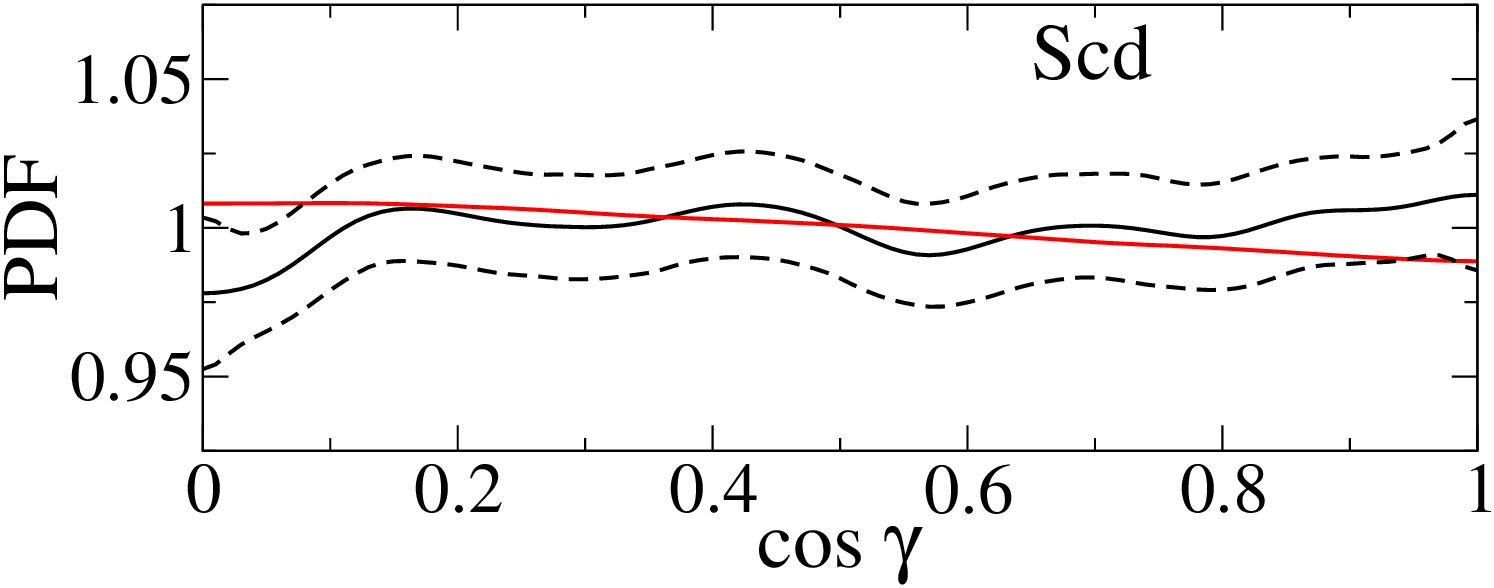}
\caption{Same as Fig.~\ref{cos_gamma_ei_mny} but for E, S0, Sab, and Scd galaxies (\citealt{Huertas-Company2011} classification) with no additional selection;
$\nu = 7.0$~$h^{-1}\mathrm{Mpc}$.}
\label{cos_gamma_hc_none_ny70}
\end{figure}

Next we divided galaxies into four types: E, S0, Sab, and Scd using the 
\citet{Huertas-Company2011} classification probabilities. We only selected those galaxies that have the probability of being in the particular type 
$>0.5$ and repeated the search for the most sensitive value of $\nu$.
E and S0 galaxies showed a maximum perpendicular alignment signal at $\nu = 7.0 \dots 12.0$~$h^{-1}\mathrm{Mpc}$, as expected.
Sab and Scd galaxies gave the strongest signals at $\nu = 7.0$~$h^{-1}\mathrm{Mpc}$, while the Sab sample showed perpendicular and the Scd
sample parallel alignment. The alignment PDFs of these samples at $\nu = 7.0$~$h^{-1}\mathrm{Mpc}$ are shown in
Fig.~\ref{cos_gamma_hc_none_ny70}. One can find the sizes of the samples used here and below in Table~\ref{sample_sizes_ny70}.
\begin{table}
\caption{Sizes of different galaxy samples for the neighbourhood radius $\nu = 7.0$~$h^{-1}\mathrm{Mpc}$.}
\label{sample_sizes_ny70}
\centering
\begin{tabular}{@{} l | l l l l @{}}
\hline\hline
   Selection                                    & E        & S0       & Sab      & Scd \\
\hline
   all                                          & $84877$  & $71800$  & $135585$ & $76893$  \\
   $N_{\mathrm{rich}} = 1$                      & $42724$  & $37053$  & $72970$  & $45429$  \\
   $2 \le N_{\mathrm{rich}} \le 3$              & $22101$  & $17482$  & $31182$  & $18046$  \\
   $N_{\mathrm{rich}} > 10$                     & $7532$   & $7370$   & $13611$  & $4877$   \\
   $4 \le N_{\mathrm{rich}} \le 10$, satellite  & $7627$   & $7632$   & $15705$  & $7622$   \\
   $4 \le N_{\mathrm{rich}} \le 10$, main       & $4893$   & $2263$   & $2117$   & $919$    \\
   $D \le \alpha_{\mathrm{D}}$                  & $18906$  & $27627$  & $65134$  & $38311$  \\
   $D > \alpha_{\mathrm{D}}$                    & $65971$  & $44173$  & $70451$  & $38582$  \\
   $L \le \alpha_{\mathrm{L}}$                  & $48385$  & $35770$  & $93746$  & $41115$  \\
   $L > \alpha_{\mathrm{L}}$                    & $36492$  & $36030$  & $41839$  & $35778$  \\
\hline
\end{tabular}
\tablefoot{
$N_{\mathrm{rich}}$ is number of galaxies in group.
The critical values $\alpha_{\mathrm{D}}$ for environmetal density $D$ and $\alpha_{\mathrm{L}}$ for $r$-band luminosity $L$ are given in Sects.~\ref{align_dens}~and~\ref{align_lum}.
}
\end{table}

To make sure that using the favoured scale does not hide any alignment effect for different samples, we calculated all subsequent PDFs of mutual alignment for
$\nu = 2,7,12,24,48$~$h^{-1}\mathrm{Mpc}$. We observed that the alignment signal depends only quantitatively on the selected scale, and we did not detect flips of the signal 
for a particular type of galaxies. The strongest signal was always found at $\nu = 7.0, 12.0$~$h^{-1}\mathrm{Mpc}$.

\subsection{Alignment with DALE depending on luminosity and environmental density}

We estimated DALE using all galaxies at $\nu = 7.0$~$h^{-1}\mathrm{Mpc}$. Four morphological types,
E, S0, Sab, and Scd, were selected for the following analysis using the \citet{Huertas-Company2011} classification probabilities as described in Sect.~\ref{diff_scales}.

\begin{figure}
\centering
\includegraphics[width=\hsize,clip]{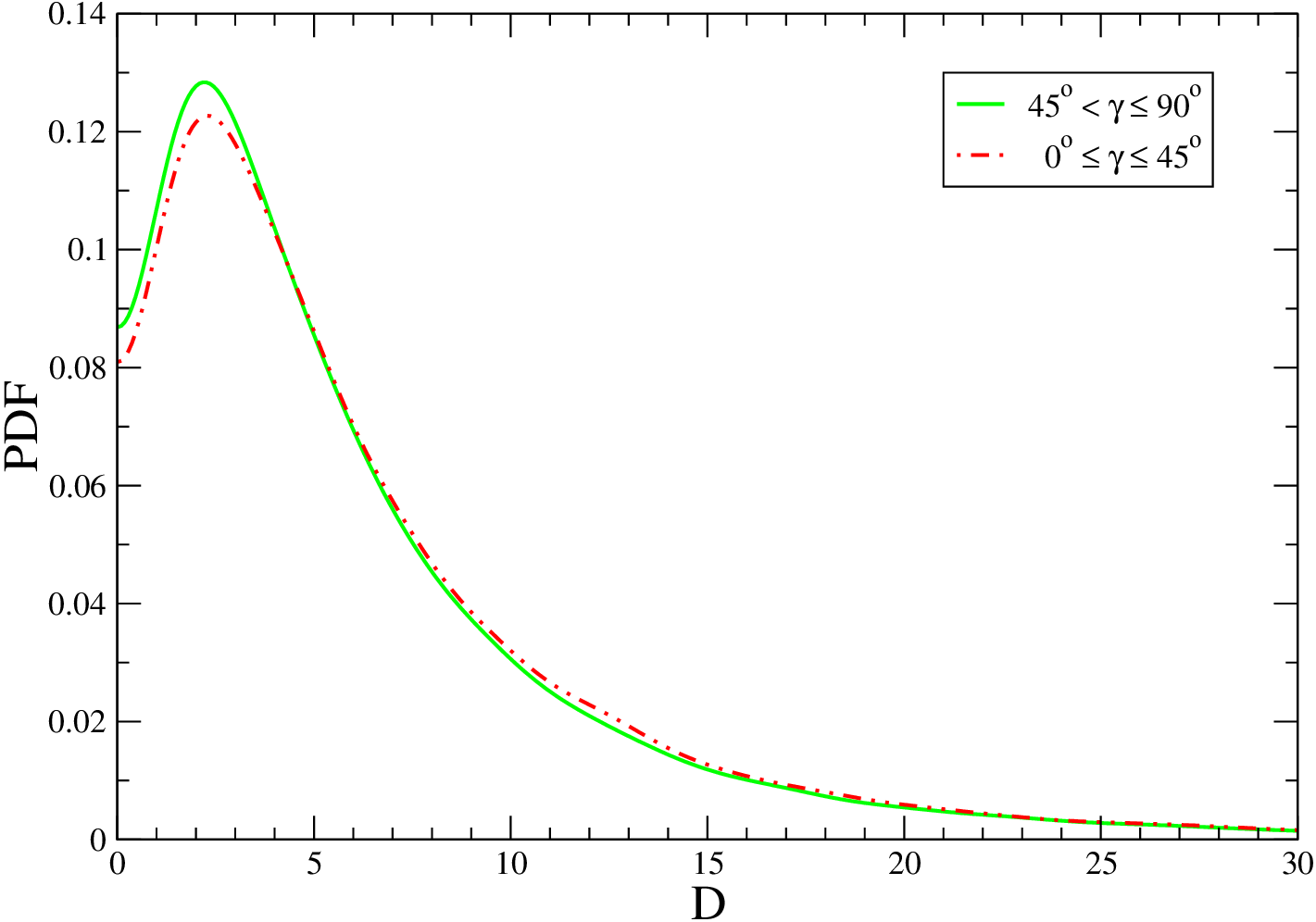}
\caption{PDFs of environmental density $D$ for Sab galaxies with rather parallel and perpendicular alignment; $\nu = 7.0$~$h^{-1}\mathrm{Mpc}$.
The relative concentration of Sab galaxies with preferred perpendicular alignment is higher at $D$ values at less than $\alpha_{\mathrm{D}} \approx 4.3$.
To present the results more clearly, we truncated the range of $D$ at $30$.}
\label{den4_pdf_hc}
\end{figure}

As discussed in Sect.~\ref{dep_par}, there are two complementary methods to study the dependence of alignment on observational parameters.
First, we compared the distributions of the normalised environmental density (for the $2$, $4$ and $8$~$h^{-1}\mathrm{Mpc}$ smoothing scales) and $r$-band luminosity ($L$)
for galaxies with $0\degr \le \gamma \le 45\degr$ and $45\degr < \gamma \le 90\degr$. The effects of the environmental
densities with different smoothing were similar, but the $4$~$h^{-1}\mathrm{Mpc}$ smoothed environmental density ($D$) introduced the strongest effect on
galaxy orientations. The results of the K-S test for the compared distributions of $D$ and $L$ are presented in Table~\ref{ks_den4_lumr}.
For example, the PDFs of $D$ for Sab galaxies with different alignment are shown in Fig.~\ref{den4_pdf_hc}.
As selection effects may affect this approach, we tried to estimate $\alpha$ visually from every plot of PDFs of $D$ and $L$ regardless of the K-S test. It was not
possible in the case of Scd galaxies, where we just chose $\alpha$ values that gave us subsamples with nearly equal size.
Second, we created subsamples using the $\alpha$ values and checked whether there are real alignment differences.

\begin{table}[htb]
\caption{Results of the K-S test for two distributions of environmental density $D$ and $r$-band luminosity $L$ composed for two samples with
alignment angle $0\degr \le \gamma \le 45\degr$ and $45\degr < \gamma \le 90\degr$.}
\label{ks_den4_lumr}
\centering
\begin{tabular}{l l l}
\hline\hline
   Morphology & $D$ & $L$ \\   
\hline
   E      & $2$                     & $1$ \\
   S0     & $1$                     & $2$ \\
   Sab    & $1$                     & $0$ \\
   Scd    & $0$                     & $0$ \\
\hline
\end{tabular}
\tablefoot{`0' denotes that the null hypothesis~-- the two distributions of the parameter are identical~-- can be rejected neither at $95$\% nor $90$\% significance
level; `1' designates that $H_{0}$ can be rejected at $95$\% significance level; and `2' expresses that $H_{0}$ can be rejected at $90$\% significance level.
}
\end{table}

\subsubsection{Galaxy alignment and environmental density}\label{align_dens}

\begin{figure}
\centering
\includegraphics[width=0.49\hsize,clip]{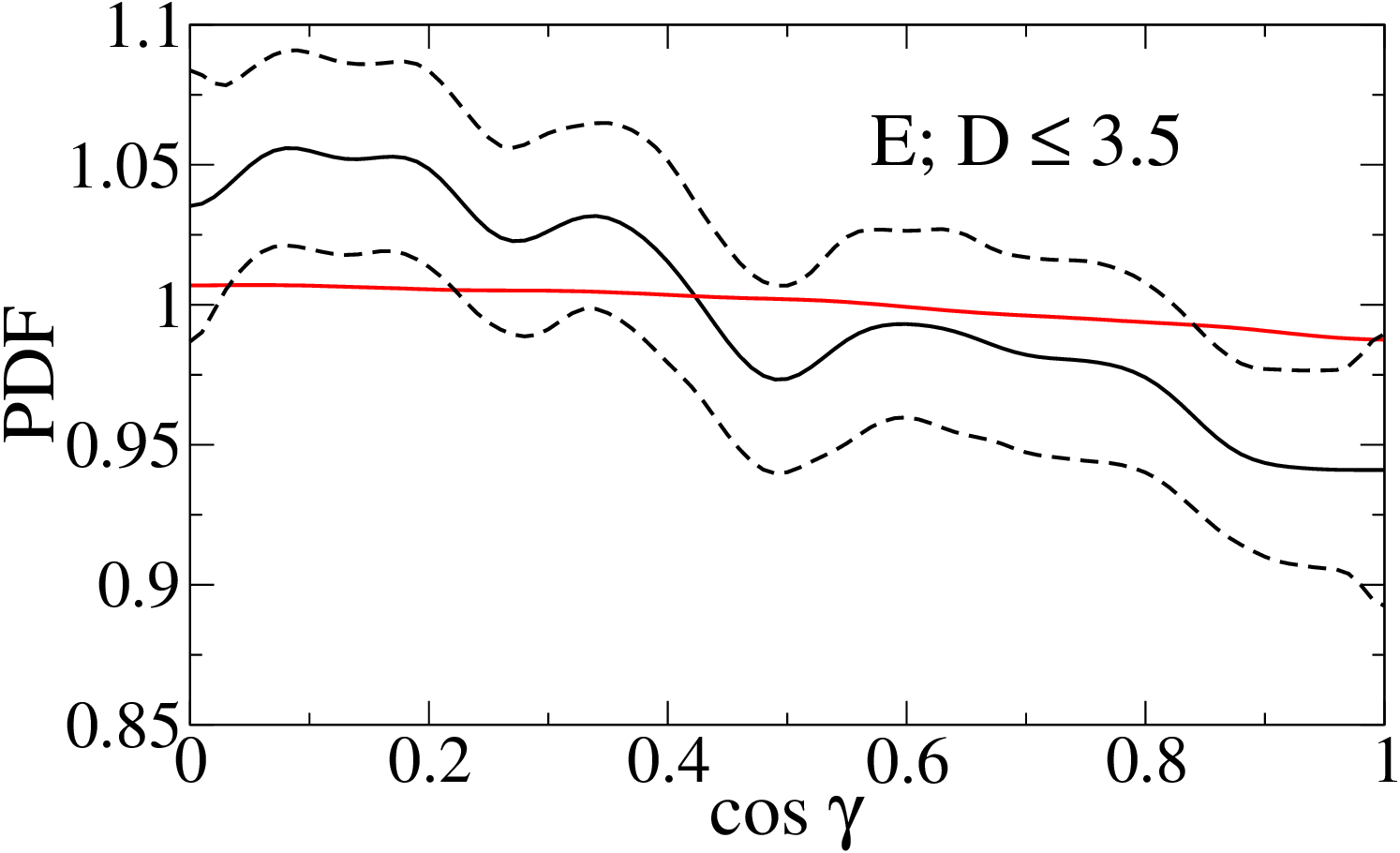}
\includegraphics[width=0.49\hsize,clip]{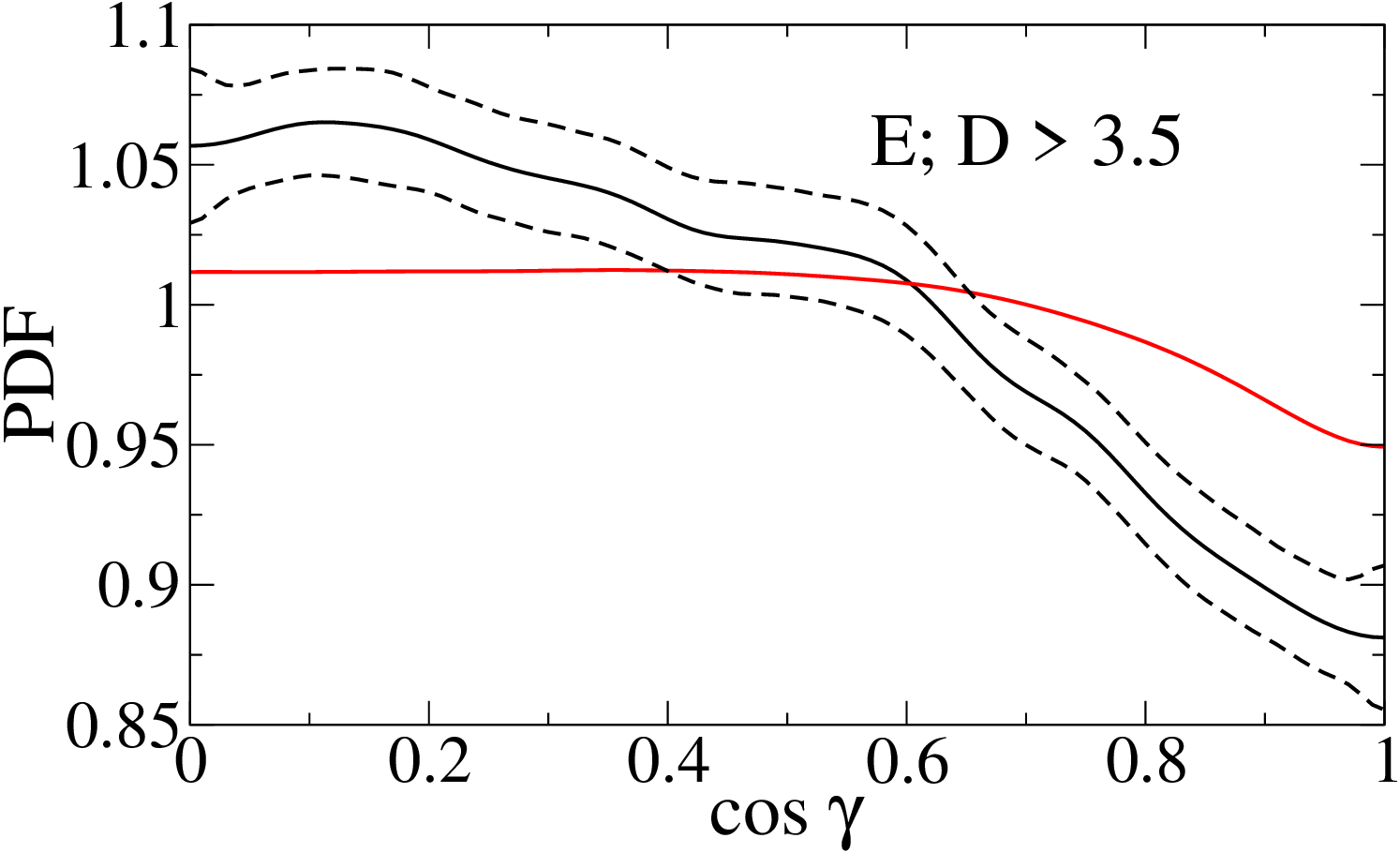}

\includegraphics[width=0.49\hsize,clip]{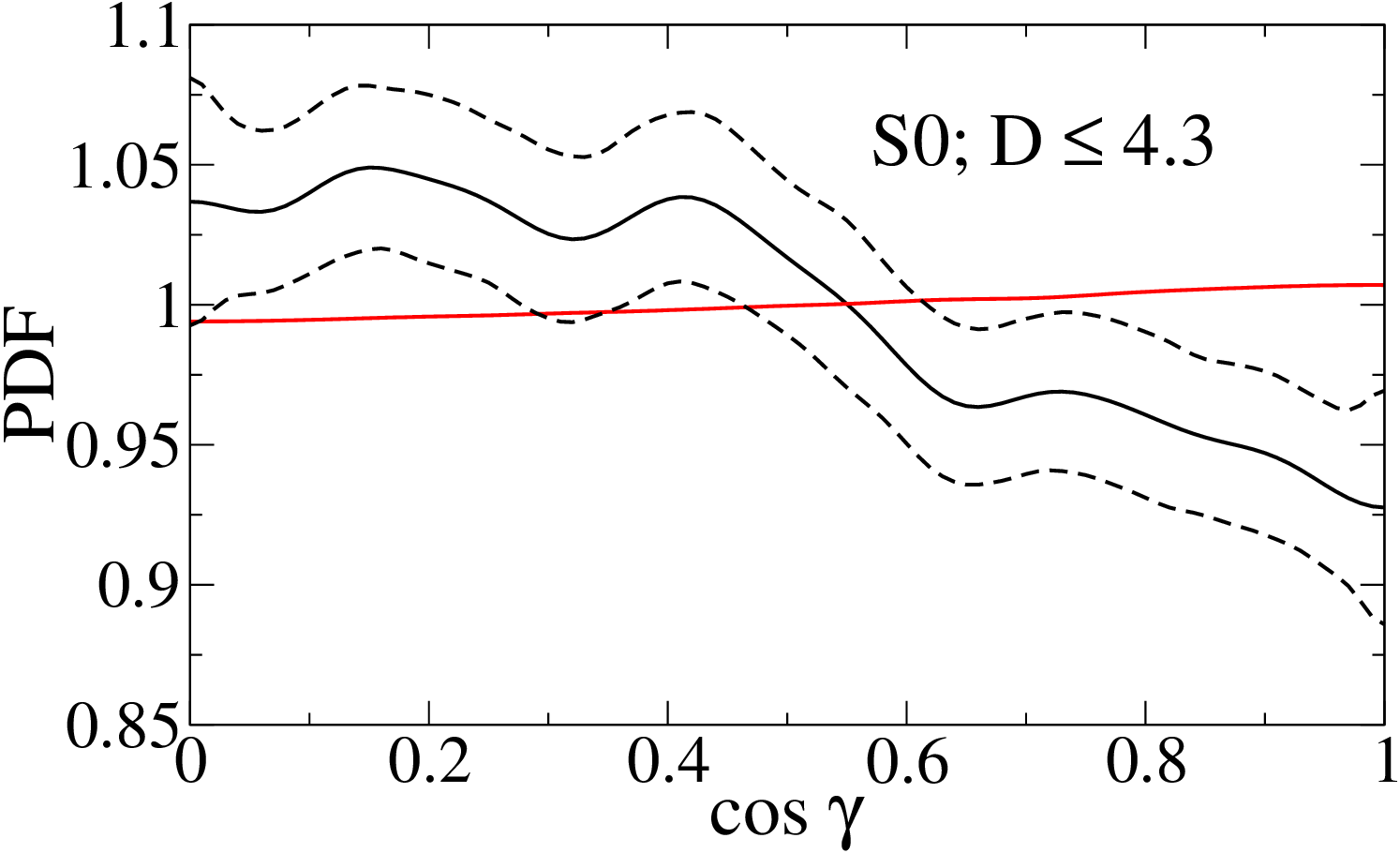}
\includegraphics[width=0.49\hsize,clip]{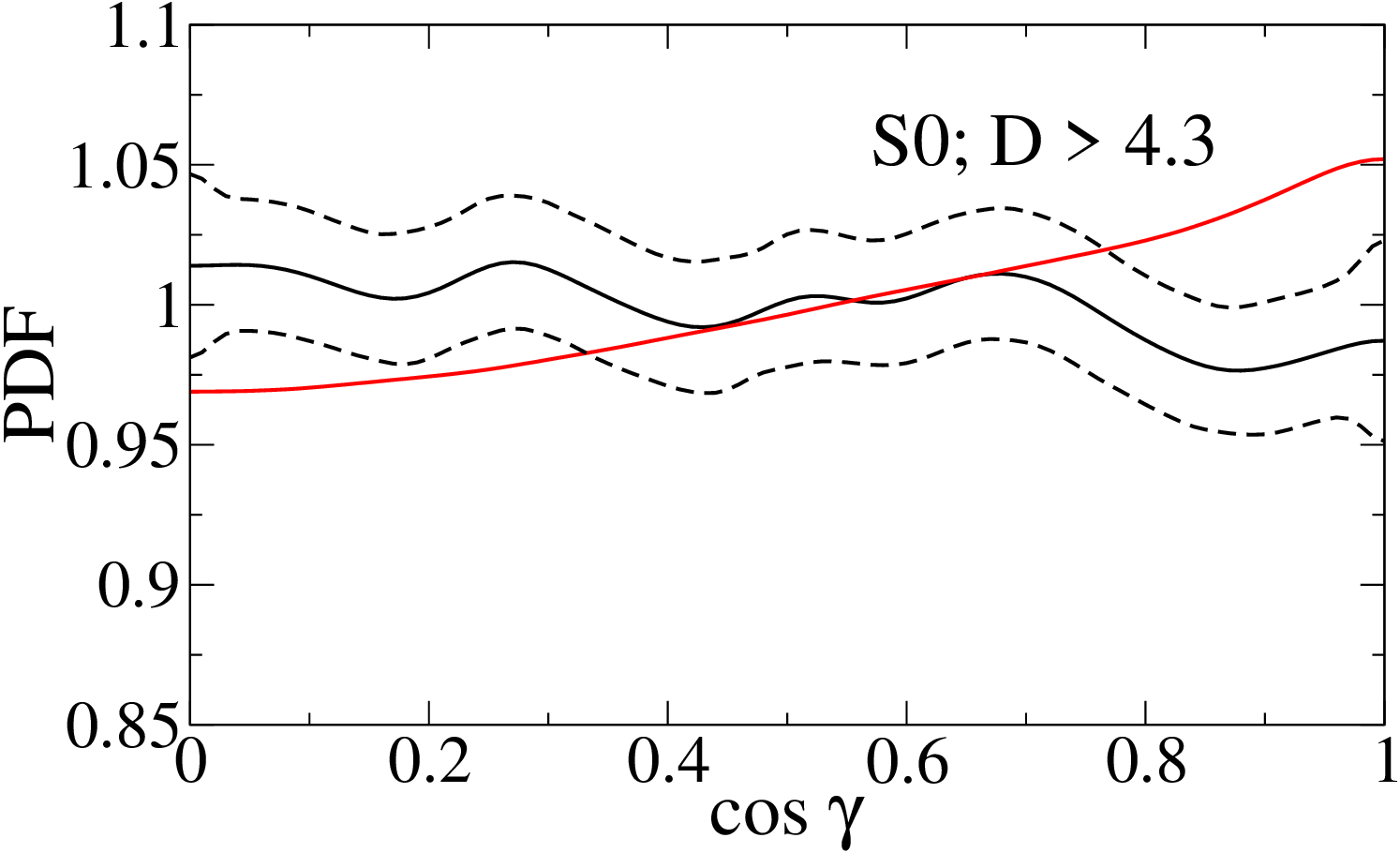}

\includegraphics[width=0.49\hsize,clip]{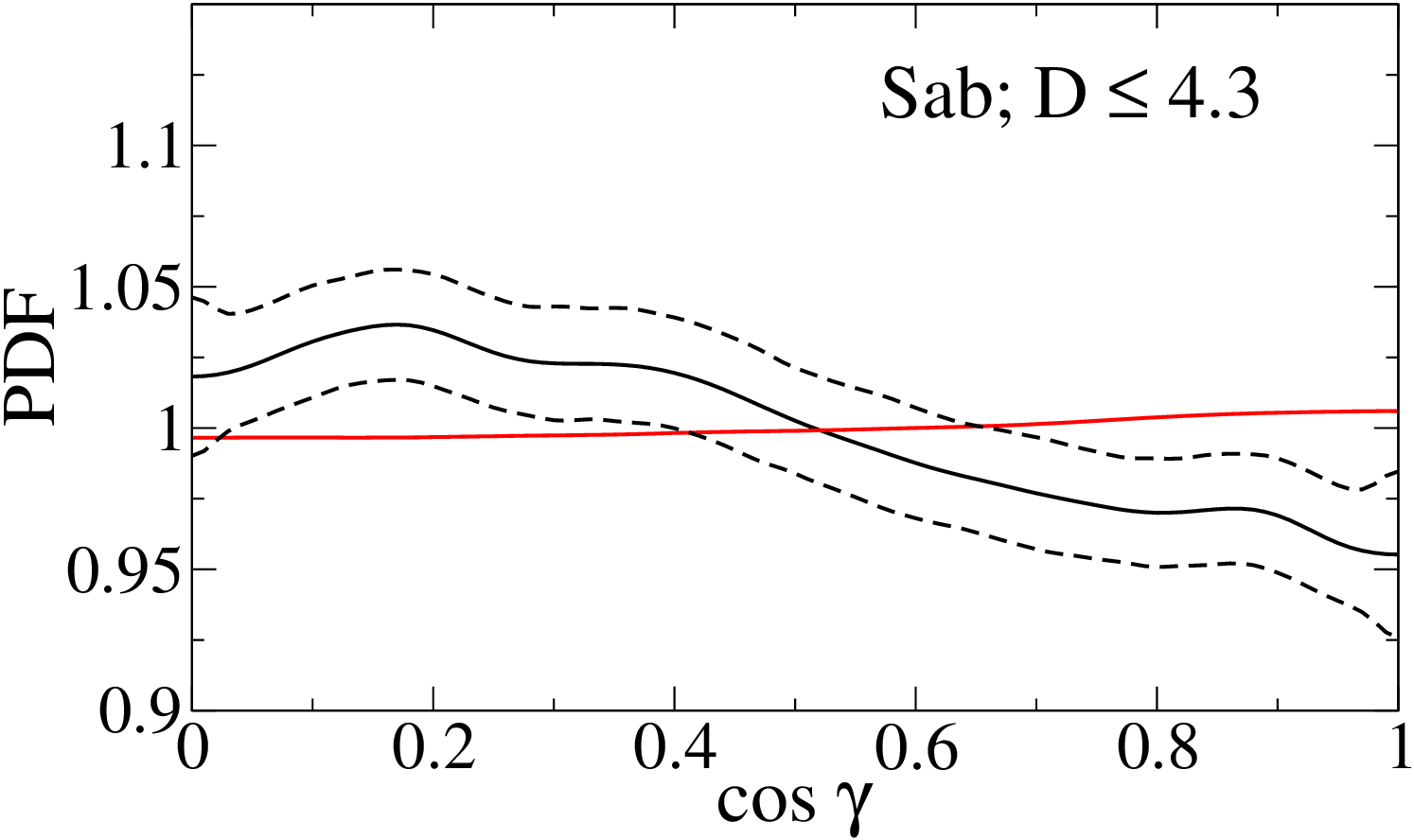}
\includegraphics[width=0.49\hsize,clip]{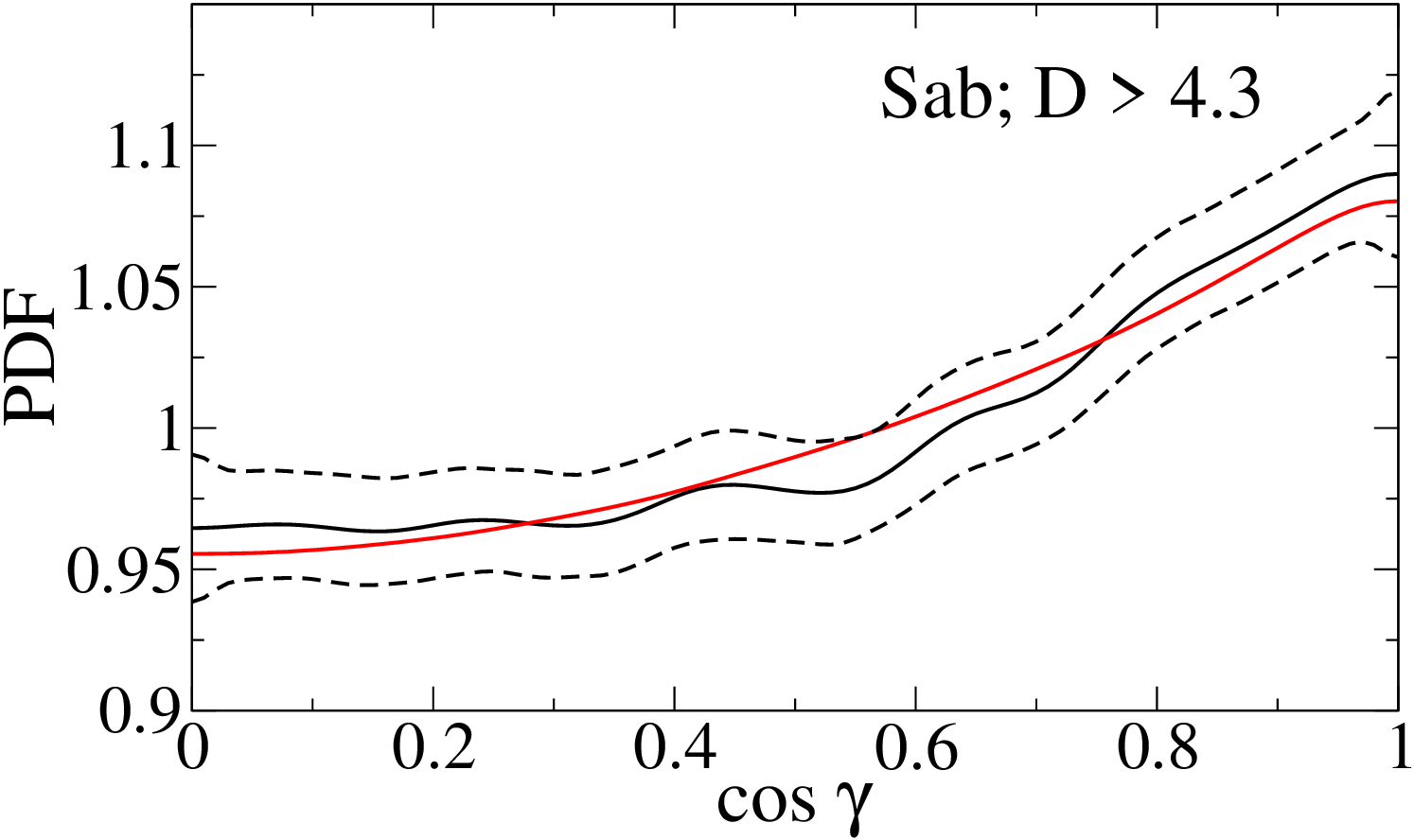}

\includegraphics[width=0.49\hsize,clip]{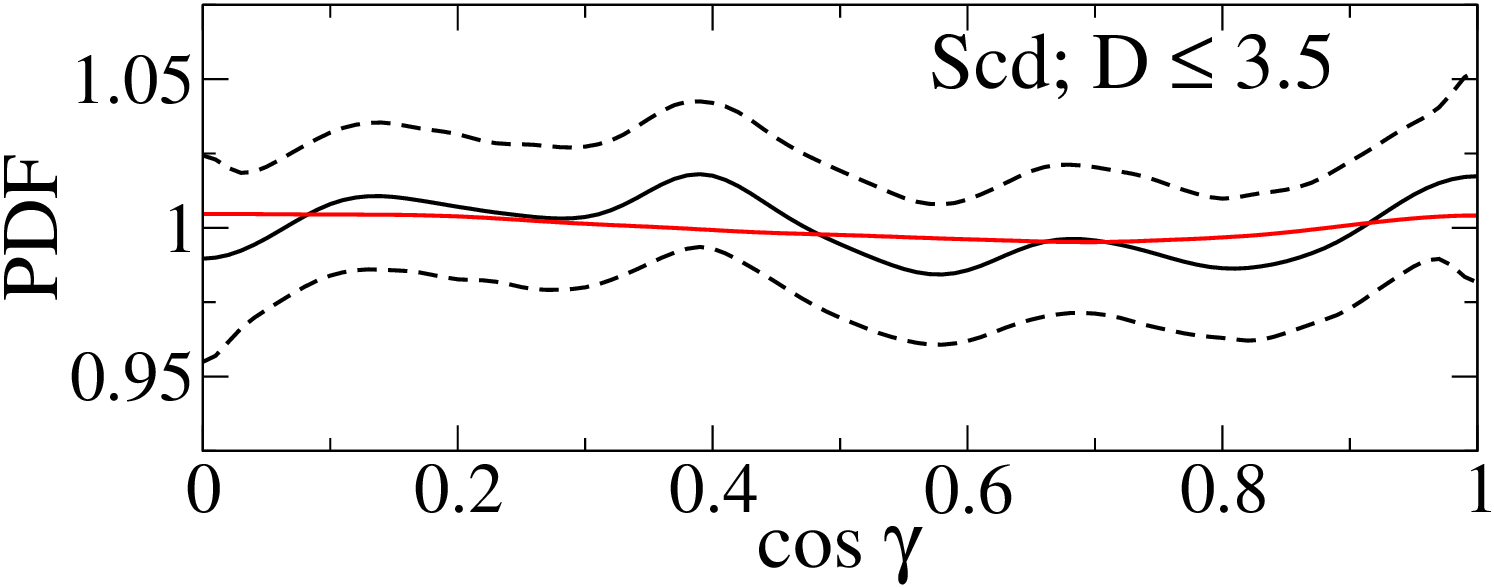}
\includegraphics[width=0.49\hsize,clip]{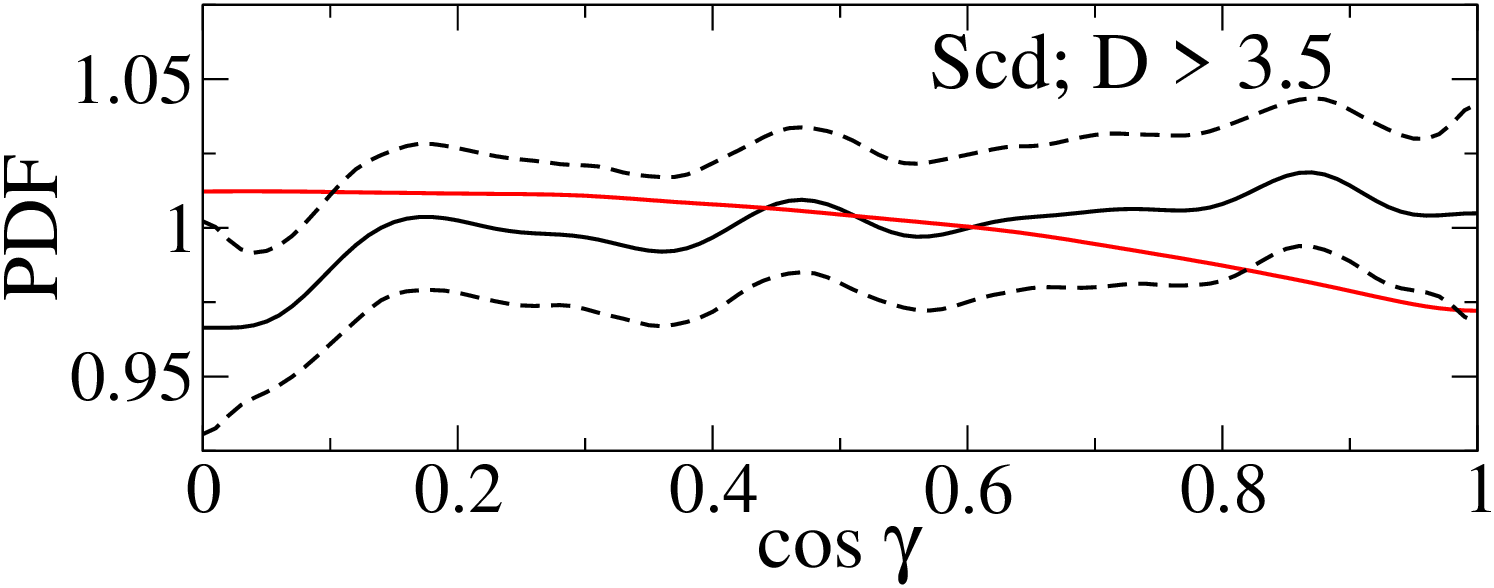}
\caption{Same as Fig.~\ref{cos_gamma_ei_mny} but for E, S0, Sab, and Scd galaxies selected by the environmental density $D$; $\nu = 7.0$~$h^{-1}\mathrm{Mpc}$.}
\label{pdf_den4}
\end{figure}

The estimated critical value of $D$ for E galaxies is $\alpha_{\mathrm{D}} \approx 3.5$. We can see in Fig.~\ref{pdf_den4} that
the perpendicular alignment signal of E galaxies is stronger in areas with higher environmental density, where $D > 3.5$.

We estimated $\alpha_{\mathrm{D}} \approx 4.3$ for S0 galaxies. However, we did not find a clear dependence of the alignment signal of S0 galaxies on
the environmental density (see Fig.~\ref{pdf_den4}). It is possible that the distributions of $D$ were different because of selection effects.

The distributions of $D$ were most different in the case of Sab galaxies. Based on Fig.~\ref{den4_pdf_hc},
we can expect a stronger perpendicular alignment signal for Sab galaxies in areas with lower environmental density than
$\alpha_{\mathrm{D}} \approx 4.3$. We can see in Figs.~\ref{pdf_den4} and~\ref{cos_gamma_hc_nrich_le_1_ny70} that if we fix $D > 4.3$ there are no signs of alignment signal,
but if $D \le 4.3$ the perpendicular alignment signal for Sab galaxies is even stronger than it is for the single galaxy sample.

Comparison of distributions of $D$ did not indicate any dependence of orientational preferences of Scd galaxies on the environmental density.
Regardless of this lack of indication, we split the Scd sample at $D = 3.5$, which gave us subsamples with a nearly equal size.
We can see in Figs.~\ref{cos_gamma_hc_none_ny70} and~\ref{pdf_den4} that the parallel alignment signal of the Scd population comes mainly from areas with higher environmental
density; we cannot detect alignment of Scd galaxies in sparser areas. The distributions of $D$ for Scd galaxies were obviously similar because of selection effects. 

\subsubsection{Galaxy alignment and luminosity}\label{align_lum}

\begin{figure}
\centering
\includegraphics[width=0.49\hsize,clip]{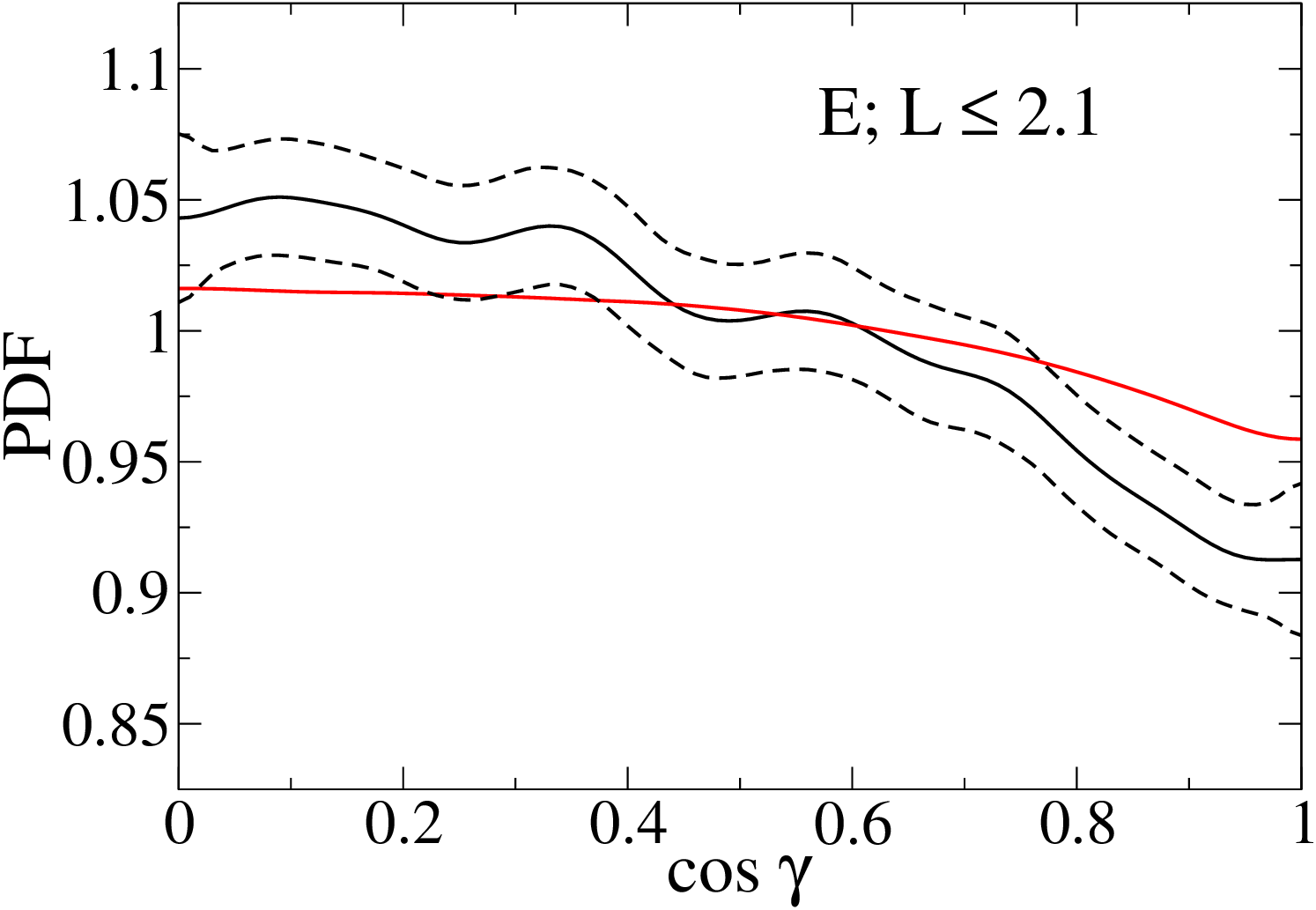}
\includegraphics[width=0.49\hsize,clip]{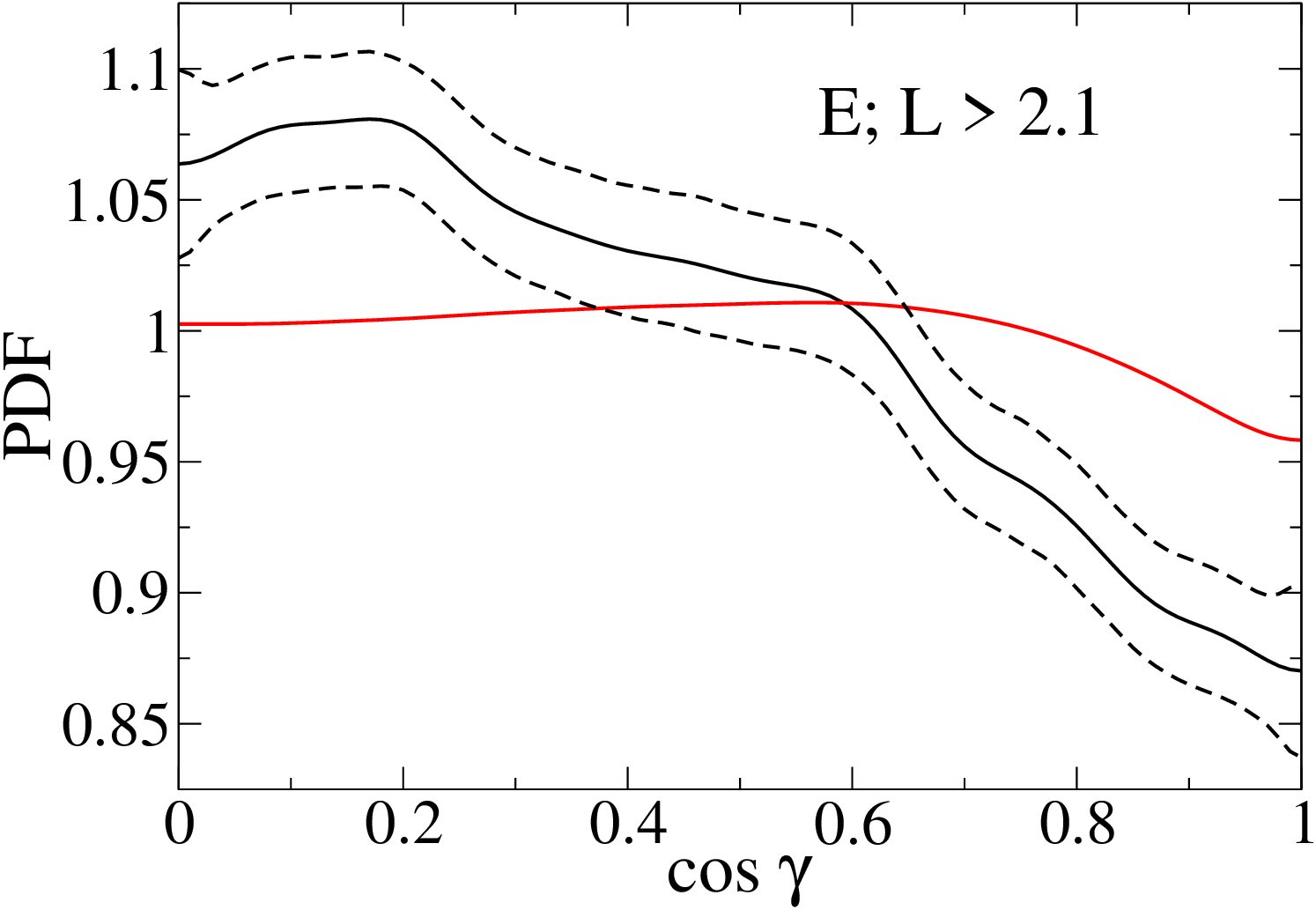}

\includegraphics[width=0.49\hsize,clip]{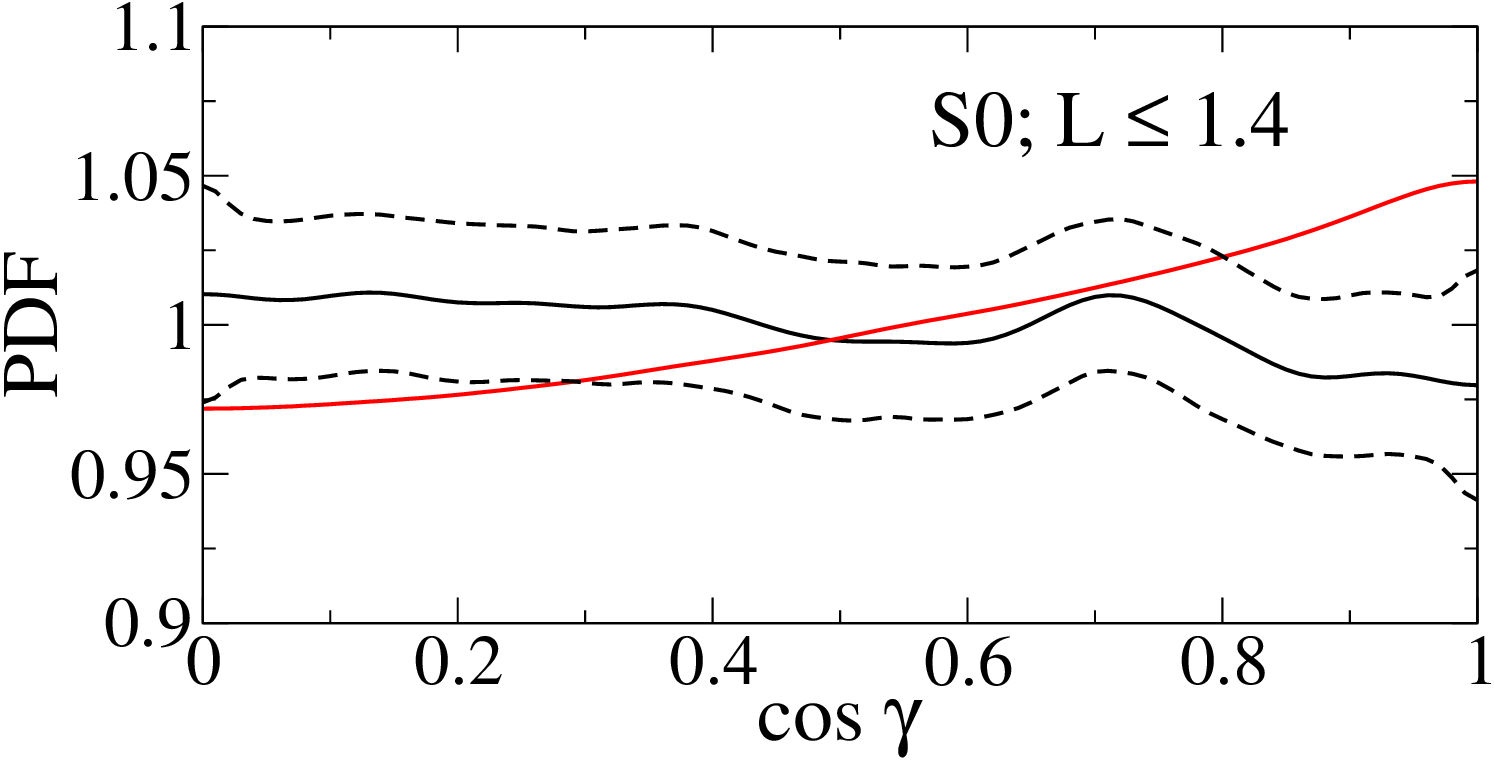}
\includegraphics[width=0.49\hsize,clip]{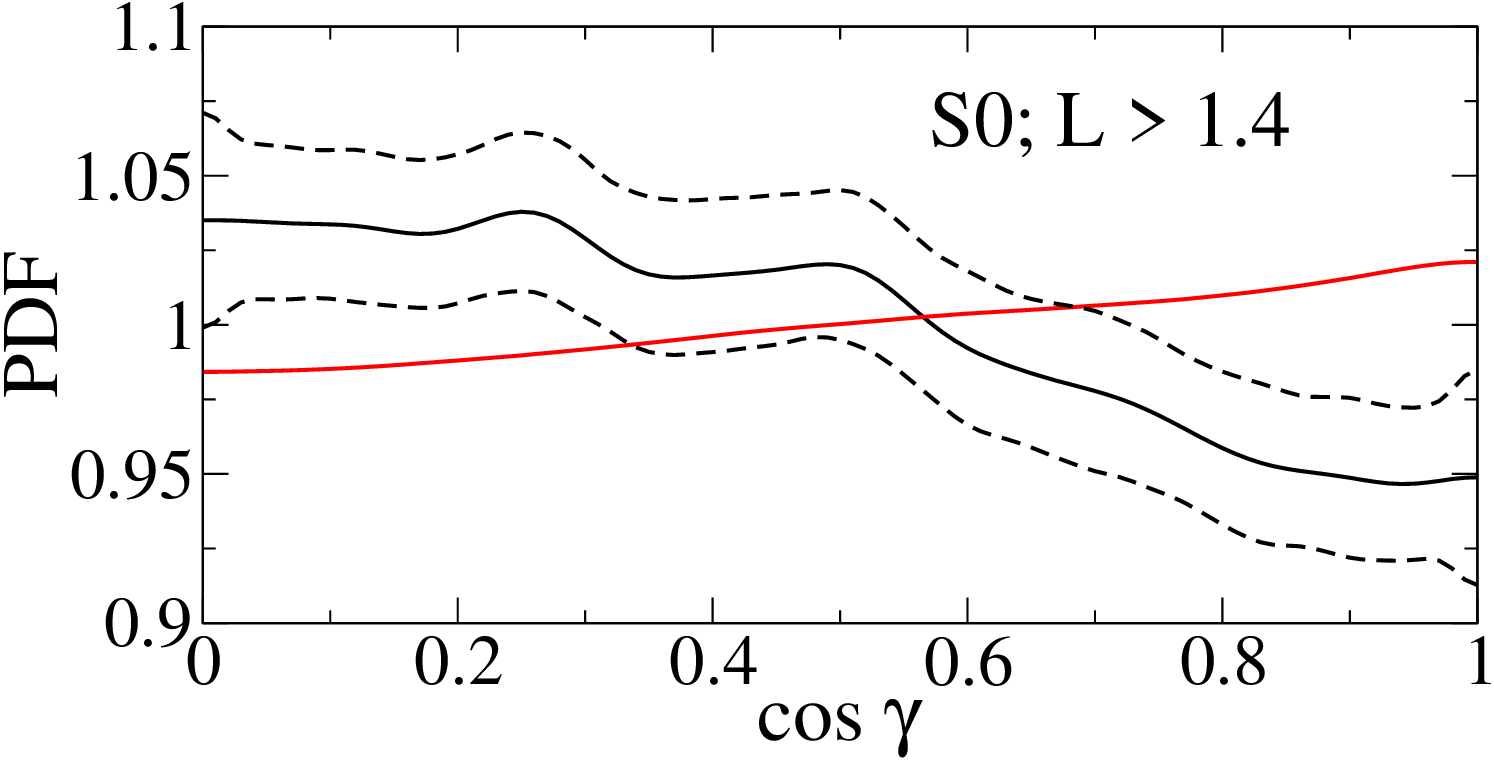}

\includegraphics[width=0.49\hsize,clip]{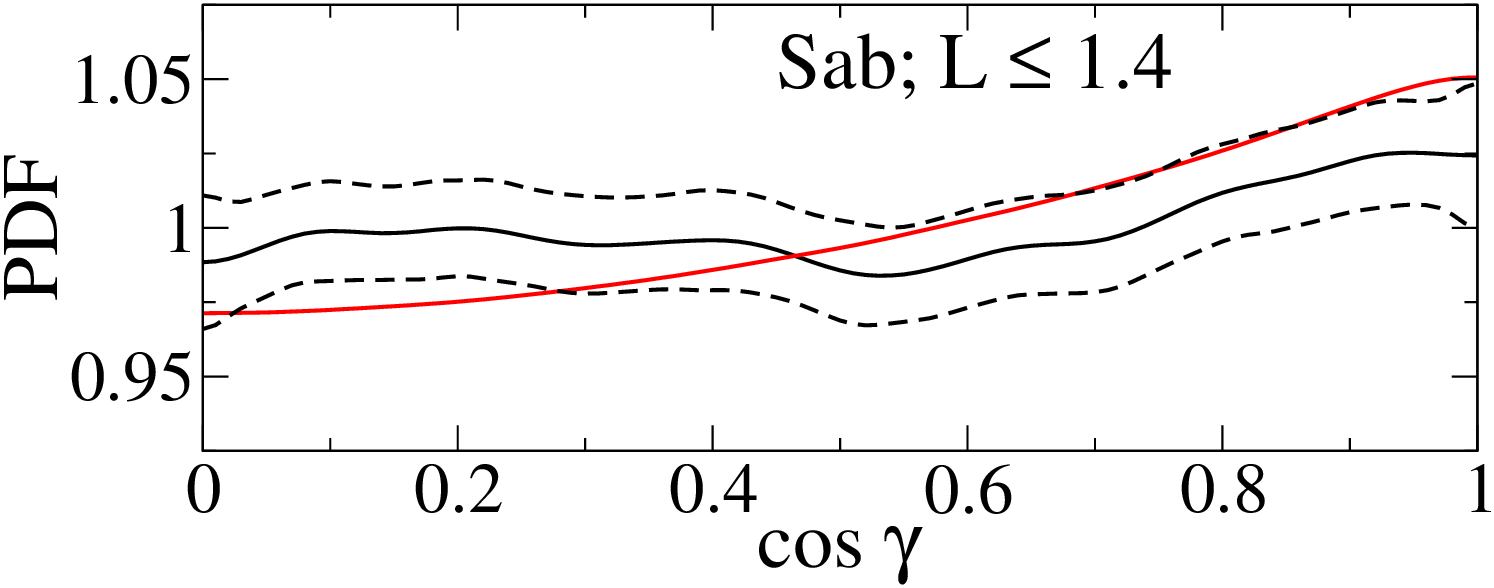}
\includegraphics[width=0.49\hsize,clip]{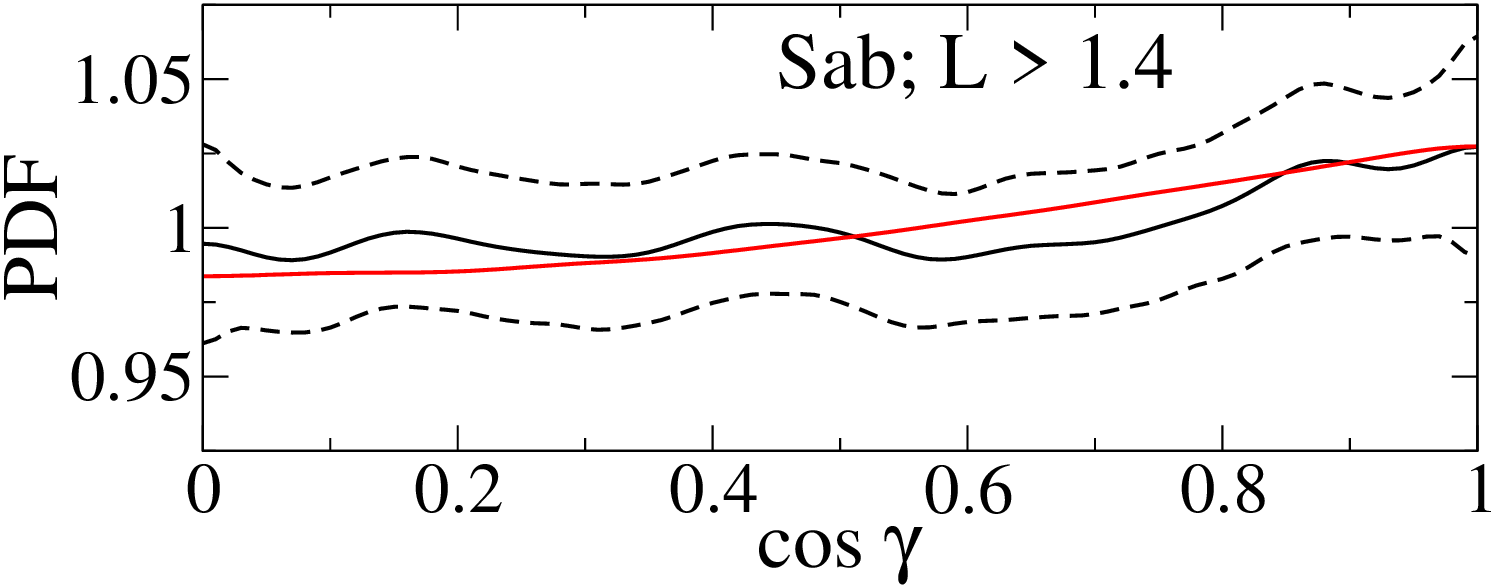}

\includegraphics[width=0.49\hsize,clip]{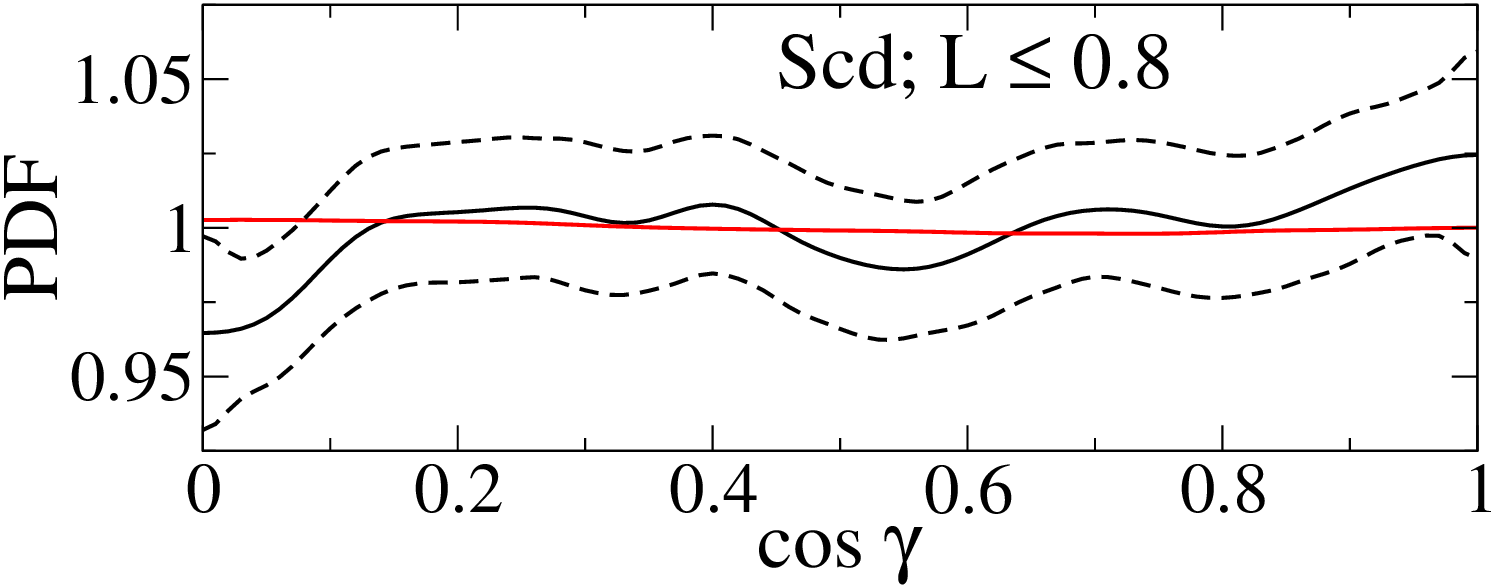}
\includegraphics[width=0.49\hsize,clip]{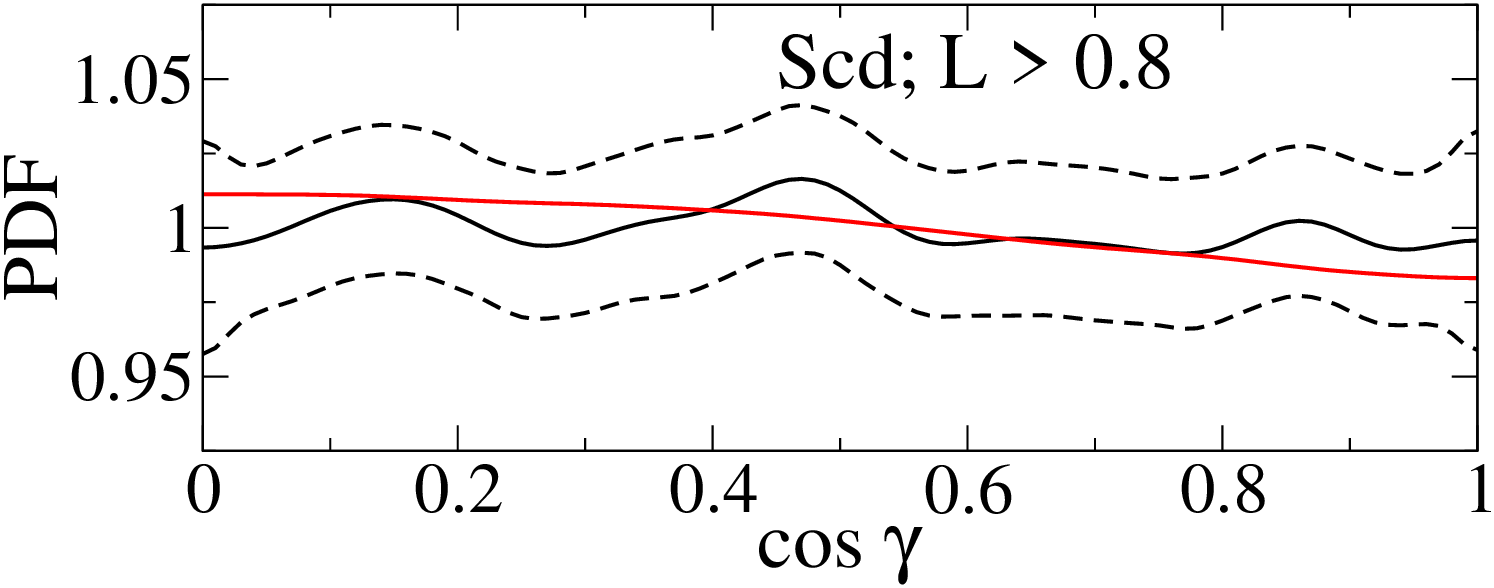}
\caption{Same as Fig.~\ref{cos_gamma_ei_mny} but for E, S0, Sab, and Scd galaxies selected by the $r$-band luminosity $L$; $\nu = 7.0$~$h^{-1}\mathrm{Mpc}$.}
\label{pdf_lumr}
\end{figure}

\begin{figure}
\centering
\includegraphics[width=0.49\hsize,clip]{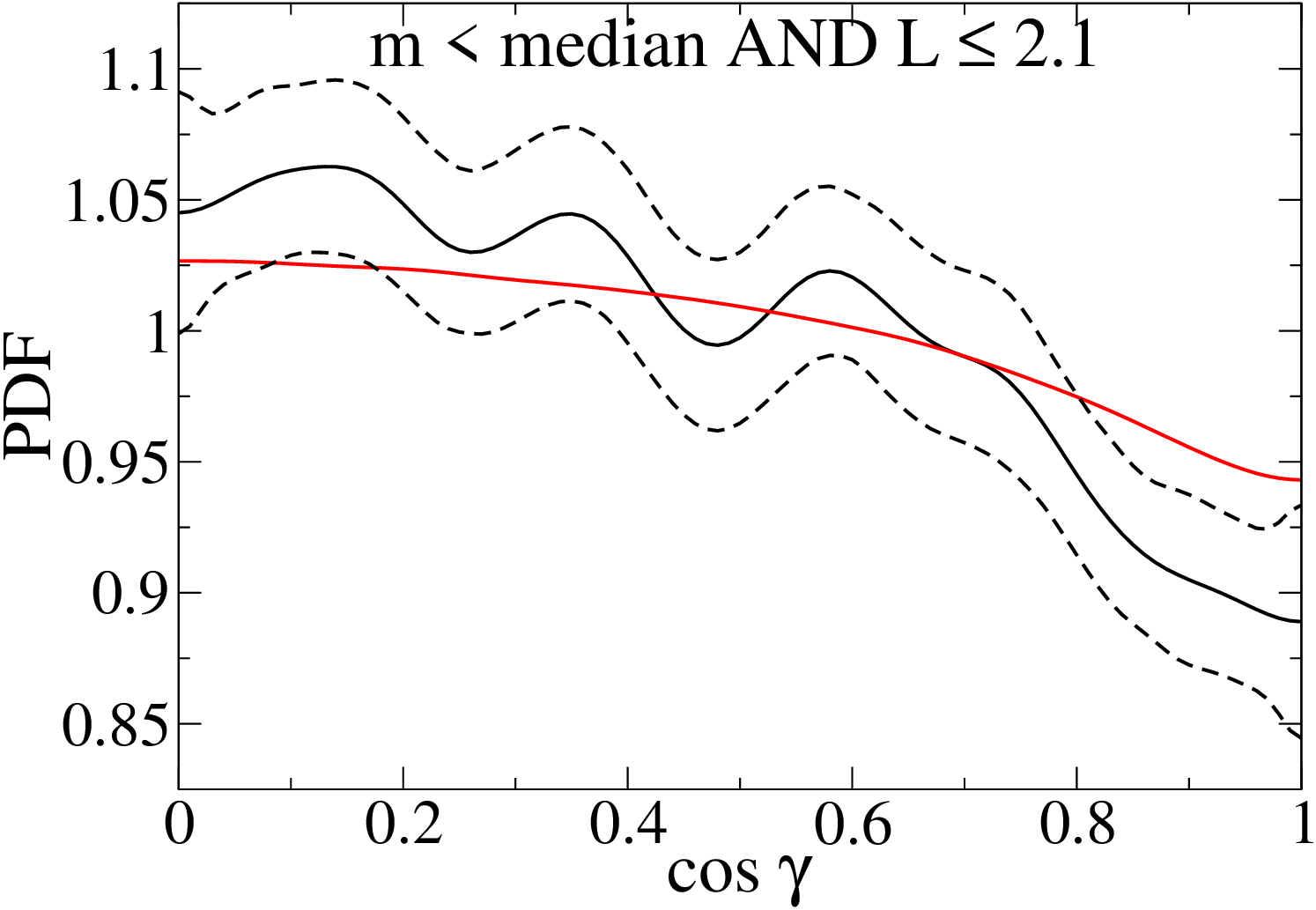}
\includegraphics[width=0.49\hsize,clip]{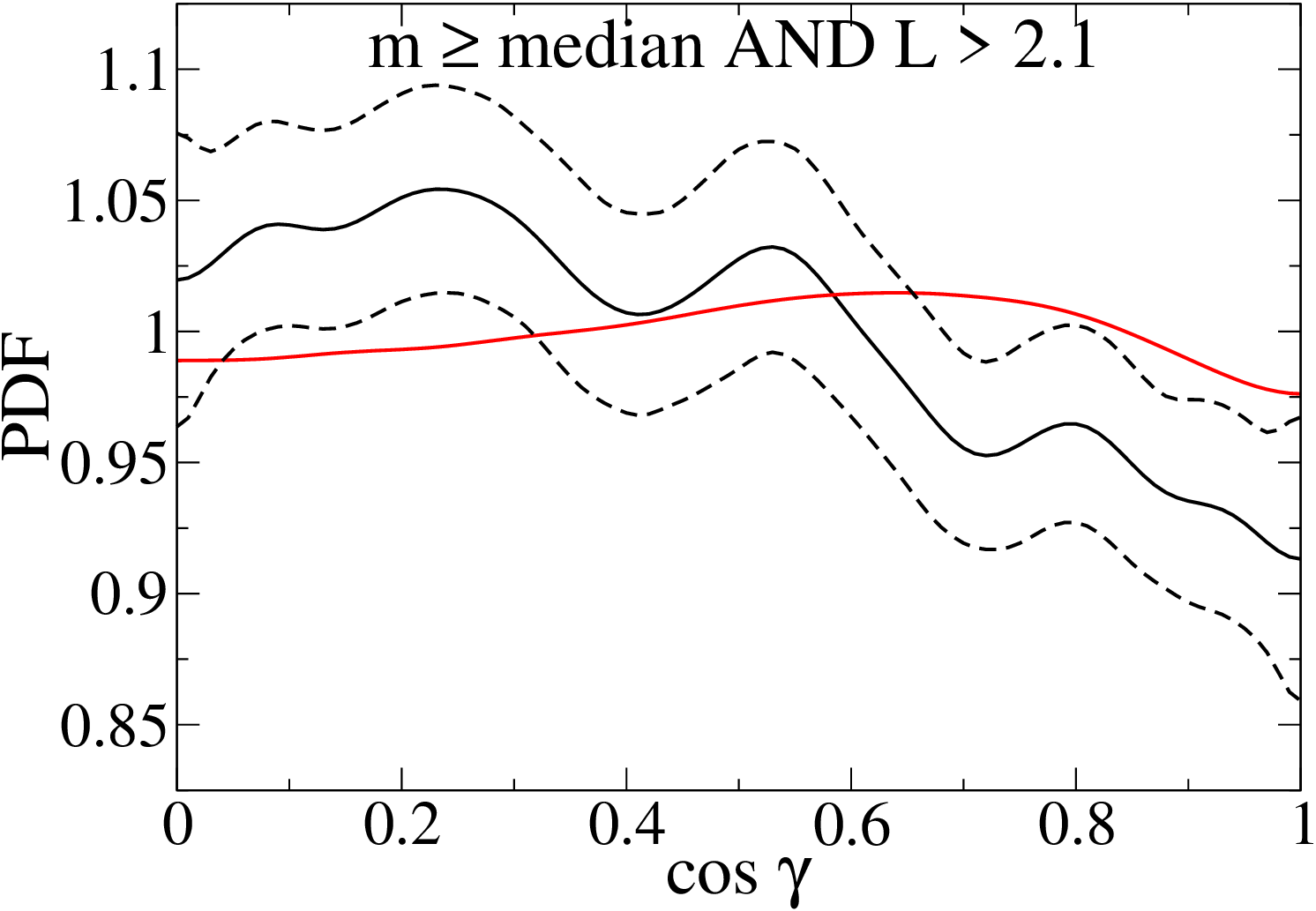}
\caption{Same as Fig.~\ref{cos_gamma_ei_mny} but for minor axes of E galaxies filtered by $r$-band magnitude $m$ and luminosity $L$; $\nu = 7.0$~$h^{-1}\mathrm{Mpc}$.}
\label{pdf_E_magr_lumr}
\end{figure}

We can see in Fig.~\ref{pdf_lumr} that E galaxies with a higher $r$-band luminosity $L$ show a stronger perpendicular alignment signal.
The estimated critical luminosity $\alpha_{\mathrm{L}} \approx 2.1$.
We are aware that the growth of the perpendicular alignment signal of E galaxies with their luminosity may be caused by difficulties in estimating orientation at high 
apparent magnitudes. To address this, we composed two samples: first, E galaxies with the apparent $r$-band magnitude ($m$) that is smaller than the median, but with the luminosity
$L \le 2.1$; and second, E galaxies with $m \ge$ median, but $L > 2.1$. The size of the first sample is $21027$ and the second sample contains $14244$ E galaxies.
Galaxies in the first sample have more reliable modelled parameters as they are apparently brighter. It has been tested that
the strength of the alignment signal for E galaxies depends strongly on $m$. If there were no dependence on
$L$, we could expect the first sample to give stronger alignment signal than the second sample.
We can see in Fig.~\ref{pdf_E_magr_lumr} that the second sample gives us the alignment signal with at least the same strength as the first sample. This suggests that E
galaxies with $L > 2.1$ give physically stronger perpendicular alignment signal than E galaxies with $L \le 2.1$.

We estimated the critical luminosity for S0 galaxies $\alpha_{\mathrm{L}} \approx 1.4$.
One can see in Fig.~\ref{pdf_lumr} that the perpendicular alignment signal for S0 galaxies increases with $L$.
However, in this case we could not rule out that the difference in the strength of the alignment signal may be induced by a more correctly estimated orientation at low magnitudes.

Because of selection effects, the distributions of $L$ of Sab galaxies were not significantly different.
Despite this, we found that the perpendicular alignment signal for Sab galaxies vanishes at a higher luminosity than $\alpha_{\mathrm{L}} \approx 1.4$ (see Fig.~\ref{pdf_lumr}).

Selection effects did not allow us to detect any difference between the distributions of $L$ of Scd galaxies.
However, we split the Scd sample at $L = 0.8$. We can see in Fig.~\ref{pdf_lumr} that the lower luminosity sample of Scd galaxies shows
a parallel alignment signal, while the higher luminosity sample does not give a measurable signal.

\subsection{Alignment of galaxies at $\nu = 7.0$~$h^{-1}\mathrm{Mpc}$ in groups with different richness}
\subsubsection{Single galaxies}
\begin{figure}
\centering
\includegraphics[width=0.49\hsize,clip]{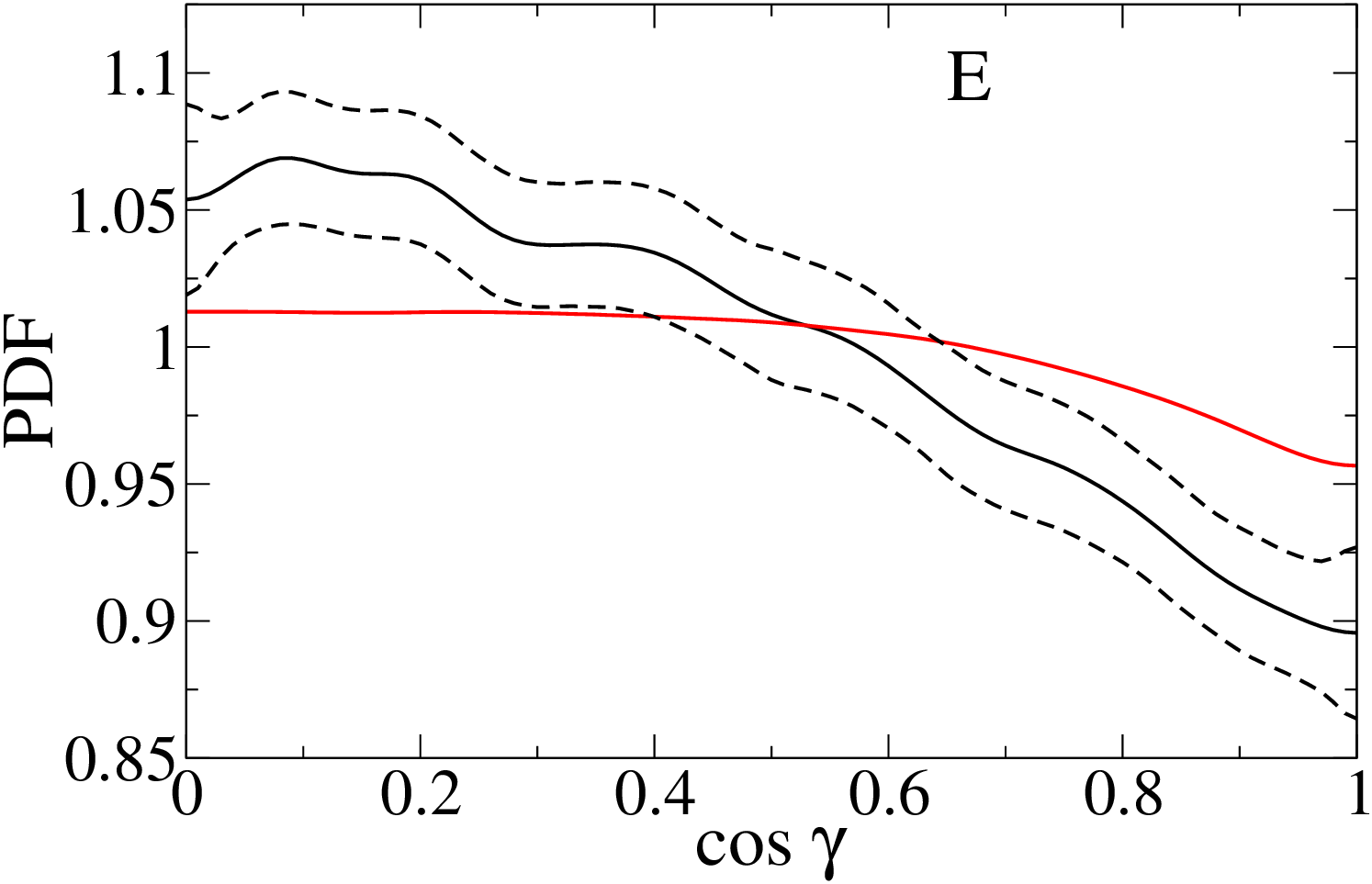}
\includegraphics[width=0.49\hsize,clip]{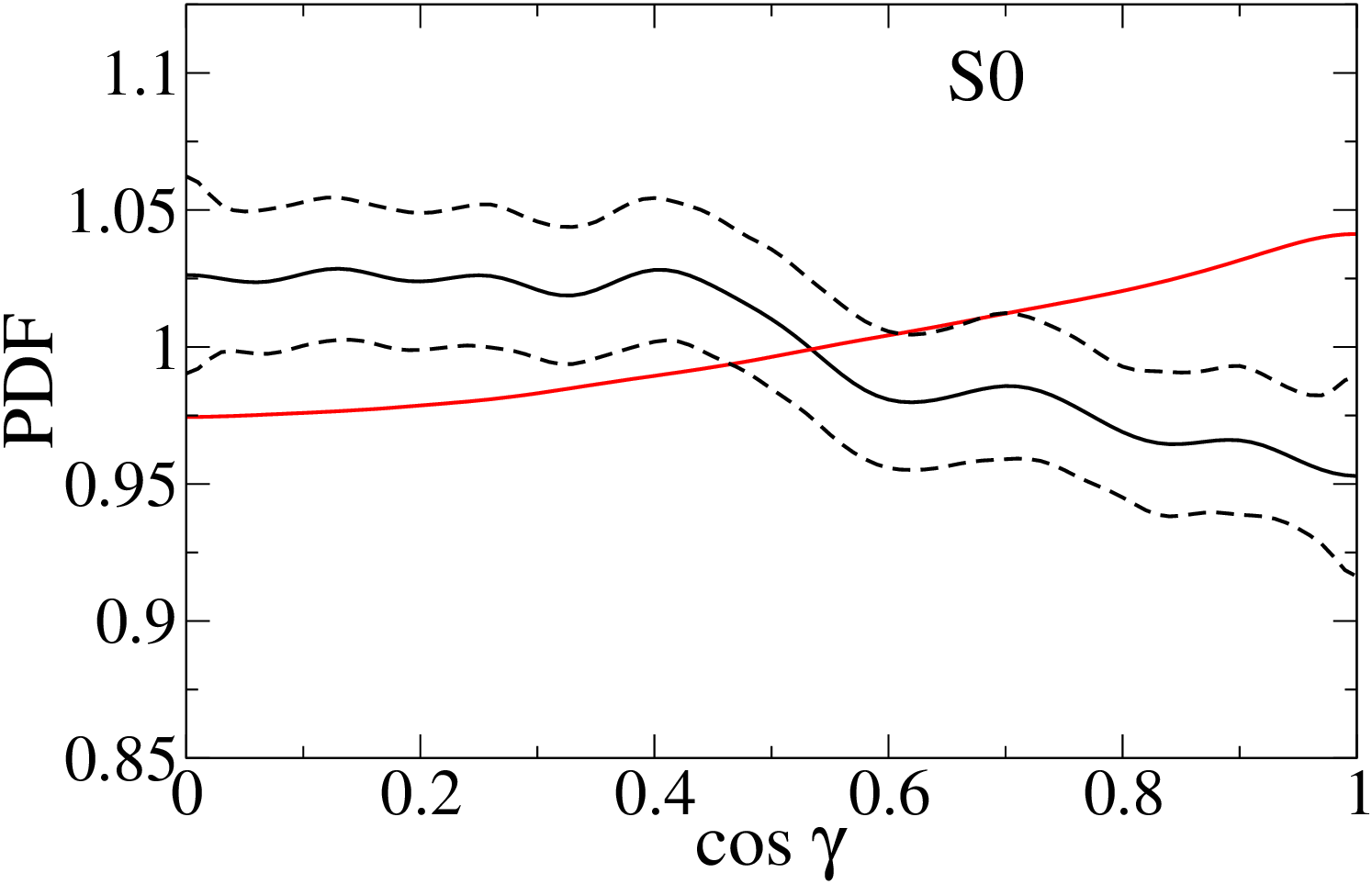}

\includegraphics[width=0.49\hsize,clip]{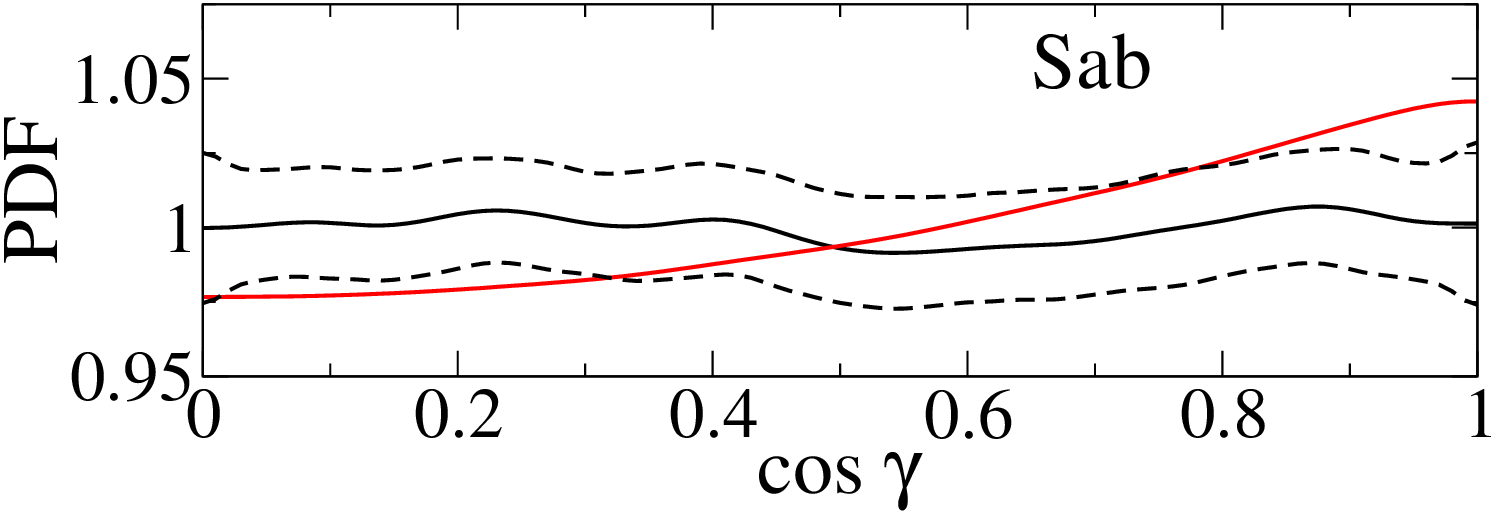}
\includegraphics[width=0.49\hsize,clip]{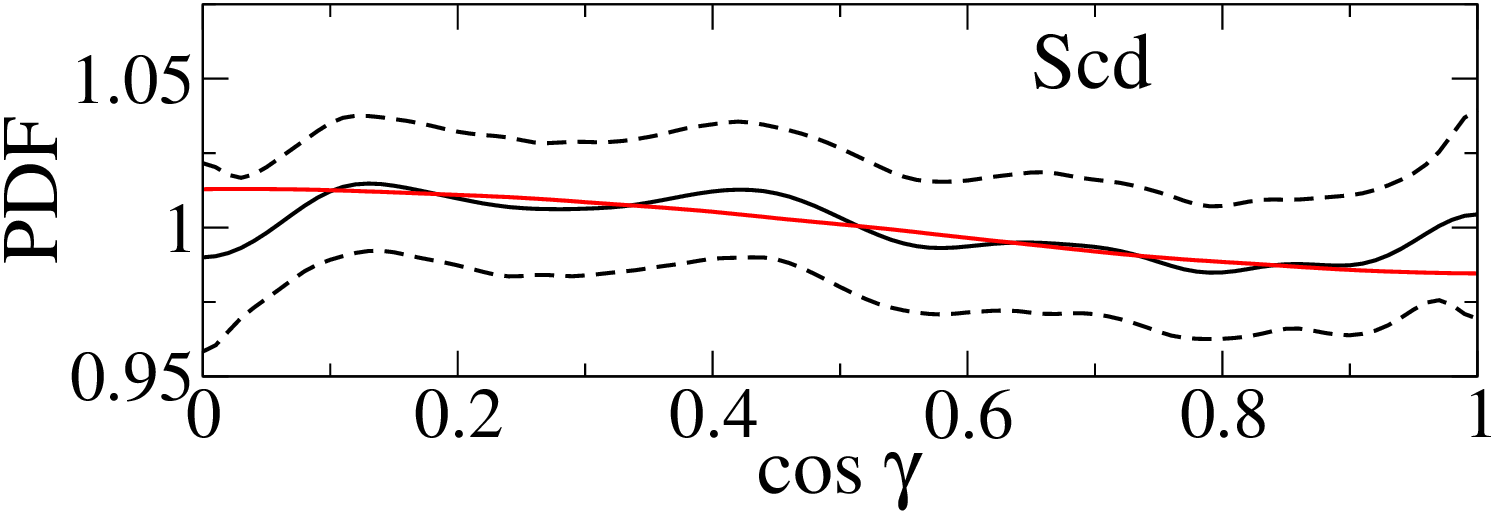}
\caption{Same as Fig.~\ref{cos_gamma_ei_mny} but for single E, S0, Sab, and Scd galaxies; $\nu = 7.0$~$h^{-1}\mathrm{Mpc}$.}
\label{cos_gamma_hc_nrich_le_1_ny70}
\end{figure}

We can see in Figs.~\ref{cos_gamma_hc_none_ny70} and~\ref{cos_gamma_hc_nrich_le_1_ny70} that single E and S0 galaxies show a strong perpendicular alignment signal
that is comparable to the signal of unfiltered samples.

The sample of single Sab galaxies shows a stronger perpendicular alignment signal than the unfiltered sample. On the other hand, the sample of single Scd galaxies
does not have significant alignment. This cannot be a result of a smaller sample size because the next sample of Scd galaxies (in $2 \dots 3$ member groups) shows alignment
regardless of an even smaller sample size.

\subsubsection{Galaxies in $2 \dots 3$ member groups}
\begin{figure}
\centering
\includegraphics[width=0.49\hsize,clip]{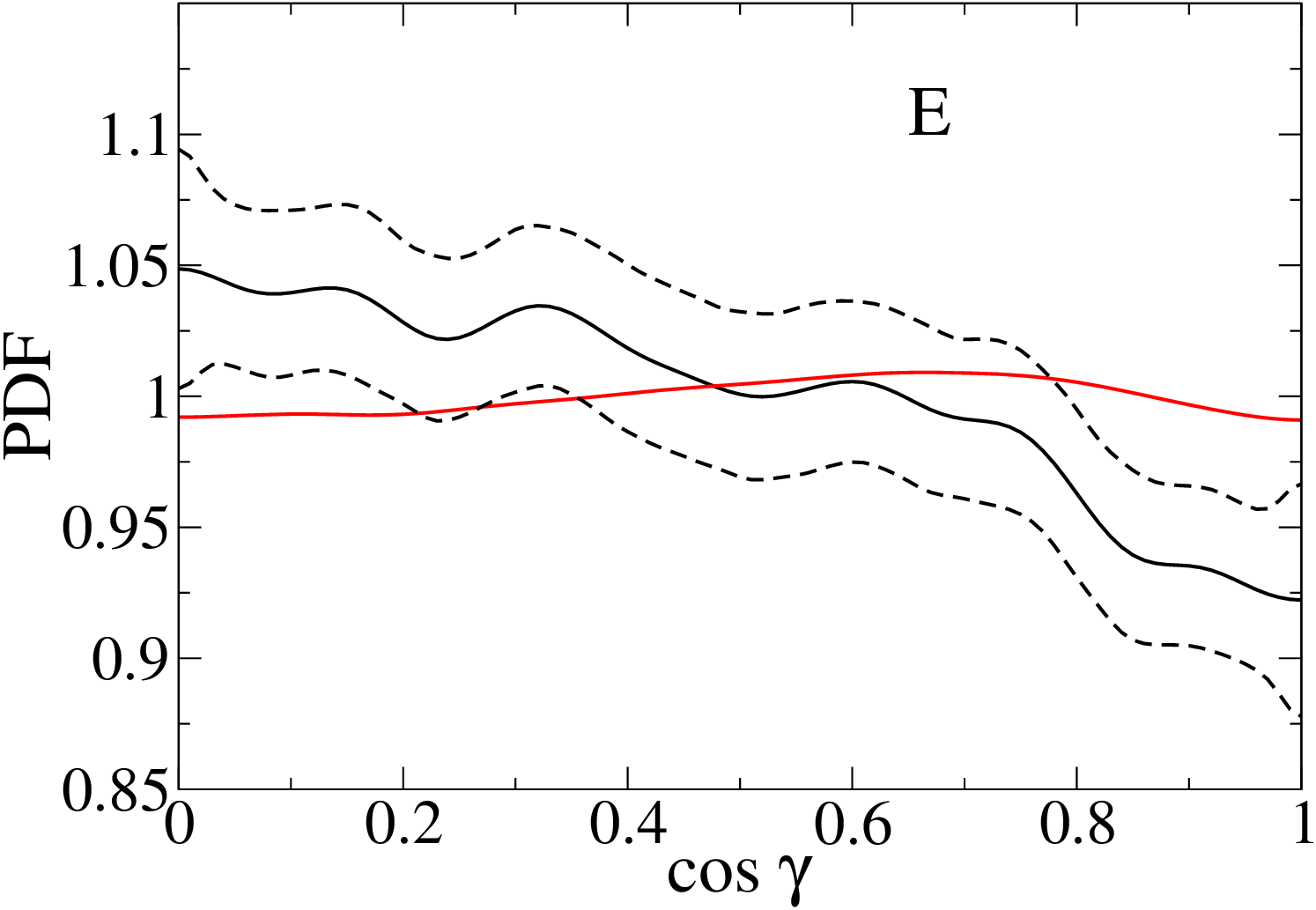}
\includegraphics[width=0.49\hsize,clip]{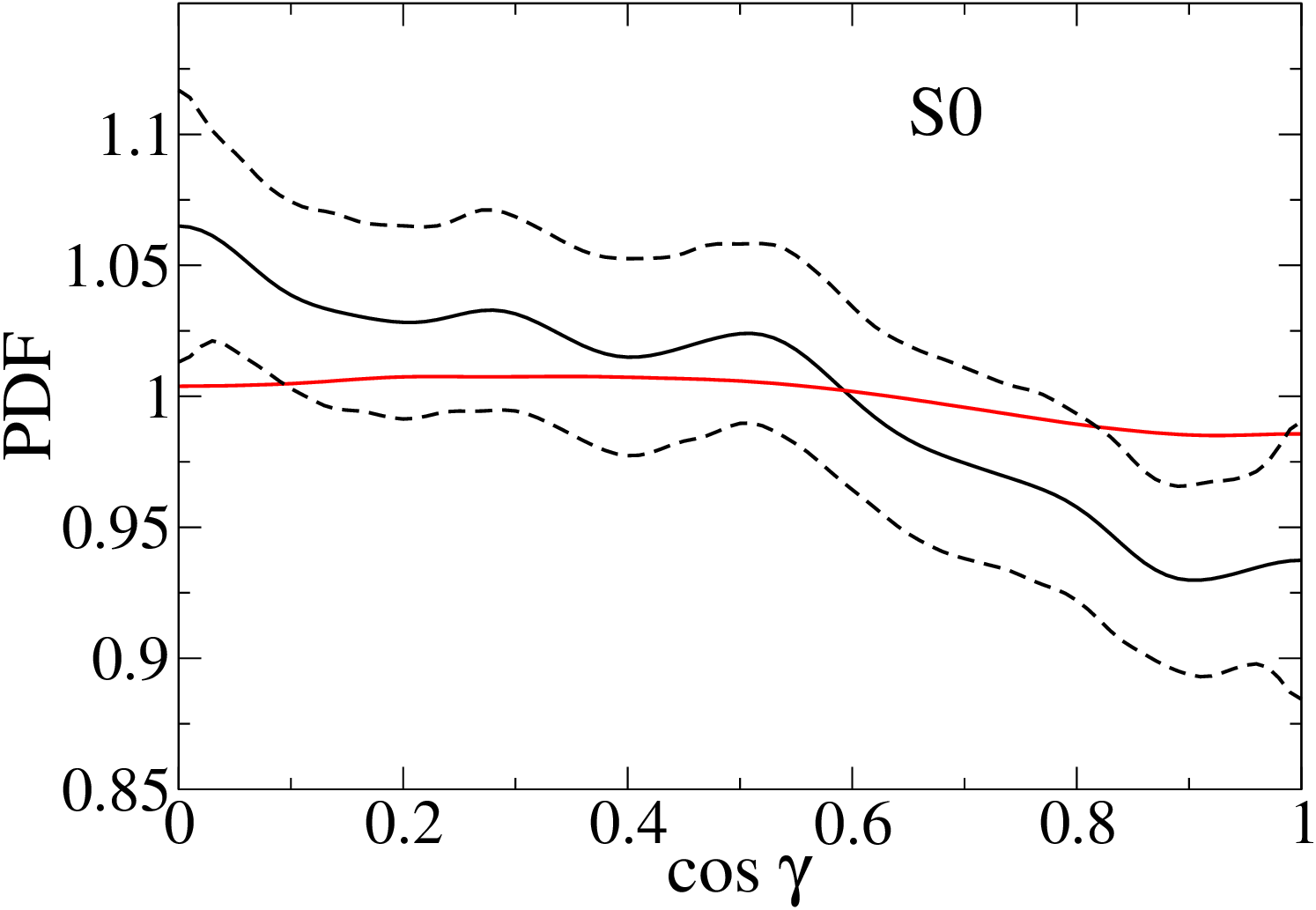}

\includegraphics[width=0.49\hsize,clip]{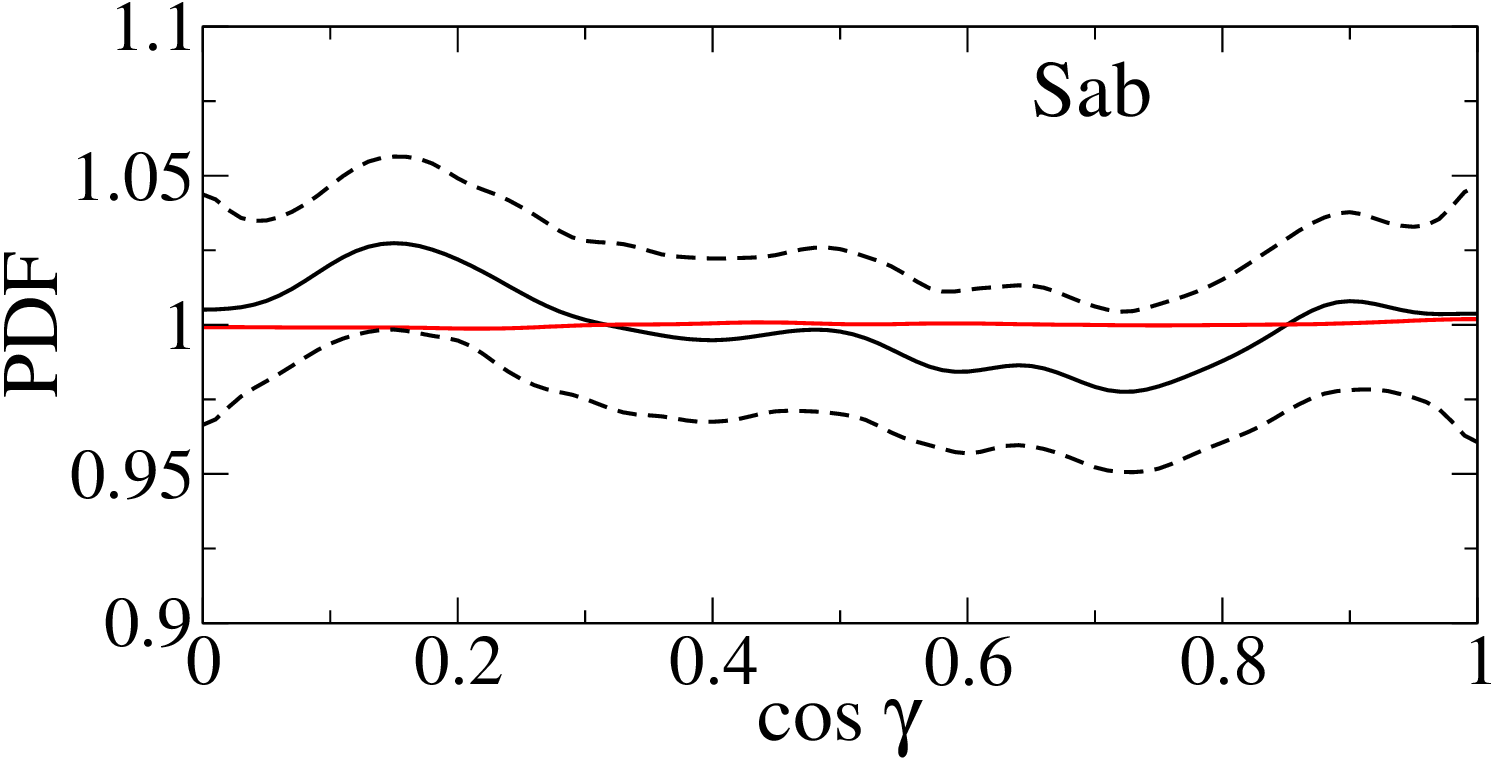}
\includegraphics[width=0.49\hsize,clip]{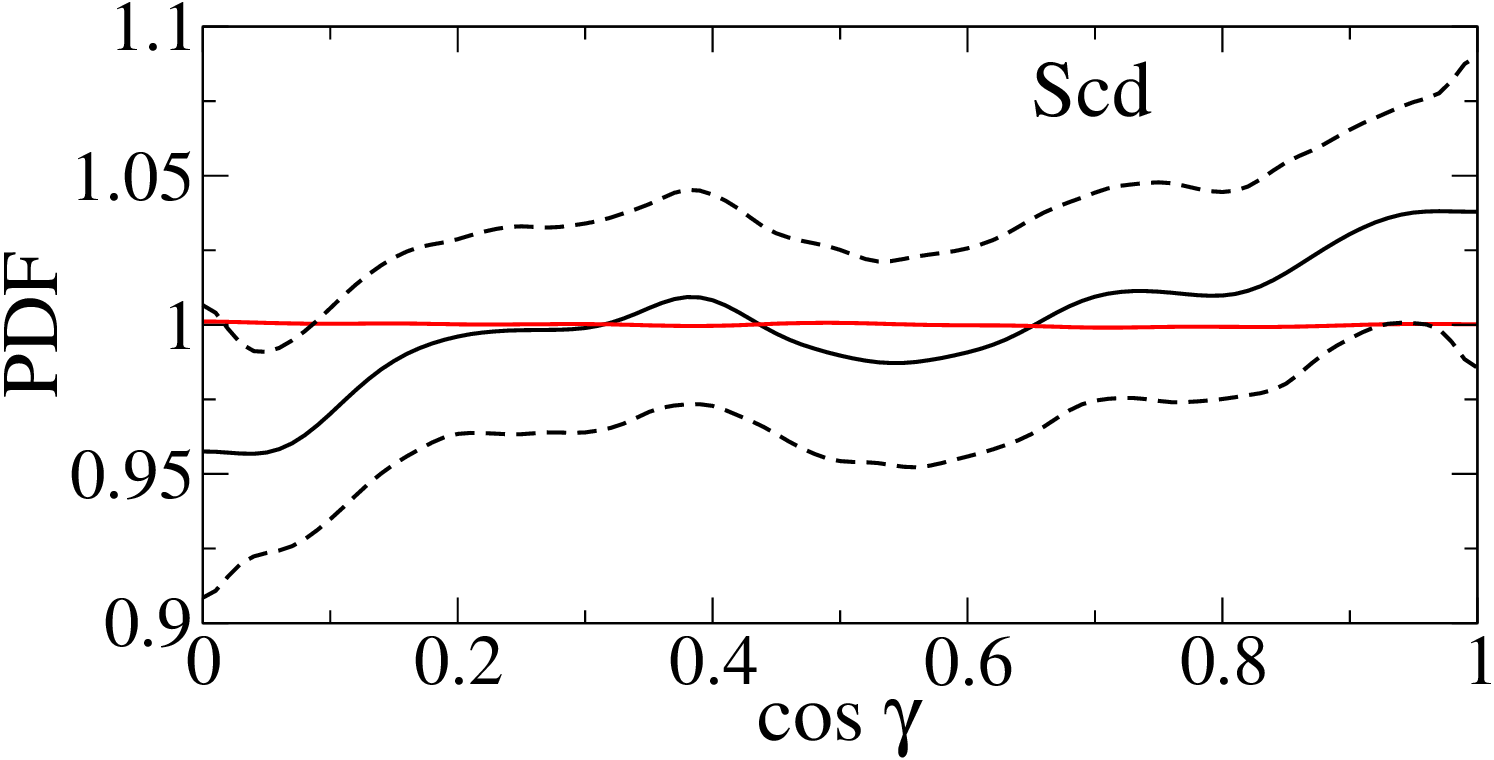}
\caption{Same as Fig.~\ref{cos_gamma_ei_mny} but for E, S0, Sab, and Scd galaxies in $2 \dots 3$ member groups; $\nu = 7.0$~$h^{-1}\mathrm{Mpc}$.}
\label{cos_gamma_hc_nrich_2to3_ny70}
\end{figure}

E and S0 galaxies in $2 \dots 3$ member groups give a clear perpendicular alignment signal (see Fig.~\ref{cos_gamma_hc_nrich_2to3_ny70}).
Sab galaxies in $2 \dots 3$ member groups do not show significant alignment, while Scd galaxies in these groups show a parallel alignment signal that is comparable to
the signal of the unfiltered sample of Scd galaxies. The loss or weakening of alignment of Sab galaxies in $2 \dots 3$ member groups can be considered a real effect, as the
size of this sample is larger than for the corresponding Scd sample and we do not expect more complicated modelling of Sab galaxies in $2 \dots 3$ member groups
compared to single Sab galaxies.

\subsubsection{Galaxies in $4 \dots 10$ member groups}
\begin{figure}
\centering
\includegraphics[width=0.49\hsize,clip]{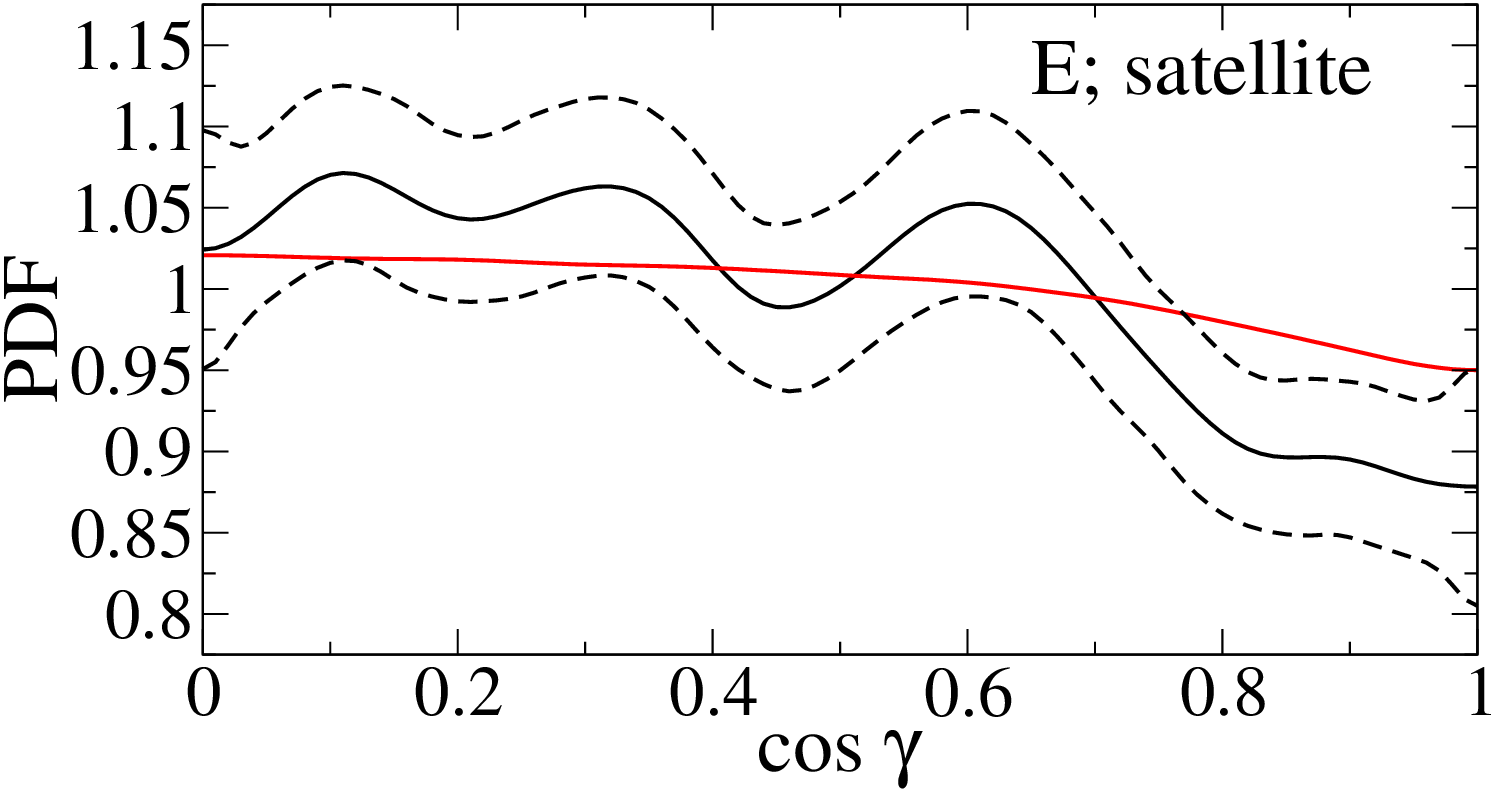}
\includegraphics[width=0.49\hsize,clip]{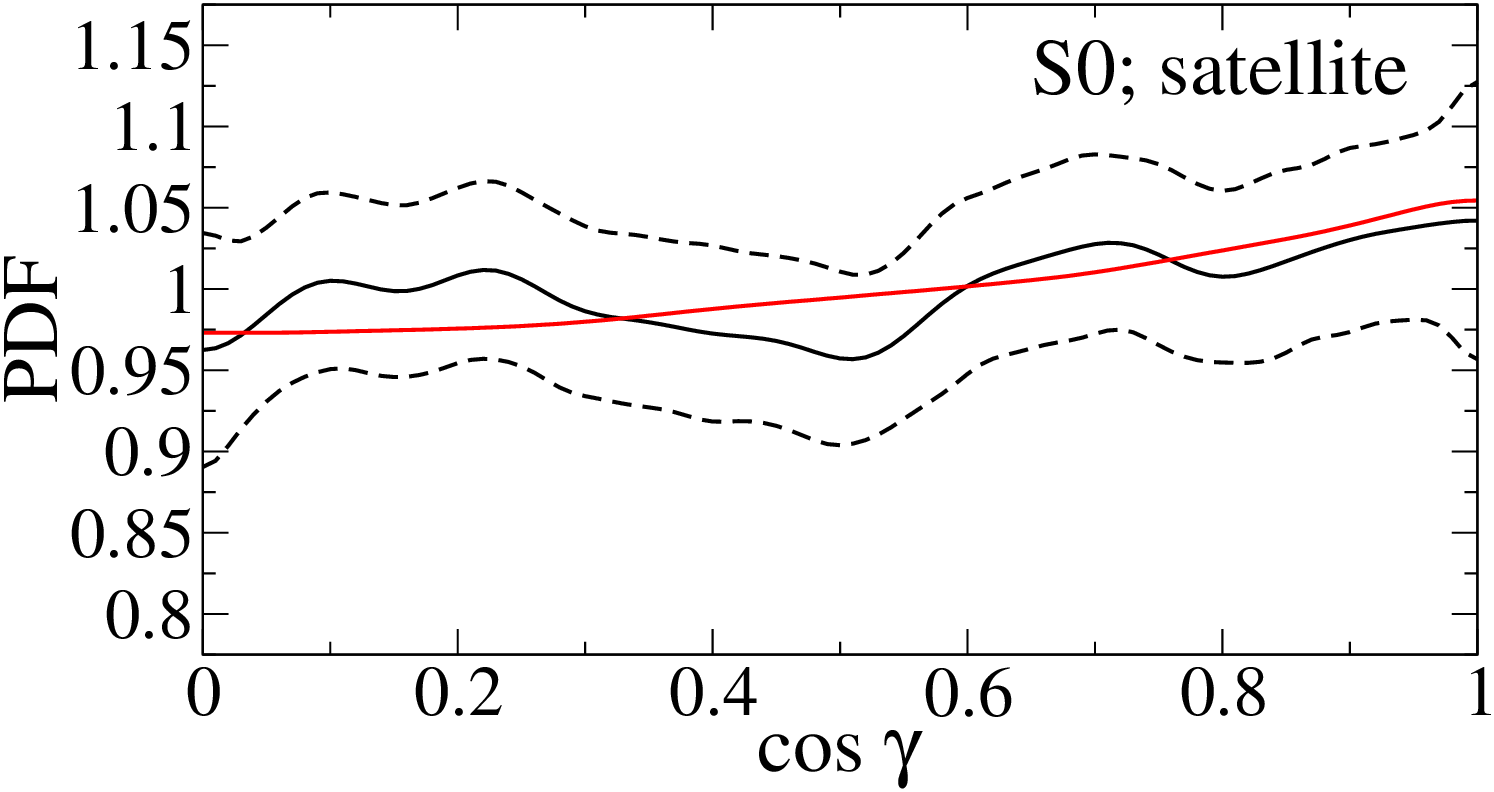}

\includegraphics[width=0.49\hsize,clip]{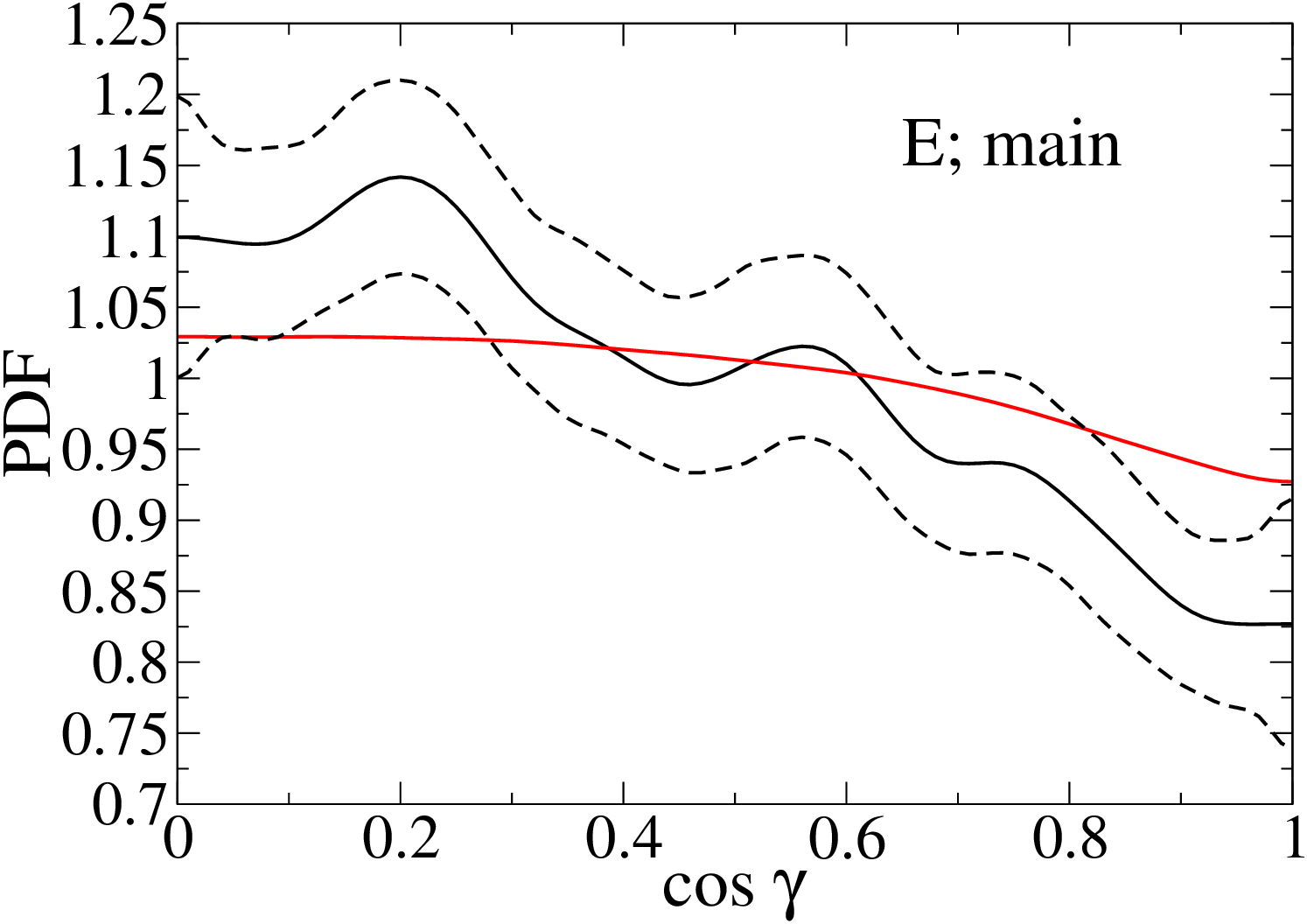}
\includegraphics[width=0.49\hsize,clip]{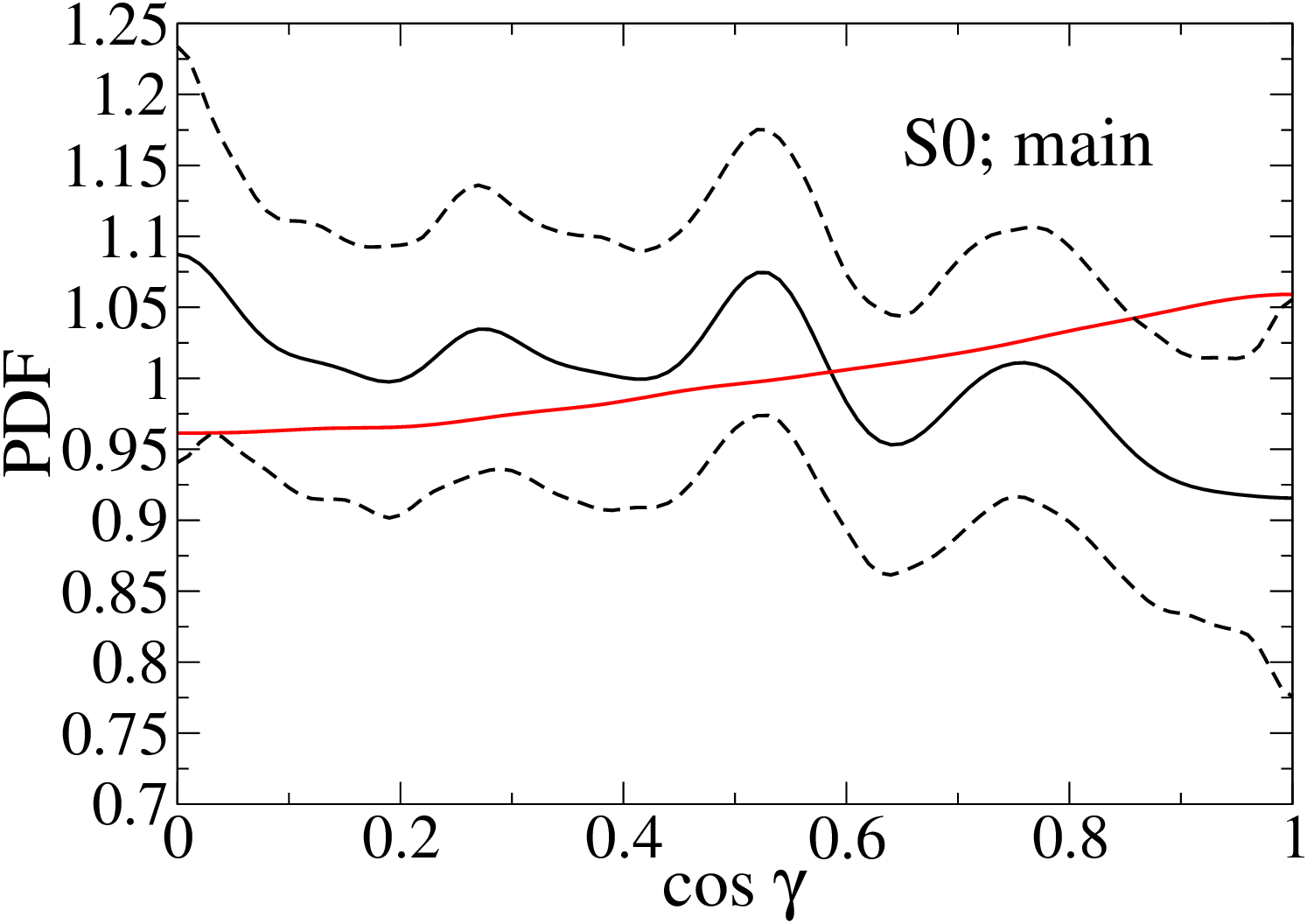}
\caption{Same as Fig.~\ref{cos_gamma_ei_mny} but for E and S0 galaxies in $4 \dots 10$ member groups; samples of main galaxies are in the bottom row;
satellite galaxies are in the upper row; and $\nu = 7.0$~$h^{-1}\mathrm{Mpc}$.}
\label{cos_gamma_hc_nrich_4to10_ny70}
\end{figure}

We can see in Fig.~\ref{cos_gamma_hc_nrich_4to10_ny70} that satellite E galaxies in $4 \dots 10$ member groups
show a perpendicular alignment signal. Satellite S0 galaxies in these groups do not have significant alignment preferences.
Meanwhile, E and S0 galaxies, which are the main galaxies of the given groups, show a perpendicular alignment signal.

\begin{figure}
\centering
\includegraphics[width=\hsize,clip]{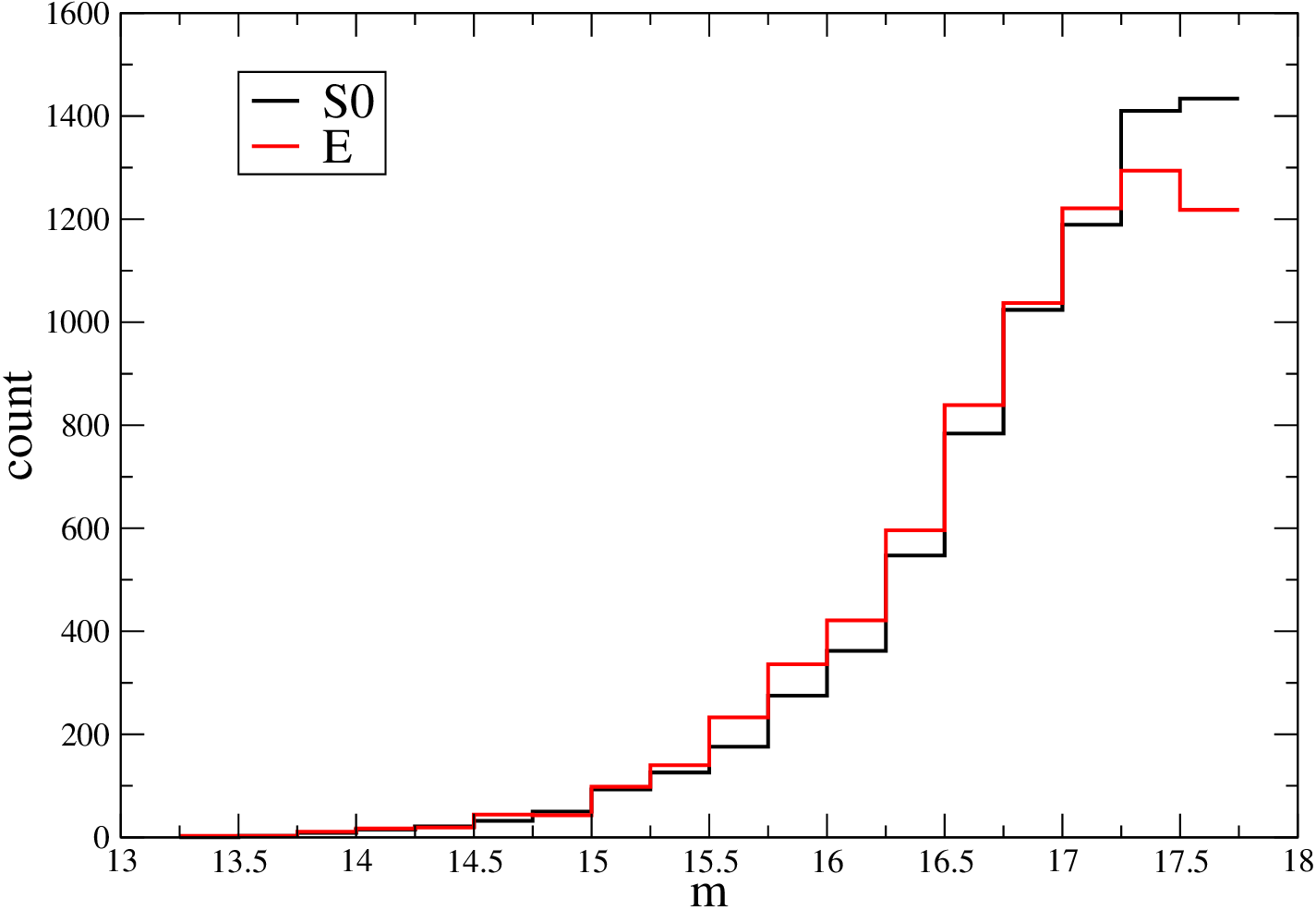}
\caption{Histograms of apparent $r$-band magnitude $m$ of E (red) and S0 (black) satellite galaxies in $4 \dots 10$ member groups.}
\label{E_S0__nrich_4to10_rank_gt_1__magr_histogram}
\end{figure}

We have seen that the unfiltered samples, as well as the $2 \dots 3$ member groups and the single samples of E and S0 galaxies show nearly identical alignment
signals. To understand whether the loss of the alignment signal of satellite S0 galaxies in $4 \dots 10$ member groups is a real effect or an observational bias, we refer to
Fig.~\ref{E_S0__nrich_4to10_rank_gt_1__magr_histogram}. We can consider that the histograms of apparent $r$-band magnitude ($m$) of satellite E and S0 galaxies in $4 \dots 10$ member
groups are not sufficiently different to induce a significant difference in the alignment analysis. The median values of $m$ are $16.99$ and $17.07$ for the E and S0 samples,
respectively. The sizes of the E and S0 samples are nearly equal: $7627$ and $7632$. We can expect that the data from modelling of the S0 sample has a higher quality, as for the
E sample we have no information about the inclination angle.
We can conclude that the absence of the alignment signal of the satellite S0 galaxy sample in $4 \dots 10$ member groups is physical.

Sab and Scd galaxies did not show alignment in $4 \dots 10$ member groups.

\subsubsection{Galaxies in groups with more than ten members}
\begin{figure}
\centering
\includegraphics[width=0.49\hsize,clip]{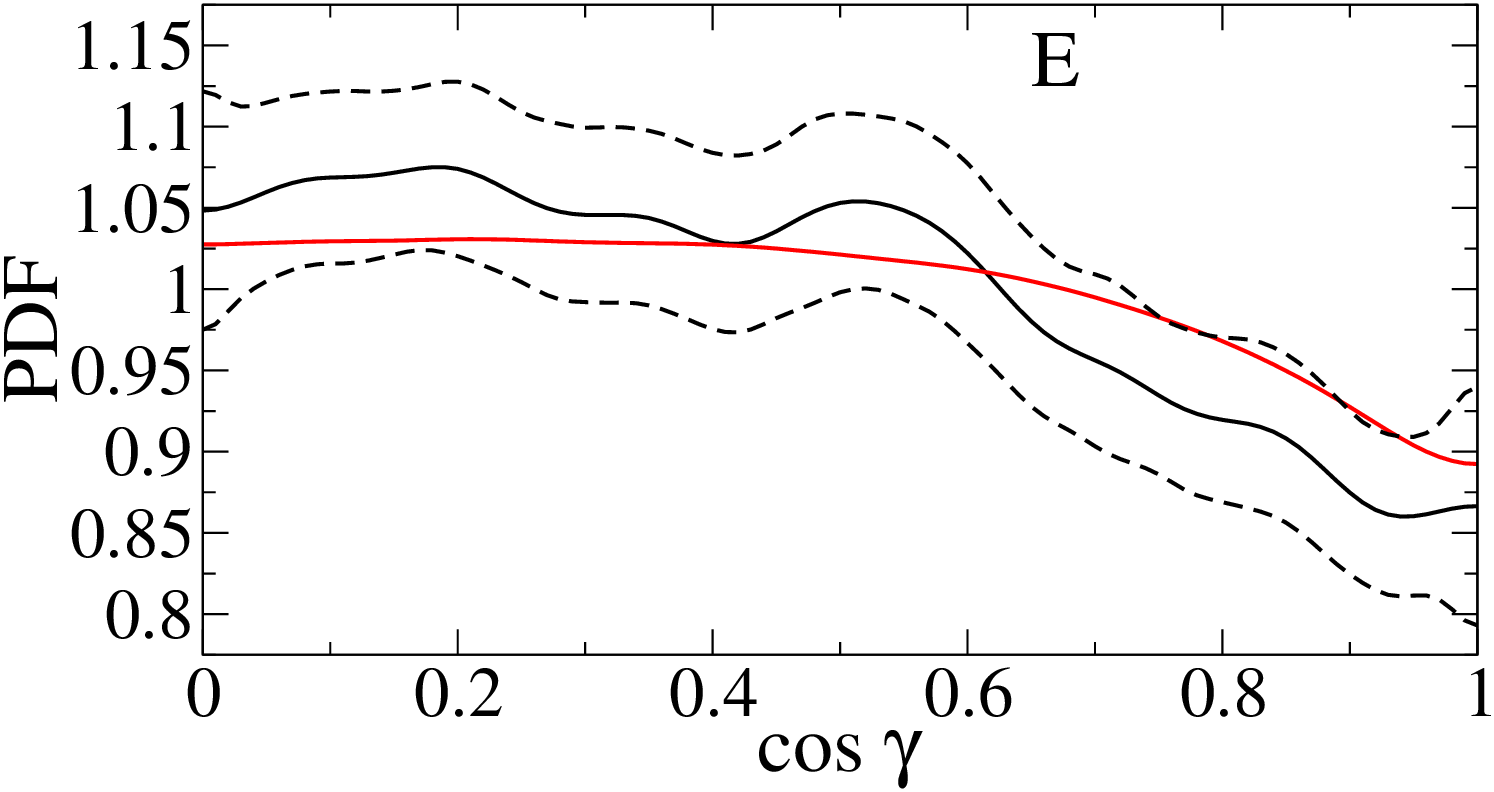}
\includegraphics[width=0.49\hsize,clip]{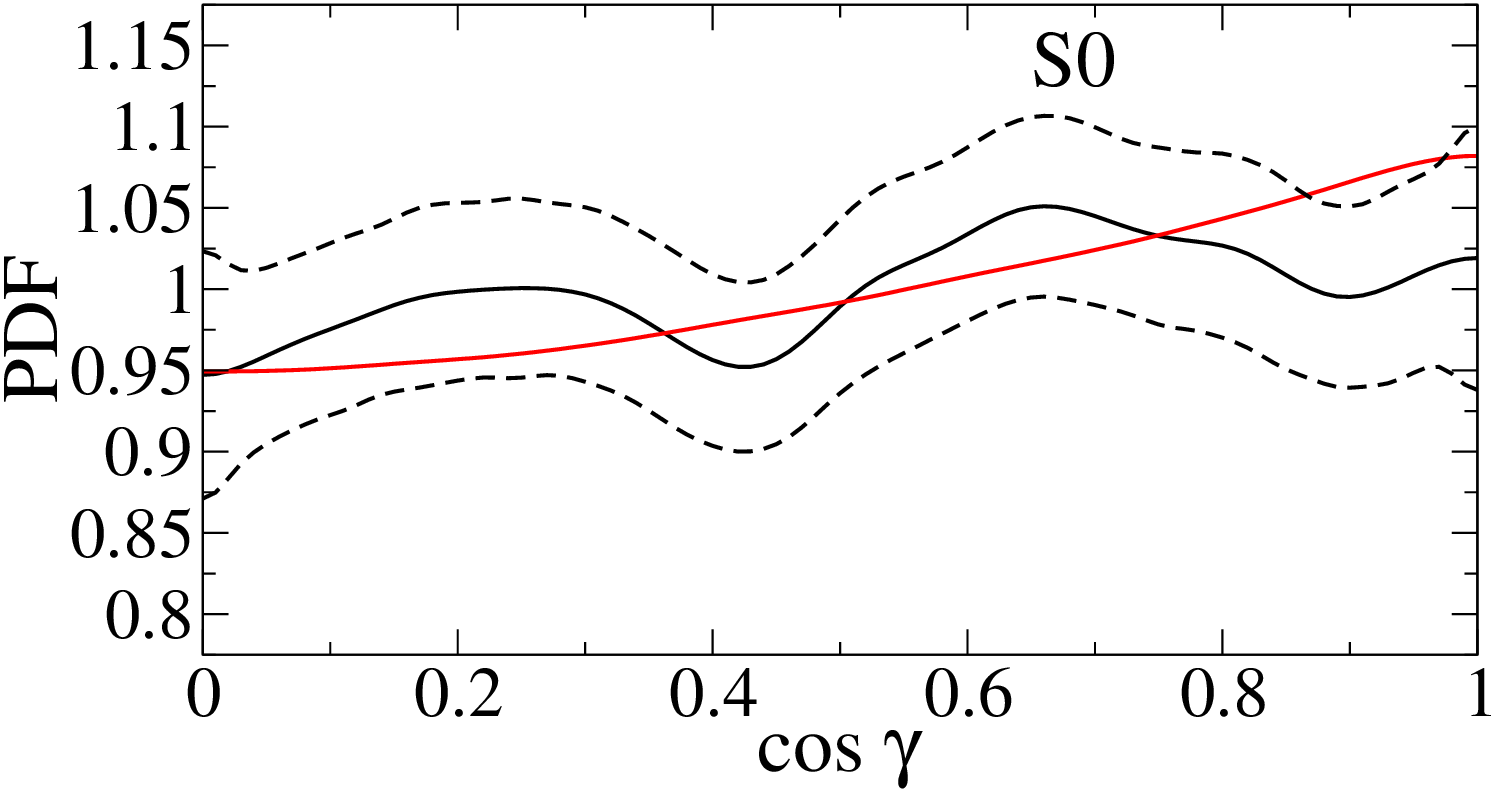}
\caption{Same as Fig.~\ref{cos_gamma_ei_mny} but for E and S0 galaxies in richer than $10$ member groups; $\nu = 7.0$~$h^{-1}\mathrm{Mpc}$.}
\label{pdf_E_S0_nrich_gt_10}
\end{figure}

E and S0 galaxies in groups with richness $> 10$ show no or a remarkably weakened perpendicular alignment signal (see Fig.~\ref{pdf_E_S0_nrich_gt_10}). The fading of the perpendicular
alignment signal for E galaxies in these groups can be considered physical, as the sample of satellite E galaxies in $4 \dots 10$ member groups with nearly equal
sample size gave us a stronger alignment signal. Moreover, we cannot expect better modelling quality for the sample of satellite E galaxies in $4 \dots 10$ member groups.
We also did not detect significant alignment of early-type main galaxies in groups with richness $> 10$.

We did not detect significant alignment of Sab and Scd galaxies in richer than ten member groups. The sample of Scd galaxies in these groups showed signs of negligible parallel
alignment that was less significant than the signal for S0 galaxies.

\subsection{Summary of the results}
The main results of alignment analysis for E, S0, Sab, and Scd galaxies are summarised in Table~\ref{summary_table}. In this table, we show for the neighbourhood
radius $\nu = 7.0$~$h^{-1}\mathrm{Mpc}$ whether the alignment of minor axes of E galaxies and spin axes of S0, Sab, and Scd galaxies relative to the DALE is parallel or
perpendicular, or no alignment was detected.
\begin{table}
\caption{Alignment of minor axes of E galaxies and spin axes of S0, Sab, and Scd galaxies relative to DALE for the neighbourhood radius $\nu = 7.0$~$h^{-1}\mathrm{Mpc}$.}
\label{summary_table}
\centering
\begin{tabular}{@{} l | c c c c @{}}
\hline\hline
   Selection                                    & E           & S0          & Sab        & Scd \\
\hline
   all                                          & $\perp$     & $\perp$     & $\perp$    & $\parallel$  \\
   $N_{\mathrm{rich}} = 1$                      & $\perp$     & $\perp$     & $\perp$    & $\bigcirc$  \\
   $2 \le N_{\mathrm{rich}} \le 3$              & $\perp$     & $\perp$     & $\bigcirc$ & $\parallel$  \\
   $N_{\mathrm{rich}} > 10$                     & $\bigcirc$  & $\perp$     & $\bigcirc$ & $\bigcirc$  \\
   $4 \le N_{\mathrm{rich}} \le 10$, satellite  & $\perp$     & $\bigcirc$  & $\bigcirc$ & $\bigcirc$  \\
   $4 \le N_{\mathrm{rich}} \le 10$, main       & $\perp$     & $\perp$     & $\bigcirc$ & $\bigcirc$  \\
   $D \le \alpha_{\mathrm{D}}$                  & $\perp$     & $\perp$     & $\perp$    & $\bigcirc$  \\
   $D > \alpha_{\mathrm{D}}$                    & $\perp$     & $\perp$     & $\bigcirc$ & $\parallel$  \\
   $L \le \alpha_{\mathrm{L}}$                  & $\perp$     & $\perp$     & $\perp$    & $\parallel$  \\
   $L > \alpha_{\mathrm{L}}$                    & $\perp$     & $\perp$     & $\bigcirc$ & $\bigcirc$  \\
\hline
\end{tabular}
\tablefoot{
$\parallel$~-- parallel, $\perp$~-- perpendicular, $\bigcirc$~-- no alignment.
$N_{\mathrm{rich}}$ is number of galaxies in group.
The critical values $\alpha_{\mathrm{D}}$ for environmetal density $D$ and $\alpha_{\mathrm{L}}$ for $r$-band luminosity $L$ are given in Sects.~\ref{align_dens}~and~\ref{align_lum}.
}
\end{table}


\section{Discussion}\label{disc_sect}

\subsection{The most sensitive value of $\nu$}
First we address the question of why the strongest alignment signal of E, S0, Sab, and Scd galaxies is 
detected at $\nu = 7 \dots 12$~$h^{-1}\mathrm{Mpc}$ scale. At first sight one may find in Table~\ref{sample_sizes} that the sizes of our samples are largest at this scale.
Larger samples may increase the detectability of the alignment signal compared to small samples with poor statistics, as the width of the confidence interval shrinks with larger sample
size.
However, this is not the case for Fig.~\ref{cos_gamma_ei_mny}, where the perpendicular alignment signal of early-type galaxies is stronger at
$\nu = 2.0$~$h^{-1}\mathrm{Mpc}$ than at $\nu = 24.0$~$h^{-1}\mathrm{Mpc}$ despite a larger sample size at $\nu = 24.0$~$h^{-1}\mathrm{Mpc}$.

At small scales a relatively bigger part of the measured alignment signal comes from the vicinity of clusters of galaxies, as our algorithm fails to find neighbours in sparser areas.
We may measure very noisy DALE estimates in the clusters, when using small $\nu$. On the other hand, our signal begins to vanish at larger scales from noise caused by
neighbouring structures.
\citet{Tempel2014a} found that galaxies are not distributed randomly along filaments, but they have preferred separations of $\sim 7$~$h^{-1}\mathrm{Mpc}$. Using larger $\nu$, we do
not improve the DALE estimation along filaments much, but include more noise from their vicinity.
Therefore, $\nu = 7 \dots 12$~$h^{-1}\mathrm{Mpc}$ is the most sensitive parameter value for our DALE estimation method.

\subsection{Predictions for the alignment of galaxies}
We can expect that single galaxies and galaxies in small, $2 \dots 3$ member groups have the strongest alignment with DALE \citep[see also][]{Hu2006}.
It is sensible to expect more massive mergers in higher density regions and for more massive early-type galaxies. On the 
other hand, less massive late-type galaxies located in lower density regions, in general, should not have encountered significant mergers. This picture has also been suggested by 
simulation studies.

Following the results of simulations by \citet{Dubois2014} and \citet{Welker2014} and their interpretation, we can expect a clear perpendicular alignment signal for massive early-type
galaxies. Meanwhile, predictions for the alignment of late-type galaxies are not so straightforward. \citet{Hahn2010} show that spins of massive disc galaxies should have been aligned
parallel to DALE at $z \sim 0$, but low-mass disc galaxies may have preserved their initial tidally induced orientation. \citet{Dubois2014} and \citet{Welker2014} predict parallel spin
alignment for low-mass blue disc galaxies at $z > 1.2$, but their simulation does not cover our range of redshifts. The theory of constrained tidal torques by \citet{Codis2015b} allows both
parallel and perpendicular tidal spin alignments, depending on the location of a galaxy relative to the saddle point in a filament.
Considering the results by \citet{Aumer2014}, late-type disc galaxies may still have encountered mergers that may have also changed their alignment by $z \sim 0$.

\subsection{Late-type galaxies}
As the perpendicular alignment signal of Sab galaxies comes from considerably bigger samples than the parallel alignment signal of Scd galaxies, we can see that the negligible
perpendicular alignment of the whole late-type sample \citep[by the classification of][]{Tempel2012} is the result of mixing Sab and Scd alignment signals.

The detected perpendicular alignment of spins of Sab galaxies with DALE agrees with the results by \citet{LeeErdogdu2007}. Their results indicate
weakening of the perpendicular alignment of spins of disc galaxies with DALE when later types are considered. However, they did not detect parallel alignment for Scd-type galaxies.
Their results also suggest a stronger perpendicular alignment of spins of disc galaxies with DALE in denser areas. As the last result was obtained using all disc galaxies, it can probably
be explained with a higher S0 content in their sample from higher density regions.

\citet{Faltenbacher2009} did not detect an alignment of blue galaxies; this is probably because they used two-dimensional analysis and did not separate Sab and Scd types.
\subsubsection{Sab galaxies}
We observed that the perpendicular alignment signal of Sab galaxies comes from areas with lower environmental density, single galaxies, and galaxies with low
brightness.
We can expect that the Sab galaxies with described properties have suffered only minor interactions with their neighbours and environment. The existence of the perpendicular alignment
signal in the case of these Sab galaxies can be consistently interpreted as evidence of their primordial tidally induced alignment.

Our result agrees qualitatively with the simulation by \citet{Hahn2010}, although they reported this perpendicular alignment to be clearly detectable at $z \gtrsim 0.5$.
We can confirm the result and interpretation of \citet{Jones2010}, who also reported that spins of less massive spirals in areas with
lower environmental density have a dominantly perpendicular alignment with the direction of filaments. Our result can be considered more solid, as \citet{Jones2010} used a carefully
selected, but small sample. We can agree also with \citet{Zhang2013} whose results indicate a weak perpendicular alignment of spins of blue galaxies and filaments.
\citet{Zhang2015} reported a perpendicular alignment of spins of disc galaxies and filaments, which is present for both main and satellite galaxies of groups of different mass. This partial
disagreement can be explained, as their results may include signal from S0 galaxies due to different classification, and they use a different randomisation procedure to suppress
selection effects.

As the detected alignment signal of Sab galaxies is weak,
the interactions may still have perturbed the initial alignment. \citet{Bournaud2005} found that minor mergers with the stellar mass ratio of the main and companion galaxies larger
than $10:1$ result in disturbed spiral galaxies. According to \citet{Aumer2014}, gas-rich minor mergers (up to the mass ratio $3:1$) as well as misaligned gas infall (relative to the
initial disc) can build up the bulge of disc galaxies and change their spin orientation.
Simulations by \citet{Welker2014} also confirm that minor mergers can change the spin orientation of a galaxy considerably, through converting the orbital angular momentum into spin
of the merger remnant. Considering (environmental) properties of Sab galaxies with perpendicular alignment, we can expect that gas infall to filaments may have perturbed their primordial
orientation and built up the morphology, while being not strong enough to erase or reverse the alignment completely.

It is sensible to expect that interactions with neighbours and environment shuffled the original alignment of Sab galaxies even more in denser areas and groups, and for galaxies
with a higher luminosity. As we did not find preferred alignment in the case of Sab galaxies with these properties, we can propose that minor mergers may have dominated gas infall to
filaments in building up the final angular momentum and morphology. Minor
mergers need not be presumably as collimated as more massive mergers or gas infall into the LSS. We do not see perpendicular alignment of Sab galaxies as a result of more massive mergers
available in denser areas and in groups because the result is probably not a Sab galaxy.

\subsubsection{Scd galaxies}
We observed that the bulk of the parallel alignment signal of Scd galaxies comes from higher density regions (within their population), from $2 \dots 3$ member groups and
from galaxies of low brightness. We can understand the critical dark halo mass $\sim 10^{12} M_{\mathrm{\odot}}$ estimated by \citet{Codis2015b} at $z \sim 0$, as the maximum parallel
tidal alignment signal should not be related to less massive spiral galaxies and the detected alignment of Scd galaxies is not of tidal origin.
Our result supports the idea that infall of matter to filaments plays a key
role in the evolution of Scd galaxies and explains the observed signal. We can expect that there has been more gas to accrete and a faster dynamical evolution should be
possible in higher density regions \citep[see][]{Einasto2005}. We can also expect that galaxies in small groups may accrete stronger gas flows than single galaxies
owing to a stronger gravitational pull.
Meanwhile, spins of galaxies with lower luminosity (mass) can be more easily changed, and galaxies with lower luminosity have presumably suffered less from mergers.
Our results agree with \citet{Aumer2014}, who reported that infall of gas with misaligned angular momentum (relative to the initial disc) can build up the bulge and change the
orientation of the spin axis of a disc galaxy. This leads us to propose that the detected alignment signal of Scd galaxies may be largely caused by
spin flip from a tidally induced perpendicular alignment to a parallel alignment that is determined by matter infall.

\citet{Cebrian2014} observed that $z<0.12$ late-type galaxies in lower density regions are larger for a given mass than their counterparts in higher
density regions. Their result is consistent with the bulge growth scenario by \citet{Aumer2014} and supports our interpretation as well. \citet{Cebrian2014} found also that
the scatter of the stellar mass-size relation for low-mass late-type galaxies is significantly larger in areas with lower density than in higher density regions. Their result
tells us that the low redshift sample of late-type galaxies must have encountered mergers and/or misaligned gas infall not only in high density regions, but
also in sparser areas.

Our result agrees with \citet{Dubois2014} and \citet{Welker2014}, who predict parallel alignment already occurs for low-mass blue disc galaxies at $z > 1.2$.
We cannot confirm the more detailed prediction by \citet{Hahn2010} and observational results
by \citet{Tempel2013a} who showed that spins of luminous spiral galaxies have pronounced parallel alignment with filaments at $z \sim 0$. The contradiction with the result of
\citet{Tempel2013a}
can be explained by the different morphological classifications used. We show that the perpendicular alignment signal of less luminous Sab galaxies slightly outweighs the parallel
alignment signal of Scd galaxies if these samples are mixed together. The negligible residual perpendicular signal was not noticed by \citet{Tempel2013a},
which is probably because they used a much smaller sample size. If they filtered out the Sab signal by choosing bright late-type galaxies, the parallel alignment of Scd galaxies might become
detectable.

We did not find a sample of single low-mass Scd galaxies with fossil perpendicular alignment in areas with low environmental density.
We can speculate that the infall of matter to filaments may have been relatively more intensive in the case of Scd galaxies and may
have perturbed primordial spins of single low-mass Scd galaxies more than those of corresponding Sab galaxies, but the infall may still have been insufficient to produce
detectable parallel alignment.

Our interpretation does not exclude mergers with Scd galaxies. However, it seems highly probable that we need a steady infall of matter to a
filament that dominates merger events for a sufficiently long period of time to produce a Scd galaxy in the particular location. 

\subsection{Early-type galaxies}
The detected perpendicular alignment of spins (or minor axes) of early-type galaxies with DALE agrees well with previous studies
\citep[e.g.][]{LeeErdogdu2007,Faltenbacher2009,Li2013,Tempel2013a,Tempel2013b,Zhang2013,Pahwa2016}.
There is also agreement with \citet{Wang2009} for main galaxies of groups.

\citet{Camelio2015} showed that tidal forces at the present epoch cannot (re)build the observed ellipticities and alignment of elliptical galaxies. The present alignment of
elliptical galaxies is either a fossil signal from the era of gravitational collapse of DM haloes or is caused by later mergers.
Roughly all of our results for early-type galaxies can be explained with an unlikely hypothesis that we observe the fossil tidal alignment under the condition of
suppressed matter infall and no remarkable mergers. However, the pure fossil alignment hypothesis cannot explain why the satellite S0 galaxies have no preferred alignment
in $4 \dots 10$ member groups and the satellite E galaxies show perpendicular alignment in these groups, while single S0 galaxies show an even stronger perpendicular alignment
signal than single E galaxies. We conclude at the end of Sect.~\ref{egal} that major mergers along DALE should be considered to explain the alignment of satellite
E galaxies in $4 \dots 10$ member groups.
The discussion below tries to connect our alignment results with various simulation studies that show significant role of mergers in the evolution of early-type galaxies.
\subsubsection{S0 galaxies}
The observed perpendicular alignment signal of S0 galaxies does not depend dramatically on the environmental density and does not vanish in small groups and with higher luminosity,
as in the case of the fossil tidal alignment signal of Sab galaxies.
Simulation studies suggest that S0 galaxies may have encountered mergers that could have erased their memory of initial orientation.
Hence, we can consider that mergers are responsible for the present alignment of S0 galaxies, at least in higher density regions, for
luminous galaxies, and in small groups. As the alignment signal
of S0 galaxies is strong, we can think of more massive mergers than in the case of late-type galaxies. We can expect that more massive mergers align better with DALE,
and they have a stronger effect on the orientation of the spin axis and the morphology of the merger remnant.

Observational results of \citet{Bruce2014} show that there is a continuous growth of bulges of massive galaxies at $z=3$ to $z=1$, and the fraction of pure bulges (E galaxies) is very
low by $z=1$. They report that massive early-type galaxies at $z=1$ are similar to the present S0 galaxies. They propose that the model of violent disc instabilities (VDI)
can be consistent with their result of gradual bulge growth.
As discussed by \citet{Bruce2014}, we cannot explain the gradual growth of bulges under the assumption of major mergers.
On the other hand, the scenario of \citet{Hopkins2008b}, in which galaxies need to grow through higher mass ratio mergers before the
major mergers become effective, together with the bulge growth scenario by \citet{Aumer2014}, can explain both our results and the \citet{Bruce2014} results.
\citet{Bournaud2005} have shown that relaxed remnants of mergers of intermediate stellar mass ratio $4.5:1$ to $10:1$ can be considered candidates of S0 galaxies.
In light of that, processes other than major mergers should dominate quenching of star formation in the population of massive galaxies at $z=3$ to $z=1$. AGN feedback
\citep[see e.g.][]{Genel2014} and AGB heating \citep[][]{Conroy2015} can be considered. However, our results alone do not exclude major mergers as a possible path to S0 galaxy
formation \citep[see also][]{Querejeta2015}.

On the other hand, we found that the satellite S0 galaxies have no preferred alignment in $4 \dots 10$ member groups.
We can speculate that they may not be massive enough to allow effective mergers along the present DALE in these groups, while mergers within or between their subhaloes caused the
S0 morphology.

Considering alignment properties, we can see that single and small group S0 galaxies have more in common with E galaxies than with spiral galaxies. It is debatable whether
one can explain the observed strong perpendicular alignment signal of single and small group S0 galaxies, while considering them the result of the isolated evolution of spiral
galaxies with fossil tidal alignment.

\subsubsection{Elliptical galaxies}\label{egal}
We observed that the strong perpendicular alignment signal of E galaxies gets even stronger in denser environments and with higher $r$-band luminosity. The perpendicular alignment
signal was clearly detected for single galaxies and for galaxies in up to ten member groups. Our results agree with several earlier studies that report stronger alignment of more
massive early-type galaxies \citep[see][]{Faltenbacher2009,Okumura2009,Li2013,Zhang2013,Huang2016}.

Simulations by \citet{Bournaud2005} show that E galaxies can be products of major mergers of disc galaxies with stellar mass ratios $1:1$ to $3:1$. \citet{Hopkins2008a,Hopkins2008b}
state that red E galaxies are produced by major mergers with mass ratios $3:1$ or smaller; we thereby obtain rapid star formation and depletion of gas. Observational results
\citep[see e.g.][]{Luo2014} promote mergers as the main cause of starburst. Our alignment analysis agrees
well with the major merger scenario, as the most massive mergers should take place along DALE.

On the other hand, \citet{Bournaud2005} also propose an alternative scenario with repeated mergers with higher mass ratio to produce E galaxies.
\citet{Hopkins2008b} explain that galaxies grow through higher mass ratio mergers before the major mergers become effective, and galaxies in denser regions reach their
major merger regime faster than galaxies in sparser areas. We can consider an alternative E galaxy formation process involving less massive and less collimated but
repeated mergers would have larger effect in lower density regions. This can explain weaker alignment of E galaxies with lower luminosity and in environment with lower density.

We found that alignment of E galaxies weakens notably in $> 10$ member groups.
\citet{Lietzen2012} reported that growth of the fraction of passive early-type galaxies stops at the group richness $\approx 10$. Simulations by \citet{Hopkins2008a}
show that major mergers with a central galaxy are efficient only in small groups, while minor mergers with a central galaxy act in both small and large groups and
major mergers between satellite galaxies are suppressed in large groups.
Therefore we can expect much fewer major mergers along DALE in the case of $> 10$ member groups. The alternative scenario of producing E galaxies via subsequent
intermediate-mass ratio mergers between group members may be relevant in this case. Although this explains both our result and the \citet{Lietzen2012} result, poorly defined
DALE estimation due to possible crossing filaments at the locations of clusters can be also considered here.

Simulations \citep[see e.g.][]{Tenneti2015} predict alignment of the shapes of satellite galaxies in the direction of the main galaxy of a group (radial alignment). Several studies
indicate that galaxy groups and clusters are generally elongated along DALE \citep[see e.g.][]{West1989,Plionis1994,Wang2009,Tempel2015b}. The detected perpendicular alignment
signal of the satellite E galaxies in $4 \dots 10$~member groups could be interpreted as an indirect confirmation of the radial alignment. However, the hypothesis of tidal radial
alignment cannot
explain why the detected alignment signal practically vanishes in $> 10$ member groups, or, as alignment signals of E and S0 galaxies are generally similar, why we detected no alignment
of satellite S0 galaxies in $4 \dots 10$~member groups. Mergers seem to be a more accurate explanation, if we consider that major mergers along DALE may not be suppressed in the case of
satellite E galaxies in $4 \dots 10$~member groups, but are suppressed for satellite S0 galaxies in these groups and for virtually all galaxies in $> 10$ member groups.

\section{Conclusions}\label{concl_sect}

We developed a new method for estimating the local direction of a web-like spatial structure. We used the method to study alignment of galaxies of different types in SDSS-DR8.
Our main results are the following.

\emph{The method.} Our method for estimating DALE can be used to study alignment of galaxies. Our approach excludes very few galaxies from the analysis, if
     compared to more refined alignment studies. This allows us to use more of the available orientational data and to suppress effects due to small sample sizes. The alignment detection
     sensitivity of our method is well comparable to the more complicated methods, if applied to large data sets.
     For an example, the DALE and directions of filaments extracted with the Bisous model are mostly parallel, as can be seen in Appendix~\ref{Bisous_DALE_sect}.

\emph{Sab and Scd galaxies.} Single Sab galaxies, Sab galaxies with low luminosity, and Sab galaxies in low density environments have their spin axes aligned preferentially perpendicular
     to DALE.
     The major part of the parallel alignment signal between the spins of Scd galaxies and DALE comes from a dense environment, from $2 \dots 3$ member groups, and
     from galaxies with low luminosity.

\emph{S0 and elliptical galaxies.} The observed perpendicular alignment of the spins of S0 galaxies and DALE does not depend strongly on the density of environment nor luminosity.
     It was detected for single and $2 \dots 3$ member group galaxies and for main galaxies of $4 \dots 10$ member groups.
     The strength of the S0 alignment signal is mostly comparable to the strength of the signal for E galaxies.
     Minor axes of E galaxies show pronounced perpendicular alignment with DALE. Their alignment increases with environmental density and luminosity.
     The perpendicular alignment was clearly detected for single galaxies and for members of groups with richness up to $10$.

Our results confirm the existence of a population of low redshift Sab galaxies that still have an initial tidally induced alignment. We propose that minor mergers may have dominated matter
infall to filaments in the process of destroying the primordial Sab alignment in dense areas, in groups, and for high luminosity galaxies.
Matter infall to filaments and accretion of this matter is needed to produce a Scd galaxy.

When comparing results of simulations and our study, we can explain alignment of S0 galaxies with mergers along DALE. These mergers should be more massive
than in the case of late-type galaxies.
The alignment of E galaxies can be explained with major mergers along DALE. However, less massive (but repeated) mergers may be important in low density areas and for galaxies with
low luminosity.

We conclude that mergers with different mass ratios can have a significant role in the evolution of Sab, S0, and E galaxies. We cannot exclude mergers in the case of Scd
galaxies, but the detected parallel alignment of their spins and DALE can be explained most likely with infall of matter to filaments.
We cannot exclude that the fossil tidal alignment complements the alignment signal of early-type galaxies.

A question that needs more observational evidence is, how are merger/accretion processes along and perpendicular to filaments balanced at different epochs.
A comparative alignment analysis of low and higher redshift galaxies may help us understand the contribution of tidal alignment to the final alignment of early-type galaxies at the
present epoch.
More observational alignment studies and higher redshift samples from deeper observations are needed to properly address these issues.

\begin{acknowledgements}
We thank the referee for valuable suggestions that helped improve the presentation of this work and to make the concept of DALE more solid.
We are grateful to K.~Annuk and I.~Suhhonenko for ensuring reliable computational resources.

We are pleased to thank the SDSS Team for the publicly
available data releases.
Funding for SDSS-III has been provided by the Alfred P. Sloan Foundation, the Participating Institutions, the National Science Foundation, and the U.S. Department of Energy Office 
of Science. The SDSS-III web site is http://www.sdss3.org/.

SDSS-III is managed by the Astrophysical Research Consortium for the Participating Institutions of the SDSS-III Collaboration including the University of Arizona, the Brazilian 
Participation Group, Brookhaven National Laboratory, Carnegie Mellon University, University of Florida, the French Participation Group, the German Participation Group, Harvard 
University, the Instituto de Astrofisica de Canarias, the Michigan State/Notre Dame/JINA Participation Group, Johns Hopkins University, Lawrence Berkeley National Laboratory, Max 
Planck Institute for Astrophysics, Max Planck Institute for Extraterrestrial Physics, New Mexico State University, New York University, Ohio State University, Pennsylvania State 
University, University of Portsmouth, Princeton University, the Spanish Participation Group, University of Tokyo, University of Utah, Vanderbilt University, University of Virginia, 
University of Washington, and Yale University.

This work was supported by research project SF0060030s08 and institutional research funding IUT26-2, IUT40-1, IUT40-2 of the Estonian Ministry of Education and Research,
and by the European Structural Funds grants TK120, TK133.
\end{acknowledgements}






\bibliographystyle{aa} 
\bibliography{gal_fil_corr} 

\begin{appendix}
\section{Comparison of DALE with the directions of filaments from the Bisous model}\label{Bisous_DALE_sect}

Here we compare the orientations of DALE vectors with the directions of filaments detected in the galaxy distribution using the Bisous model \citep[see][]{Stoica2005,Tempel2016}.
The filaments and their directions were adopted from the catalogue of filaments from SDSS-DR8 compiled by \citet{Tempel2014b}. The radius and length of the elementary cylinder
used to locate the filaments are $0.5$~$h^{-1}\mathrm{Mpc}$ and $3 \dots 5$~$h^{-1}\mathrm{Mpc}$. The points defining the spines of the filaments are separated by
$\approx 0.5$~$h^{-1}\mathrm{Mpc}$.

We selected, for comparison, the locations of galaxies with estimated DALE that are closer than $0.5$~$h^{-1}\mathrm{Mpc}$ to a point of the spine of a filament.
The border area of the survey was excluded as in Sect.~\ref{some_problems}. Every selected DALE vector was compared to the direction of a filament estimated in the closest spine point.

\begin{figure}
\centering
\includegraphics[width=0.49\hsize,clip]{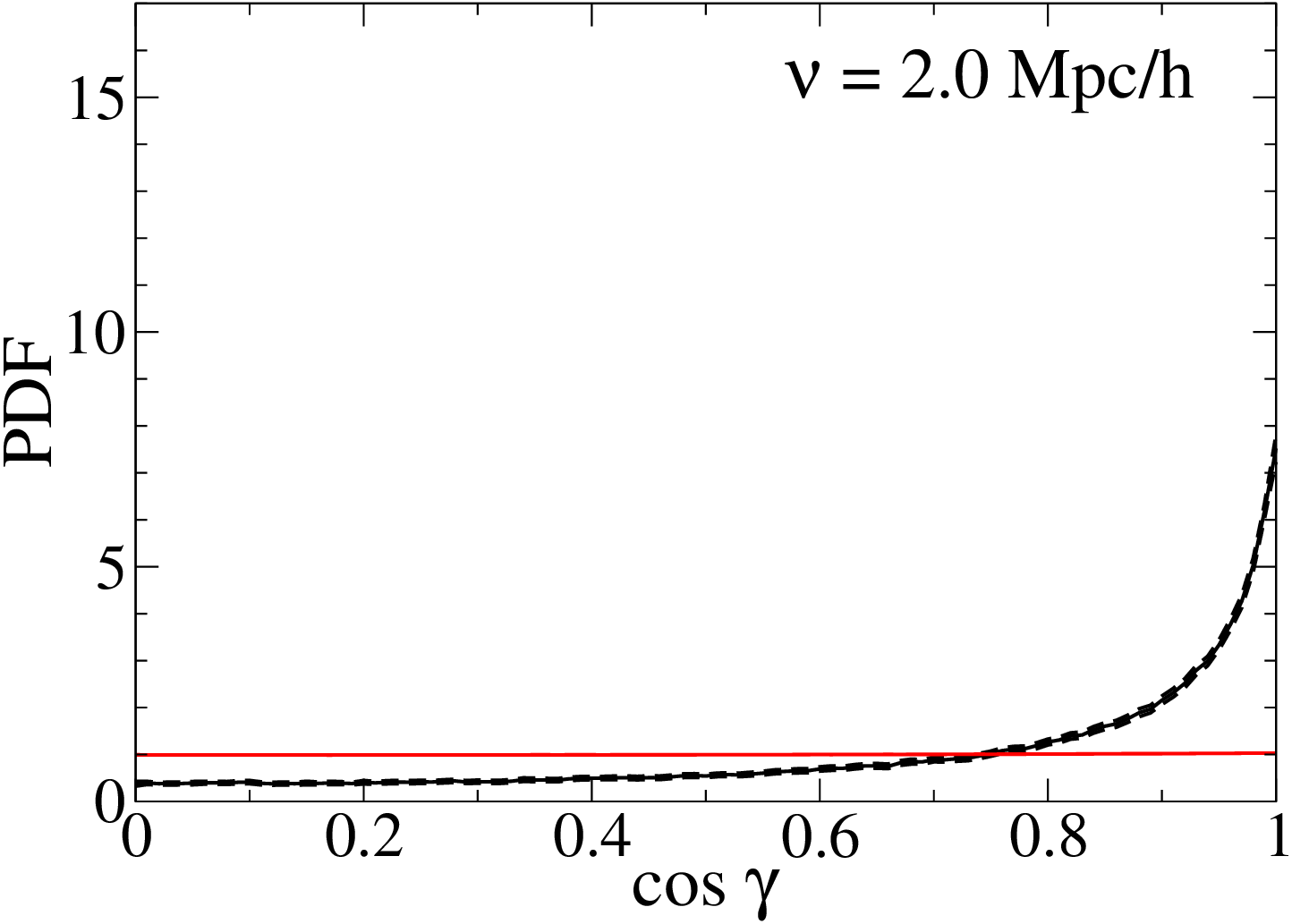}
\includegraphics[width=0.49\hsize,clip]{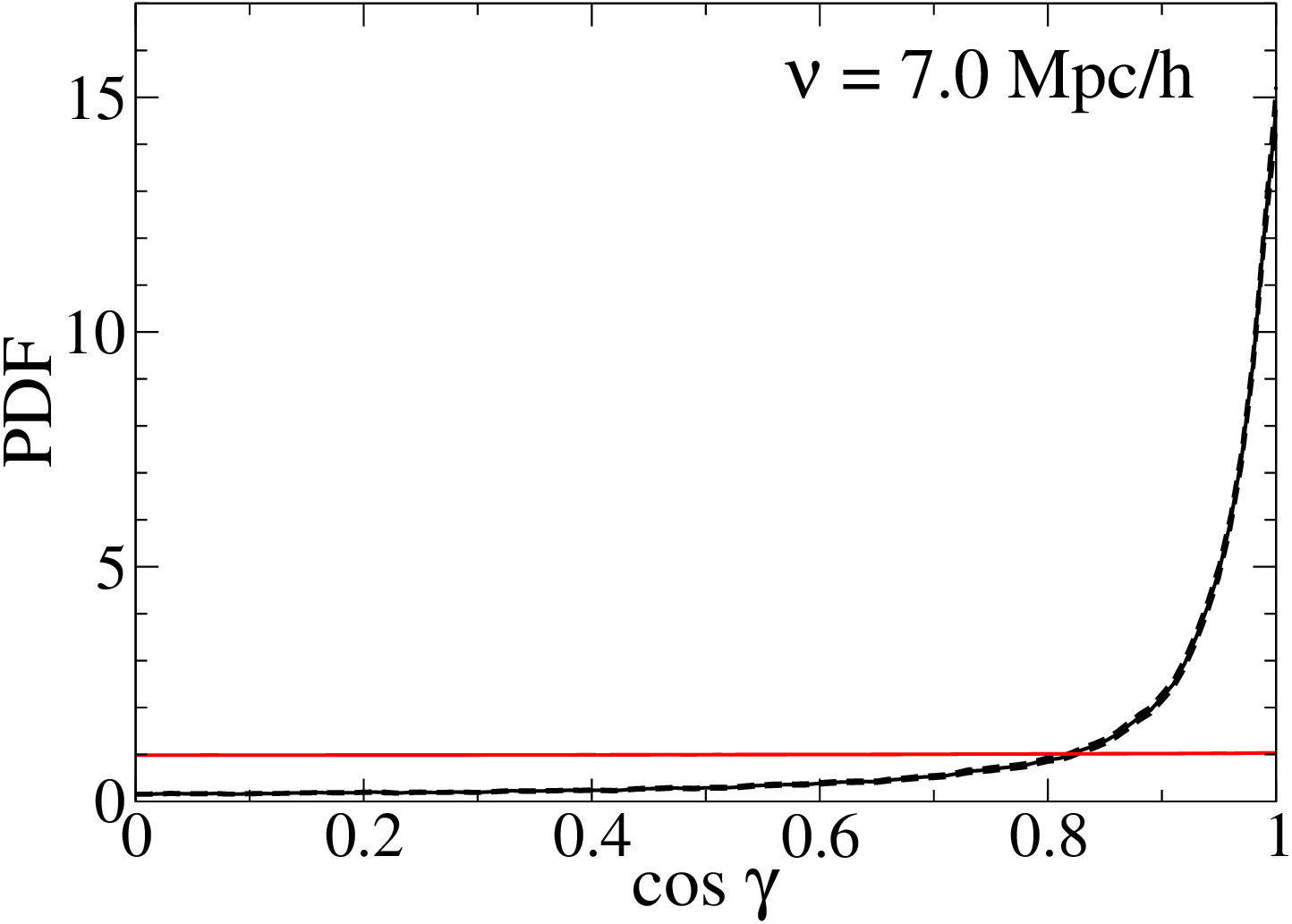}

\includegraphics[width=0.49\hsize,clip]{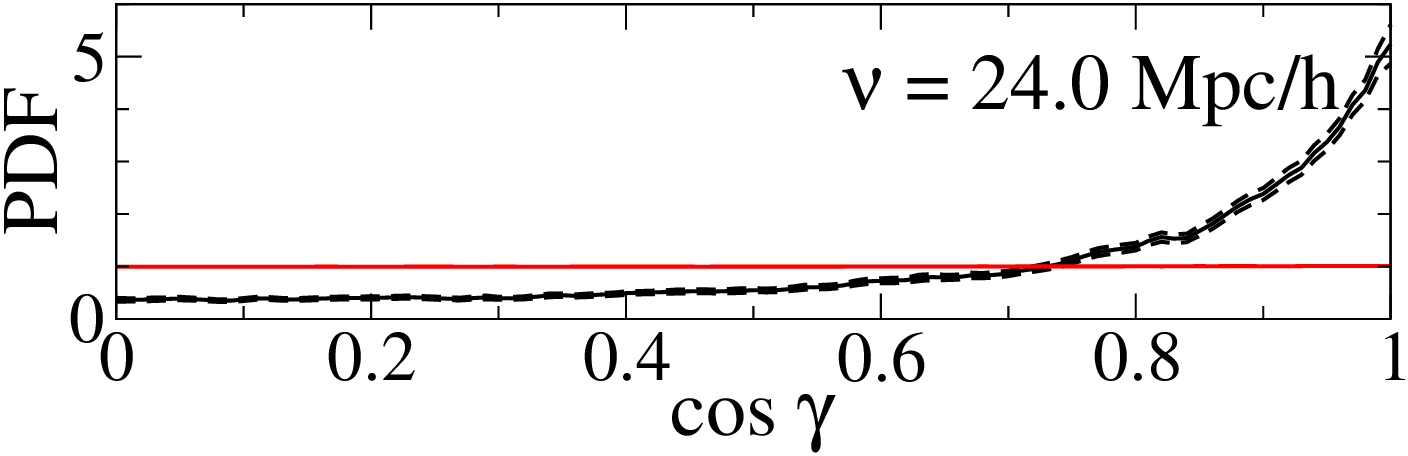}
\includegraphics[width=0.49\hsize,clip]{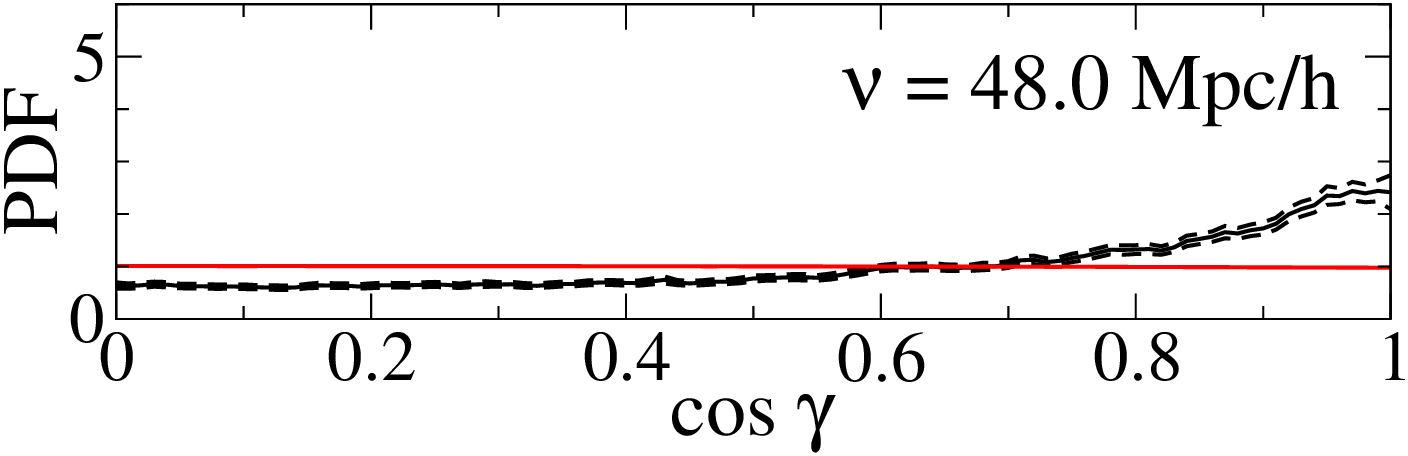}
\caption{PDFs of the cosine of the angle $\gamma$ between DALE and the directions of filaments from the Bisous model. The neighbourhood radius
$\nu = 2.0, 7.0, 24.0, 48.0$~$h^{-1}\mathrm{Mpc}$.
The observed signal with its $95$\% confidence interval is depicted in black; the selection effect is depicted in red. Excess of high (and deficiency of low) values of observed
$\cos \gamma$ compared to the PDF of the selection effect indicates parallel alignment of DALE relative to directions of filaments.}
\label{Bisous_DALE_PDF}
\end{figure}

The comparison between DALE and the directions of filaments extracted with the Bisous model was carried out for different values of the neighbourhood radius:
$\nu = 2.0$, $7.0$, $12.0$, $24.0$, and $48.0$~$h^{-1}\mathrm{Mpc}$. We converted the Cartesian data of the filaments catalogue into right ascension, declination and distance of
the points of the spines, and into inclination and position angles of the direction vectors of the filaments. The algorithms developed in Sect.~\ref{method_desc} were applied for
comparison, keeping in mind that there is no sign ambiguity of $\hat{L}_{\mathrm{z}}$ in the case of filaments. One can find the results in Fig.~\ref{Bisous_DALE_PDF} and
Table~\ref{Bisous_DALE_median}. We can see that in the filaments the majority of DALE vectors are rather parallel with the direction of a filament. The parallel alignment is strongest
for $\nu = 7.0$~$h^{-1}\mathrm{Mpc}$, but it is remarkable for all tested values of $\nu$. As the selection effects are negligible
here, the same can be concluded from the median values of $\cos \gamma$ given in Table~\ref{Bisous_DALE_median}.
\begin{table}[htb]
\caption{Median values of the cosine of the angle $\gamma$ between DALE and the directions of filaments from the Bisous model.
$N_{c}$ is the number of galaxies with estimated DALE used for the comparison at a given value of $\nu$.}
\label{Bisous_DALE_median}
\centering
\begin{tabular}{r r r}
\hline\hline
   $\nu$      & $N_{c}$      & $\mathrm{median} \{ \cos \gamma \}$ \\   
   ($h^{-1}\mathrm{Mpc}$) &  &		\\
\hline
   $2.0$      & $110900$     & $0.832$ \\
   $7.0$      & $104460$     & $0.938$ \\
   $12.0$     & $96392$      & $0.918$ \\
   $24.0$     & $78781$      & $0.823$ \\
   $48.0$     & $51598$      & $0.691$ \\
\hline
\end{tabular}
\end{table}

We can conclude that DALE can be a method of choice if one wants to estimate directions in some web-like spatial structure.
\end{appendix}

\end{document}